\documentclass{jfm}
\usepackage{graphicx}
\usepackage{natbib}
\usepackage{amsmath,amssymb,stmaryrd,array}
\usepackage{epsfig,color,soul,rotating,overpic}
\usepackage{epstopdf}
\usepackage[section]{placeins}
\usepackage{array}
\usepackage{caption}
\usepackage{subcaption}
\usepackage{wrapfig}
\usepackage{color}
\definecolor{red}{rgb}{0.8500, 0.1250, 0.0480} 

\title[Networked oscillator based modeling and control of unsteady wakes]
{Networked oscillator based modeling and control of unsteady wakes} 
\author[A. G. Nair, S. L. Brunton and K. Taira]
{
Aditya G.~Nair$^1$\thanks{Email address for correspondence: agn13@my.fsu.edu},
Steven L. Brunton$^2$
and
Kunihiko Taira$^1$
}
\affiliation{$^1$Department of Mechanical Engineering, Florida State University, Tallahassee, FL 32310, USA \\
$^2$Department of Mechanical Engineering, University of Washington, Seattle, WA 98195, USA}

\pubyear{2017}
\volume{???}
\pagerange{???--???}
\date{?; revised ?; accepted ?. - To be entered by editorial office}
\begin{document}

\maketitle

 
\begin{abstract}
A networked oscillator based analysis is performed for periodic bluff body flows to examine and control the transfer of kinetic energy. Spatial modes extracted from the flow field with corresponding amplitudes form a set of oscillators describing unsteady fluctuations. These oscillators are connected through a network that captures the energy exchanges amongst them. To extract the network of interactions among oscillators, amplitude and phase perturbations are impulsively introduced to the oscillators and the ensuing dynamics are analyzed. Using linear regression techniques, a networked oscillator model is constructed that reveals energy transfers and phase interactions among the modes. The model captures the nonlinear interactions amongst the modal oscillators through a linear approximation. A large collection of system responses are aggregated into a network model that captures interactions for general perturbations. The networked oscillator model describes the modal perturbation dynamics better than the empirical Galerkin reduced-order models. A model-based feedback controller is then designed to suppress modal amplitudes and the resulting wake unsteadiness leading to drag reduction. The strength  of the proposed approach is demonstrated for a canonical example of two-dimensional unsteady flow over a circular cylinder. The present formulation enables the characterization of modal interactions to control fundamental energy transfers in unsteady vortical flows.
\end{abstract}



\section{Introduction}
\label{sec:intro}

Oscillations play an important role in many physical and biological systems. These oscillations often result from a set of self-sustaining (autonomous) entities called oscillators. Biological oscillators, including neurons and heart cells are integral to the various rhythms and regulatory systems of the human body. Such collective rhythms arise from the coupling of multiple oscillators with the physics encapsulated by the transfer of energy between them \citep{strogatz2004sync}. There has been a rich history of studies on the collective dynamics of oscillators, in particular by \cite{kuramoto1984chemical} and \cite{strogatz2014nonlinear}.  The foundational work laid out by \citet{kuramoto1975self} elegantly describes the interactive phase dynamics between oscillators. Mutual synchronization of a system occurs when interacting oscillators affect their phases so as to spontaneously lock on to a particular frequency or phase \citep{pikovsky2003synchronization}. In the works of \cite{aizawa1976synergetic} and \cite{mirollo1990amplitude}, the oscillator phase interactions are generalized to incorporate amplitude variation effects. Amplitude and phase coupling between oscillators result in a variety of interesting physical phenomena.

Unsteady fluid flows are often characterized by temporal oscillations. In flows past bluff bodies, oscillatory behavior of the flows is exhibited through shedding of coherent vortices observed in the wake. Such periodic shedding generates unsteady forces on the body which can lead to detrimental increase in drag associated structural fatigue due to the emergence of flow-induced vibrations \citep{williamson2004vortex,sarpkaya2004critical}. In the work of \cite{roshkodrag}, the relationship between form drag and vortex shedding was explored in detail. It was demonstrated that unsteady force oscillations and drag can be reduced by mitigating the wake unsteadiness. Since then, a myriad of studies using active and passive flow control strategies have focussed on controlling bluff-body wake vortex shedding and the resulting unsteady forces, summarized in a review by \cite{choi2008control}. Although there have been tremendous breakthroughs in applying flow control for drag reduction, only a few studies make use of the fundamental energy transfer mechanisms and interactions in unsteady fluid flows and controlling the flow unsteadiness therein.

The wake unsteadiness stems from the inherent vortical oscillations. Using modal decomposition techniques, oscillations embedded in fluid flows can be extracted naturally as spatial structures (modes) and their associated temporal weights \citep{taira2017modal}. For time-periodic flows, individual coherent structures are described by conjugate mode pairs extracted from these modal decomposition techniques. These mode pairs can be viewed as a set of modal oscillators exhibiting periodic fluctuations. In particular, proper orthogonal decomposition (POD) \citep{Sirovich:QAM87, aubry1988dynamics, berkooz1993proper} and dynamic mode decomposition (DMD) \citep{Schmid:JFM10, Rowley:JFM09, kutz2016dynamic} techniques can decompose snapshot flow field data into modal oscillators based on energy and dynamics of the flow, respectively. The interactions between these modal oscillators govern the behavior of unsteady fluid flows. The general behavior of nonlinear flows can also be described by spectral analysis of the linear, infinite-dimensional Koopman operator which yields Koopman modes \citep{Mezic2005nd, Rowley:JFM09, mezic2013analysis}. 

The interactions between the modes in unsteady fluid flows can be encapsulated by reduced-order models which may dramatically reduce the computational complexity to analyze fluid flows of interest \citep{aubry1988dynamics,Noack:JFM03}. The projection of the Navier--Stokes equations onto the modal basis results in an empirical Galerkin formulation \citep{Holmes2012}. The empirical Galerkin models contain linear and quadratic nonlinear terms which govern the dynamics of the modes. However, strong nonlinear interactions between modes in unsteady fluid flow pose a significant challenge in the accuracy of reduced-order models \citep{lassila2014model, Brunton2015amr}. The study by \cite{sapsis2013attractor} uses dynamically orthogonal equations which relates the low-dimensionality of fluid flow attractors with energy exchanges between the mean and dynamical components of the flow. In the current study, we examine how modal oscillators in unsteady fluid flows interact to distribute energy in non-equilibrium conditions. This involves a perturbation-response based approach to characterize the interactions and extract both phase and amplitude coupling between modal oscillators in a fluid flow. In this approach, perturbations are introduced to the modal oscillators and the ensuing energy transfer dynamics are analyzed using a  graph-theoretic framework.

A graph $\mathcal{\boldsymbol{G}} = \{\mathcal{V},\mathcal{E},\mathcal{W}\}$ consists of a set of nodes $\mathcal{V}$ connected by edges $\mathcal{E}$ with associated edge weights $\mathcal{W}$ \citep{Newman10, Nair:JFM15}. The network nodes form the quantities of interest with the interactions between them as edges. Such a simplistic viewpoint of analyzing collections of interactions in network science has had far-reaching socio-economic as well as scientific impacts \citep{BarabasiNS16}. Network analysis is primarily concerned with extracting the interactions between quantities of interest \citep{Bollobas98, Newman10} and has found widespread applications in social sciences \citep{otte2002social}, biological sciences \citep{zhu2007getting} and many other fields \citep{dorogovtsev2013evolution}. The studies of various technological networks like the Internet and World Wide Web have improved the efficiency of global communications tremendously \citep{Newman10}. The Human Connectome project \citep{wedeen2005mapping,hagmann2007mapping,sporns2011human} provides access to big neural data to explore connectivity pathways in the brain and has led to a wealth of new discoveries. In epidemiology, network analysis has aided in forecasting epidemics and designing appropriate containment and control measures \citep{Salathe:PLOSCB10, Robinson:TPB12, dudas2016virus}. 
 
The application of network analysis has recently been extended to represent vortical interactions in fluid flows \citep{Nair:JFM15}. The network-theoretic framework is comprised of discrete point vortices as nodes and interactions between them as edges. The network representation allows for the utilization of techniques such as spectral sparsification  \citep{Spielman:SIAMJC11b} to identify key vortical interactions and maintain the spectral properties of such interactions. The application of such spectral sparsifiers has led to the development of {\it sparsified-dynamics} models, which preserve the invariants of discrete vortex dynamics. These models are capable of tracking bulk motion of vortical clusters in a computationally effective manner. Moreover, the extraction of the vortical network structure of turbulent flows has revealed the scale-free network property of decaying two-dimensional isotropic turbulence \citep{taira2016network}. The resulting framework enables the assessment of the resilience of turbulent flow structures. As such, network analysis has demonstrated the ability to capture fundamental vortical interactions in fluid flows, leveraging graph-theoretic principles. In the present work, we extend network analysis to describe and control {\it modal interactions in fluid flows}, by casting the fluid flow in terms of a networked oscillator system. In this approach, we view the modal oscillators as nodes and interactions between them as edges, highlighting complex energy transfer dynamics. 

We utilize modal decomposition techniques in conjuction with coupled oscillator models to capture the interactive physics involved in unsteady fluid flows. Our objective in this work is three-fold; (1) characterize the nonlinear energy transfer between modes and construct a networked dynamics model for amplitude and phase perturbations in unsteady fluid flows, (2) describe interactive dynamics between modes from a network-theoretic perspective and (3)  control the perturbations with respect to the limit cycle state of periodic flows as well as the full state itself. To accomplish these goals, mode pairs describing individual coherent structures in baseline (unforced and unperturbed) time-periodic flows are considered as a set of oscillators. A graph representation with the oscillators as nodes and coupling interactions between them as edges is constructed to form a set of networked oscillators.

Perturbations are introduced in the unsteady fluid flow to examine the associated energy transfer among the modes which are tracked over the networked oscillator based framework. A {\it networked oscillator} model herein describes the temporal dynamics of modes in unsteady fluid flow with a network structure embedded in it. The associated network structure is extracted using a linear regression procedure \citep{kutner2004applied}. With the network dynamics model established, we are able to study the amplitude and phase dynamics of these perturbations on the modal interaction network. Thus, we arrive at a reduced-order network model using modal oscillators that highlight interaction dynamics among the modes which can subsequently be used for the control of the flow unsteadiness. We design flow control strategies to suppress modal fluctuations using the network based insights that consequently leads to drag reduction. In what follows, we first lay the theoretical foundation of this work in \S{\ref{sec:formulation}}. We then demonstrate the strength of our approach using a canonical example of two-dimensional cylinder flow in \S{\ref{sec:cf}}. We end the paper with concluding remarks in \S{\ref{sec:cr}}.


\section{Formulation}
\label{sec:formulation}

In this study, we are interested in modeling and controlling the unsteady fluctuations of flows with well-identified coherent structures captured as spatial modes. Various modal decompositions techniques can be used to extract coherent structures in fluid flows as mentioned in \S \ref{sec:intro}. In the present work, we are interested in capturing the energy transfer dynamics between those modes. In this regard, POD provides an optimal set of minimum number of modes capturing maximum energy content of the dynamical system. We first describe the procedure to extract these modes in unsteady fluid flows using the POD technique. 

\subsection{Oscillator representation}
\label{subsec:or}

We consider a baseline case corresponding to time-periodic flow without any forcing or perturbations introduced in the Navier--Stokes equations. To perform the POD analysis, we use the method of snapshots \citep{Sirovich:QAM87}. Using this approach, the unsteady velocity field $\boldsymbol{u}$ can be approximated by a finite expansion in terms of a mean (time-averaged) velocity field $\bar{\boldsymbol{u}}$ and $N$ orthonormal spatial POD modes $\boldsymbol{\phi}_j^{\boldsymbol{u}}$ as
\begin{equation}
   \boldsymbol{u}(\boldsymbol{x},t)  \approx \bar{\boldsymbol{u}}(\boldsymbol{x}) + \sum_{j=1}^N a_j(t) \boldsymbol{\phi}_j^{\boldsymbol{u}}(\boldsymbol{x}), 
   \label{modal_decomp}
\end{equation}
where  $a_j(t) = \left<\boldsymbol{u}(\boldsymbol{x},t)-\bar{\boldsymbol{u}}(\boldsymbol{x}),\boldsymbol{\phi}_{j}^{\boldsymbol{u}}(\boldsymbol{x})\right>$ are the temporal coefficients and $\left<\cdot,\cdot\right>$ denotes the inner product over the computational domain. With $||\boldsymbol{\phi}_{j}^{\boldsymbol{u}}|| = 1$, the kinetic energy of the modes are given by $a^2_j/2$, amounting to a total modal energy of $E = \sum_{j=1}^N a^2_j/2$. As mentioned earlier, we obtain modes in conjugate pairs from POD for time-periodic flows. Each conjugate mode pair, describing periodic coherent structures in the baseline case define an oscillator in our analysis. From a set of $N$ POD modes, we obtain $M=N/2$ such oscillators. These oscillators can also be identified using the procedure described in the work of \cite{sieber2016spectral}. 

The temporal oscillations associated with each oscillator can be described by their coefficients in the complex plane. Conjugate mode pairs $(\boldsymbol{\phi}_{2j-1}^{\boldsymbol{u}},\boldsymbol{\phi}_{2j}^{\boldsymbol{u}})$ with temporal coefficients $(a_{2j-1},a_{2j})$ can be represented in the complex plane as,
\begin{equation}
z_m = a_{2j-1} + i a_{2j} = r_m \exp(i \theta_m),
\label{eq:osc1}
\end{equation}
where $m = \text{I}, \text{II} \dots M$, $j = 1,2 \dots N/2$, $r_m = |z_m|$, and $\theta_m = \angle z_m$. Throughout this work, the oscillators will be numbered in roman numerals, $\{$I, II$~ ...~M\}$ denoted by index $m$ to distinguish from mode numbering, $\{1, 2,~...~N\}$ denoted by index $j$. Each oscillator $m$ is associated with a temporal coefficient defined on the complex plane ($z_m$) which consists of an odd-number mode with coefficient $a_{2j-1}$ and an even-number mode with coefficient $a_{2j}$. That is, modes $1$ and $2$ constitute oscillator I, modes $3$ and $4$ constitute oscillator II, and so on. The temporal coefficient corresponding to the mean flow ($\bar{\boldsymbol{u}}$) is fixed to unity, $z_0 = 1$. Equivalent to Eq. (\ref{modal_decomp}), we can recover the velocity field in terms of oscillators  and their associated temporal coefficients  as
\begin{equation}
\boldsymbol{u}(\boldsymbol{x},t) = \bar{\boldsymbol{u}}(\boldsymbol{x}) + \sum_{m=1}^{M} \left[\Re(z_m(t)) \boldsymbol{\phi}^{\boldsymbol{u}}_{2m-1}(\boldsymbol{x})  + \Im(z_m(t)) \boldsymbol{\phi}^{\boldsymbol{u}}_{2m}(\boldsymbol{x})  \right], 
\label{eq3}
\end{equation}
where $\Re(z_m)$ and $\Im(z_m)$ represent real and imaginary components of the temporal coefficient $z_m$, respectively. 

In the baseline case, oscillators follow a natural limit-cycle, the generalized dynamics of which is described by the Stuart--Landau equation as 
\begin{equation}
\dot{z}^b_m = z^b_m (\lambda_m - |z^b_m|^2 + i \Omega_m^b)
\label{eq4}
\end{equation}
with $z_m^b = r_m^b \exp(i\theta_m^b)$, $\lambda_m = (\overline{r^b_m})^2$ and $\Omega_m^b$ is the oscillator frequency, where $\overline{\cdot}$ denotes the time average. Here, the superscript $()^b$ denotes the baseline case. Based on Eq. (\ref{eq4}), the conjugate mode pairs of each oscillator exchanges energy among themselves to describe the self-sustaining equilibrium behavior of each coherent structure in unsteady fluid flows. Stable limit cycles are a characteristic feature of systems exhibiting such self-sustaining oscillations \citep{strogatz2014nonlinear}. This self-sustaining behavior may not be true for systems with non-periodic or chaotic attractors, such as turbulent flows.

The baseline temporal dynamics of the individual oscillators for periodic flow are independent of each other as given by Eq. (\ref{eq4}) and does not involve any coupling between oscillators. To highlight interactions between oscillators, we introduce perturbations impulsively in DNS to the baseline temporal coefficients of oscillators at $t = t_0$. These perturbations introduced to the baseline shedding state leads to the emergence of nonlinear interactions and energy exchange among the oscillators resulting in fluctuations described by
\begin{equation}
z_m^\prime = a_{2j-1}^\prime + ia_{2j}^\prime  = \epsilon_m^\prime r_m^b \exp(i[\theta_m^b + \theta_m^\prime]),
\label{eq41}
\end{equation}
where $()^\prime$ denotes the perturbation quantity and $\epsilon_m^\prime(t_0)$ and $\theta_m^\prime(t_0)$ are the  amplitude and phase perturbations for the $m$th oscillator, respectively. It should  be noted that $\epsilon_m^\prime$ is normalized by the baseline amplitude of each oscillator. 

\begin{figure}
   \begin{center}
    \begin{tabular}{c} 
          \begin{overpic}[width=4.25in]{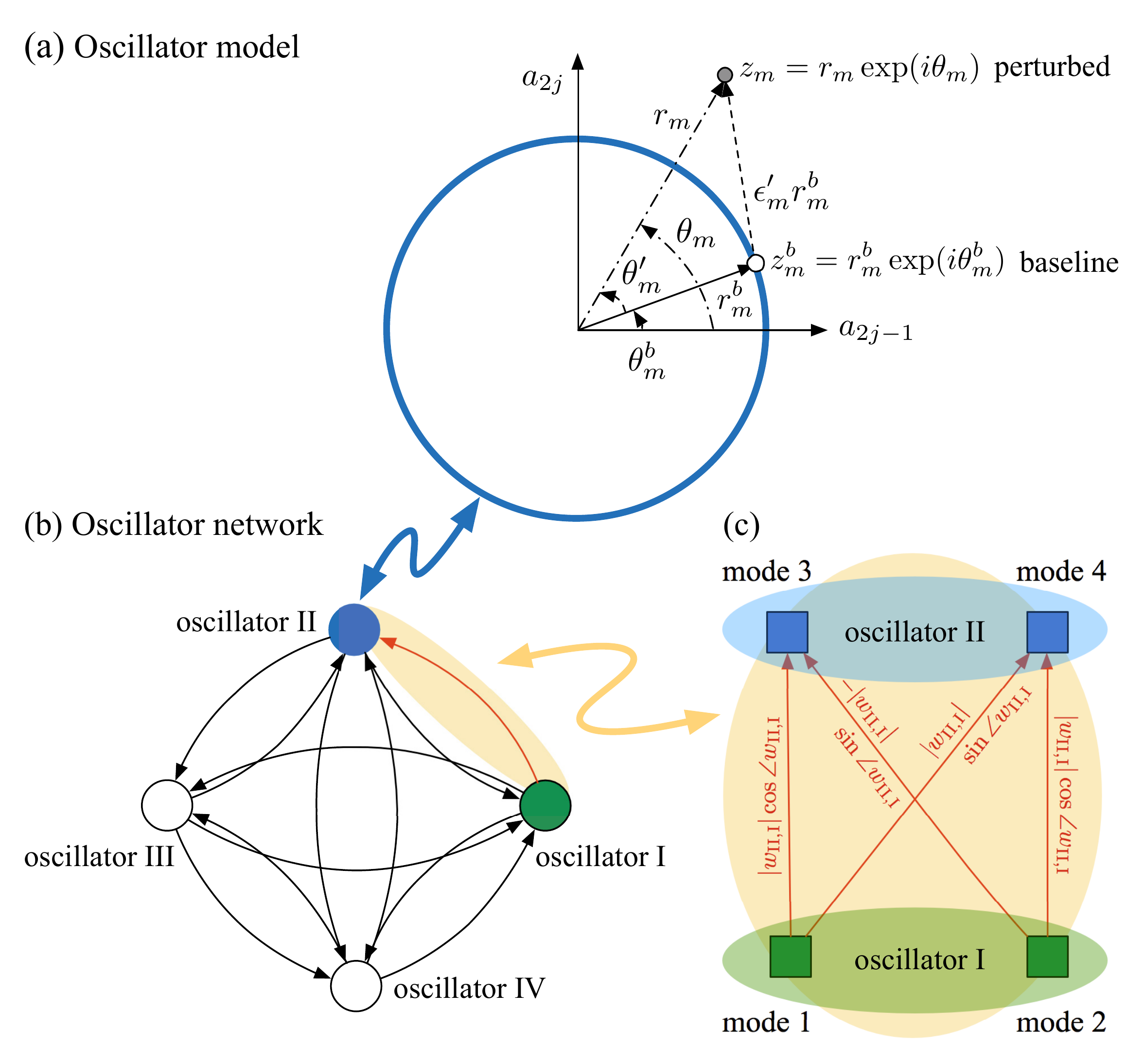}
          \end{overpic} 
    \end{tabular}
   \end{center}
   \caption{(a) Modal oscillator model in complex space and (b) oscillator network. The circle in oscillator model describes the limit-cycle trajectory of the oscillators ($z_m^b$) and the perturbed temporal coefficients ($z_m$). The nodes of the network correspond to oscillators with directed edges showing interactions. Modal interactions corresponding to edge from oscillator I to II highlighted in red is further expanded in (c).}
   \label{fig1}
\end{figure}

The total temporal coefficient for each oscillator in the perturbed case can then be described by combining the baseline temporal coefficients and fluctuation as
\begin{equation}
z_m = r_m \exp(i\theta_m) = z_m^b +  z_m^\prime.
\label{eq5}
\end{equation}
The overall oscillator model highlighting perturbations introduced is shown in figure \ref{fig1} (a). The blue circle in the figure indicates the natural limit-cycle of oscillator $m$ with baseline temporal coefficient ($z_m^b$) lying on it. The perturbations in amplitude ($\epsilon_m^\prime$) and phase ($\theta_m^\prime$) result in a total temporal coefficient ($z_m$) off the limit cycle. We need to consider two main factors to characterize the perturbations introduced for each oscillator $m$; (1) the initial amplitude of perturbation $\epsilon_m^\prime(t_0)$ and (2) the initial phase perturbation size $\theta_m^\prime(t_0)$. Once these factors are determined, using Eq. (\ref{eq41}) and (\ref{eq5}) in Eq. (\ref{eq3}), the initial velocity field for DNS of the perturbed case is prescribed. 

Once the perturbations are introduced, the perturbation energy is distributed among the oscillators according to the flow physics. The temporal dynamics of the modes are attributed to the interactions between them resulting primarily from the convective term of the Navier--Stokes equation. These interactions cause energy transfers among the modes in unsteady fluid flow. As the POD modes are orthogonal to each other, the temporal coefficients in the perturbed case can be extracted by projection, as $a_j = (\boldsymbol{u}-\bar{\boldsymbol{u}},\boldsymbol{\phi}_{j}^{\boldsymbol{u}})$. We then construct the temporal coefficients in the complex plane $z_m$ for each oscillator $m$ using Eq. (\ref{eq:osc1}). To capture the fluctuating amplitude and phase of the temporal coefficients, we track the normalized fluctuation
\begin{equation}
\zeta_m  = z_m^\prime/z_m^b = \epsilon_m^\prime \exp(i\theta_m^\prime).
\label{eq:nf}
\end{equation} 
Thus, the normalized fluctuation tracks the amplitude $\epsilon_m^\prime(t)$ and phase $\theta_m^\prime(t)$ fluctuations of the modal oscillators due to interactions.

The amplitude fluctuation of the oscillators is related to the variation in oscillator energy $E_m^\prime(t)$ compared to the baseline, given by
\begin{align}
\begin{split}
E_m^\prime = \frac{1}{2}\left(|z_m|^2 - |z_m^b|^2\right) = \frac{1}{2} [(1+\epsilon_m^\prime)^2-1](r_m^b)^2, \\ 
E_m^\prime (t_0) = \beta_m E^b,~~~~~~~~~~~~~~~~~~~~~
\end{split}
\label{eq:amp}
\end{align}
where $E_m^\prime (t_0)$ represents the oscillator energy input, $E^b$ is the total baseline modal energy given by $E^b= \sum_{j=1}^N (a_j^b)^2/2 = \sum_{m=\text{I}}^{M} (r^b_m)^2/2$ and $\beta_m$ represents the factor of input energy introduced to oscillator $m$ at initial time $t_0$ compared to the total baseline energy. If no amplitude perturbation is introduced in the simulation, $E_m^\prime (t_0) = 0$. If only a phase perturbation is introduced in the simulation, variation in oscillator energy may be observed due to oscillator interactions, although the oscillator energy input is zero. To be consistent with the tracking of normalized fluctuation $\zeta_m$ in Eq. (\ref{eq:nf}), the normalized quantities $E_m^\prime/(r^b_m)^2$ and $a_j^\prime/r^b_m$ are used to track the energy perturbations of the oscillators and amplitude perturbations of the modes, respectively while $\theta_m^\prime$ tracks the phase of the oscillators. This normalization is adopted to facilitate the tracking of the energy transfer between modes. Multiple combinations of oscillator energy input and phase perturbations can be used to quantify energy exchange for each perturbed case. 

\subsection{Networked oscillator representation}
\label{subsec:nor}

Using the oscillators extracted from POD and their normalized fluctuations extracted from the impulse response about the baseline flow, we can now create a network-theoretic representation of the unsteady fluid flow system. In the present work, we consider the oscillators (modes pairs) characterizing unsteady fluctuations to be the nodes ($\mathcal{V}$) of the network and the interaction between them as edges ($\mathcal{E}$). To characterize the interactions between oscillators, we need a model that entails coupling between oscillators. This motivates a more general networked oscillator model of $M$ linearly coupled oscillators given by
\begin{equation}
\dot{\zeta}_m =  \sum_{n=\text{I}}^{M} [\boldsymbol{A}_{\mathcal{G}}]_{mn} \left(\zeta_n - \zeta_m\right) = -\sum_{n=\text{I}}^{M} [\boldsymbol{L}_{\mathcal{G}}]_{mn} \zeta_n,
\label{eq:zeta}
\end{equation}
where $m = \text{I}, \text{II} \dots M$, $[\boldsymbol{A}_{\mathcal{G}}]_{mn}$ and $[\boldsymbol{L}_{\mathcal{G}}]_{mn}$ are the adjacency and the in-degree graph Laplacian matrices, respectively. The networked oscillator model considers propagation processes on networks and is used as a simple model for examining the spread of kinetic energy among oscillators. It should be emphasized that this is a linear model. 

The dynamics of normalized fluctuations of temporal coefficients are nonlinear in general due to convective physics in unsteady fluid flows. The nonlinear change of variables into oscillator magnitude and phase coordinates enables this linear representation of the nonlinear flow dynamics. In general, coordinate transformations that embed nonlinear dynamics in a linear framework are related closely to the Koopman operator~\citep{Koopman1931pnas,Mezic2005nd,mezic2013analysis}.  
Obtaining useful coordinate transformations that approximate intrinsic Koopman coordinates remains an open challenge in data-driven dynamical systems~\citep{kutz2016dynamic,Brunton2016plosone}.  
Current data-driven methods to identify Koopman eigenfunctions include the variational approach of conformation dynamics (VAC)~\citep{noe2013variational,nuske2016variational}, the equivalent extended dynamic mode decomposition (EDMD)~\citep{Williams2015jnls,Williams2015jcd,klus2015numerical}, and the use of delay coordinates~\citep{Brunton2017natcomm,Arbabi2016arxiv}. 
The proposed transformation into oscillator coordinates in the present study provides a promising approximate linear representation of nonlinear dynamics. 

For a set of $M$ oscillators, the above adjacency matrix $\boldsymbol{A}_{\mathcal{G}} \in \mathbb{C}^{M\times M}$ concisely describes the network connectivity given as
\begin{equation}
[\boldsymbol{A}_{\mathcal{G}}]_{mn} = 
\begin{cases}
	w_{mn}  = |w_{mn}| \exp{(i\angle{w_{mn}})} &\text{if there is an edge from } n \text{ to } m \\
         0 &\text{otherwise.}
\end{cases}
\label{adjeq}
\end{equation}
In our formulation, no self-loops are present in the network connectivity of oscillators. This is primarily attributed to the nature of the networked oscillator model relying on interactions between oscillators to describe the dynamics of oscillator $m$. This is reflected from Eq. (\ref{eq:zeta}), for $n=m$, the contribution to the dynamics of oscillator $m$ is zero. The rows of the adjacency matrix indicates the dependence of the oscillators $n$ on the dynamics of oscillators in column $m$, i.e., $w_{mn}$. The in-degree ($k_m$) of a node $m$ represents the summation of the incoming weights of the edges connected to it given by $k_m = \sum_{n=\text{I}}^M [{\boldsymbol{A}}_\mathcal{G}]_{mn}$. The (in-degree) graph Laplacian is closely related to the adjacency matrix as $\boldsymbol{L}_{\mathcal{G}} = \boldsymbol{D}_{\mathcal{G}} - \boldsymbol{A}_{\mathcal{G}}$, where $\boldsymbol{D}_{\mathcal{G}}$ is a diagonal matrix with elements equal to the in-degree of the nodes, $\boldsymbol{D}_{{\mathcal{G}}} = \text{diag}([k_m]_{m=\text{I}}^M)$. The graph Laplacian is a discrete analog of the continuous Laplacian operator. In the current study, it encodes structural properties of the networked oscillator model. Propagation (diffusion) processes on networks are neatly described by this graph Laplacian.  

Using the knowledge of time evolution of normalized fluctuations from DNS for each oscillator, (i.e., $\boldsymbol{\zeta}$ and $\dot{\boldsymbol{\zeta}}$), we use a linear regression procedure to determine the adjacency matrix weights. These weights could also be obtained through a Galerkin regression approach \citep{loiseau2016constrained}. Linear regression allows us to examine the relationships between predictive (independent) and response (dependent) variables. It enables us to find the best possible linear model fit for estimating the response of a system by regressing the dynamics onto the predictor variables. Here, the time derivative of the normalized fluctuations of the individual oscillators ($\dot{\zeta}_m$) are the response variables and $\left(\zeta_n - \zeta_m\right)$ are the predictor variables. The result of such a multiple regression over $m$ oscillators yields the adjacency matrix weights for the networked oscillator model. It must be noted that as the temporal coefficient associated with the mean flow is constrained to unity, $z_0^\prime = 0$ and does not contribute to normalized fluctuations of the modes. Thus, mean flow is considered as an isolated node in this formulation. 

The oscillator network representation is illustrated in figure \ref{fig1} (b), where the oscillators are taken to be nodes of the network. The oscillator model in figure \ref{fig1} (a) highlights the temporal coefficient dynamics associated with the nodes of the network, e.g., oscillator II (shown in blue).  The complex adjacency weights obtained from linear regression $w_{mn}$ can be decomposed into a magnitude ($|w_{mn}|$) and phase ($\angle{w_{mn}}$). The magnitude of the edge weights signify the influence of oscillator $n$ on oscillator $m$ as illustrated in figure \ref{fig1} (b). The phase of the edge weights represent the individual modal contributions in oscillator phase interactions. In particular, it highlights the phase advances or delays imposed between modes of interacting oscillators. The odd-odd mode interactions and even-even mode interactions are given by $|w_{n,m}|\cos{\angle{w_{n,m}}}$ while the odd-even and even-odd mode interactions are given by $|w_{n,m}|\sin{\angle{w_{n,m}}}$ and $-|w_{n,m}|\sin{\angle{w_{n,m}}}$, respectively. 

For comparison with the networked oscillator model, we also construct the Galerkin projection model for the incompressible Navier-Stokes equations, by projecting the equations onto the POD modes to construct the POD-Galerkin reduced-order model \citep{noack2011reduced, Holmes2012}. The resulting Galerkin system is given by
 \begin{equation}
\begin{split}
\dot{a_j} =  \sum_{k = 0}^N \psi_{jk} a_k + \sum_{k,\,l = 0}^N \chi_{jkl} a_k a_l, \quad j = 1, 2, \dots N,
\end{split}
\end{equation}
where $\psi_{jk}$ and $\chi_{jkl}$ are the linear  and quadratic coefficient terms, respectively. Here, as mentioned before, the temporal coefficient associated with the mean is fixed to unity, $a_0 = 1$. The linear term contains diffusive physics of the modes while the quadratic term contains the convective physics. The networked oscillator model captures the effect of both diffusion and convection in the network structure that is extracted. Both the network and Galerkin models require only the specification of initial condition of the temporal coefficients of the mode pairs only. 

The networked oscillator model, as discussed above, describes both amplitude and phase dynamics of collective oscillation and captures the energy exchange and phase dynamics among the modes. As discussed earlier, the present model is derived from the analysis of impulse response in DNS. In \S\ref{sec:cf}, we describe the application of the oscillator network-based formalism to a canonical problem of two-dimensional incompressible flow over a cylinder. The model can also serve as a foundation to perform network-informed feedback control of unsteady flow as will be discussed in \S\ref{subsec:ctrlflow}.


\section{Application to cylinder flow}
\label{sec:cf}

Two-dimensional flow over a circular cylinder is a canonical problem in fluid mechanics and serves as a model to capture fundamental wake dynamics for many flows. The coherent structures identified are well-reported in literature, and thus cylinder flow forms an excellent testbed for our analysis. Moreover, for the cylinder flow problem, POD modes are well characterized and are the same as DMD modes as demonstrated in the work by \cite{chen2012variants}. Thus, as an alternative to POD for the current approach, DMD can be also considered. To ensure orthogonality of the modes required for the current approach, recursive DMD can be performed to obtain orthogonal pure frequency modes \citep{noack2016recursive}. 


\subsection{Computational approach} 
\label{subsec:ca}

For this study, we perform DNS of incompressible flow past a circular cylinder using the immersed boundary projection method \citep{Taira:JCP07,Colonius:CMAME08, kajishima2017numerical} at a diameter-based Reynolds number of $Re = 100$. This method employs a  Cartesian grid with the immersed body formulation to generate the cylinder. We take advantage of the multi-domain technique to simulate cylinder flow in free space. The innermost domain is chosen as $-1\le x/d \le 29, -15 \le y/d \le 15$ with a resolution of $600 \times 600$ grid points where $d$ is the cylinder diameter. Here, $x$ is the streamwise direction and $y$ is the cross-stream direction. The outermost domain is chosen as $-16\le x/d \le 44, -30 \le y/d \le 30$, far enough so as to not affect the results in the near-field. Uniform flow is prescribed at the domain boundaries. For time integration, the method uses an implicit Crank-Nicholson scheme for the viscous term and an Adam-Bashforth discretization for the convective term. 

\begin{figure}
   \begin{center}
    \begin{tabular}{c} 
          \begin{overpic}[width=5.2in]{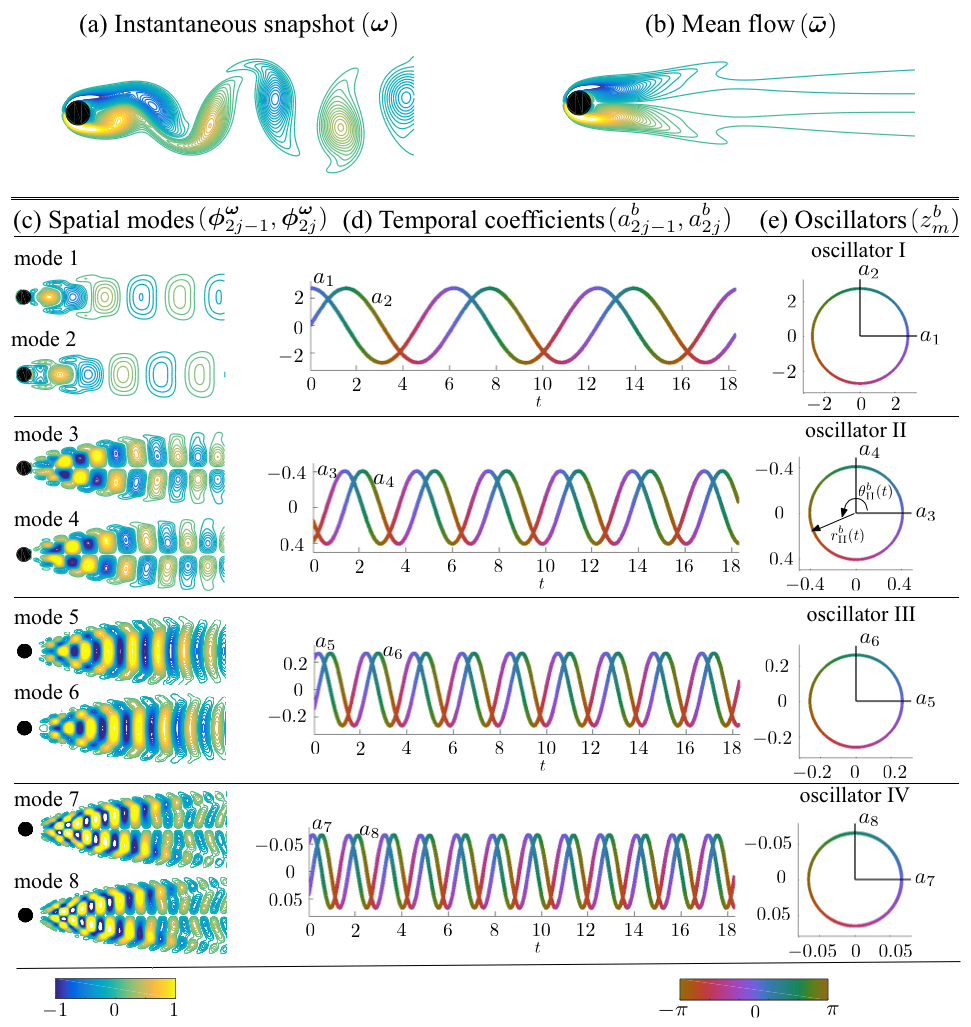}
          \end{overpic} 
    \end{tabular}
   \end{center}
   \caption{(a) Instantaneous and (b) time-averaged (mean) vorticity fields. Proper orthogonal decomposition (POD) applied to cylinder flow problem results in (c) spatial modes and (d) temporal coefficients associated with the modes. (e) Temporal coefficients of oscillators in complex plane. The colorbar of the spatial modes indicates the contour level and colorbar of the temporal coefficients and oscillators indicates phase of the oscillators varying from $[-\pi, \pi]$ over the periodic limit cycles.}
   \label{fig2}
\end{figure}

From the simulation, the drag coefficient $(C_D)$ and lift coefficient $(C_L)$ are computed as 
\begin{equation}
C_D = \frac{F_D}{\frac{1}{2}\rho U^2 d} \quad \text{and} \quad C_L = \frac{F_L}{\frac{1}{2}\rho U^2 d},
\label{forcecoeff}
\end{equation}
where, $F_D$ and $F_L$ are the drag and lift forces on the cylinder, $\rho$ is the freestream density, and $U$ is the freestream velocity field. These are obtained by integrating the boundary force representing the cylinder in the immersed boundary formulation. The Strouhal number for the flow is defined as $St = f_n d/U$, where $f_n$ is the natural shedding frequency. We obtain a Strouhal number of $St = 0.164$, drag coefficient $C_D = 1.35 \pm 0.009$ and lift coefficient $C_L = \pm 0.325$ which agree well with the literature \citep{liu1998preconditioned, Taira:JCP07, canuto2015two}. The flow exhibits vortex shedding behavior in the cylinder wake, as shown by the instantaneous vorticity field in figure \ref{fig2} (a). Such vortex shedding characterizes a von K\'{a}rm\'{a}n vortex street due to the repetitive pattern of vortices in the unsteady wake. The time-averaged (mean) vorticity field is shown in figure \ref{fig2} (b). We then perform POD with the method of snapshots \citep{Sirovich:QAM87} using the velocity field data ($\boldsymbol{u}$) gathered from DNS. The modes and temporal coefficients obtained are in agreement with those from the work by \cite{Noack:JFM03}. 

\subsection{Baseline (unperturbed flow)}
\label{sec:analysis}

The extracted spatial modes and associated temporal coefficients ($a_j^b$) are shown in figure \ref{fig2} (c) and (d), respectively, in terms of the vorticity field, $\boldsymbol{\phi}_j^{\boldsymbol{\omega}} = \nabla \times \boldsymbol{\phi}_{j}^{\boldsymbol{u}}$. We define the conjugate mode pairs as independent oscillators in our formulation. Oscillator I constitutes mode pair ($1,2$), oscillator II constitutes mode pair ($3,4$) and so on. The mode pairs (oscillators) are ordered in terms of decreasing energy content, $E_j = (a_j^b)^2/2$, shown in figure \ref{fig2} (top to bottom). As the first eight POD modes capture $99.98 \%$ of the modal kinetic energy, we choose $N=8$ for our analysis. The energy content of oscillators I, II, III and IV are $96.87 \%, 2.18 \%, 0.88 \%$ and $0.05 \%$, respectively. The oscillator temporal coefficients in the complex plane ($z_m^b$) corresponding to temporal coefficients of the individual mode pairs from Eq. (\ref{eq:osc1}) are shown in figure \ref{fig2} (e). For this canonical problem, the frequency associated with the higher-order oscillators are harmonics of the lowest-order oscillator, $\Omega_m^b = m\Omega_\text{I}^b$. The frequency of oscillator I corresponds to the Strouhal number associated with the natural shedding cycle of the flow, $\Omega_\text{I} = 0.164$. As the frequency ($\Omega_m^b$) associated with the temporal coefficients of the mode pairs increases, the size of the spatial modal structures decreases. Again, the oscillators in the baseline flow are associated with limit cycle temporal dynamics, which can be described by Eq. (\ref{eq4}), independent of each other with no coupling between them. This lack of coupling in the generalized limit-cycle dynamics does not capture interactions between oscillators in unsteady fluid flow.


\subsection{Perturbed flow}
\label{subsec:pa}

To capture interactions between oscillators, additional energy and phase perturbations are impulsively introduced to the simulation through the initial condition. These perturbations cause added fluctuations in the temporal coefficients of the modes. The projected coefficients $a_n = \left<\boldsymbol{u}-\bar{\boldsymbol{u}},\boldsymbol{\phi}_{n}^{\boldsymbol{u}}\right>$ from the perturbed case are extracted from DNS and  the normalized fluctuation $\zeta_m(t)$ is tracked using the networked oscillator model discussed in \S \ref{sec:formulation}. We expect that as the perturbation convects downstream, the normalized fluctuation of the oscillators will decay to zero and the perturbed field will return to the baseline limit-cycle behavior. However, the addition of a perturbation to the flow results in a constant phase shift of each oscillator in the final limit cycle, compared to the unperturbed limit cycle. 

In two-dimensional unsteady cylinder flow, the leading POD mode pair corresponding to oscillator I contains maximum energy and any deviations in the final limit cycle is immediately detected by the oscillator I phase ($\theta_\text{I}$). Thus, to construct the normalized fluctuation time history for each oscillator, we align the phase of oscillator I for the perturbed case and the baseline case as $\zeta_m = (z_m - z_m^b|_{\theta = \theta_\text{I}})/z_m^b|_{\theta = \theta_\text{I}}$. Once the normalized fluctuation $\zeta_m$ is constructed, we also compute its time derivative $\dot{\zeta}_m$. We then construct the library of functions ($\zeta_n - \zeta_m$) for each oscillator $m$. We perform a simple linear regression procedure to obtain the network structure $[\boldsymbol{A}_{\mathcal{G}}]_{mn}$. Once the coupling function (network structure) is obtained, we solve the linear networked oscillator model in Eq. (\ref{eq:zeta}) for prediction with a prescribed initial condition and compare the fluctuations with those obtained in DNS. While it is not necessary, one could also consider interaction terms that are quadratic or higher \citep{brunton2016discovering,loiseau2016constrained}. 


\subsubsection{Amplitude perturbations}

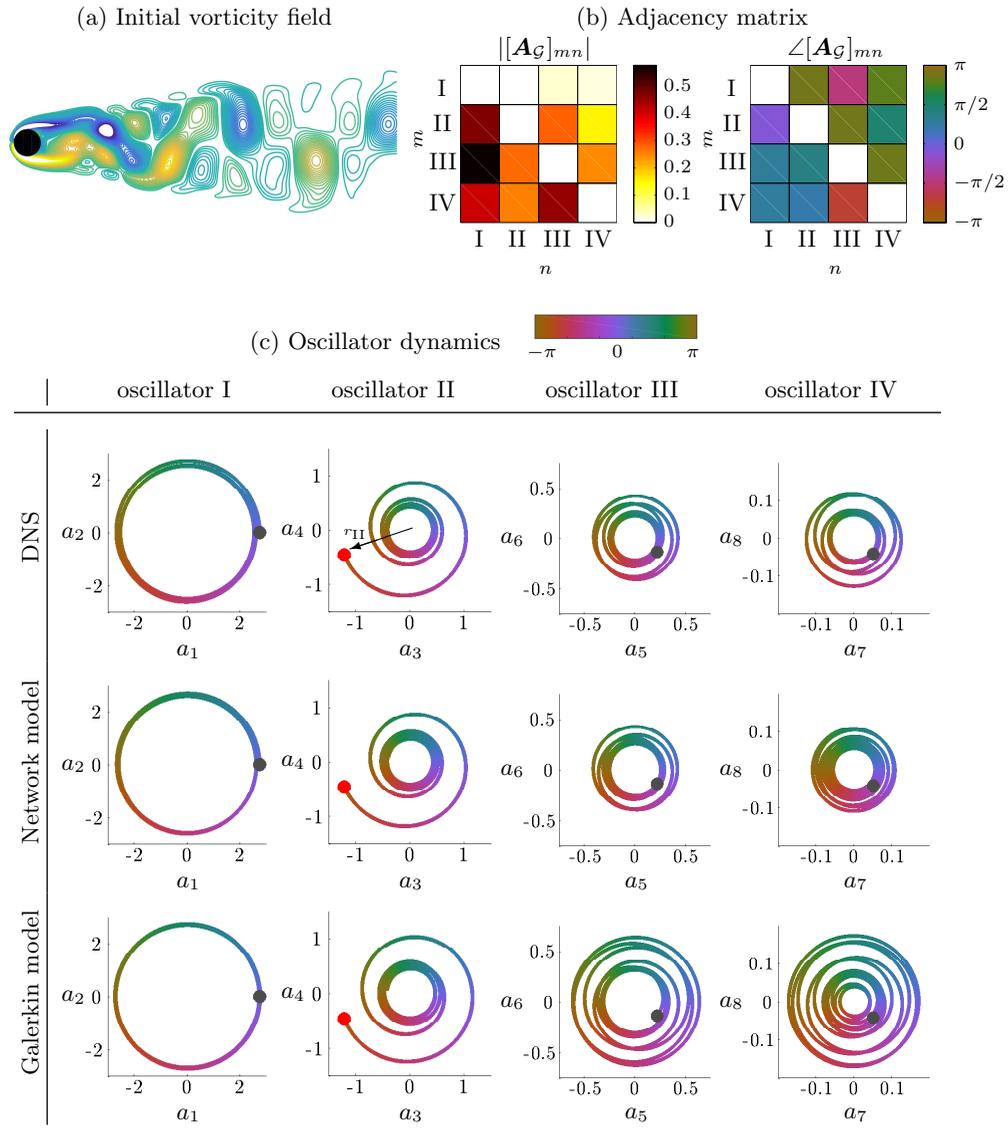
\begin{figure}  
\begin{center}  
	\begin{tabular}{cc}
	\vspace{0.1in}
	(a) Initial vorticity field & (b) Adjacency matrix \\	
	\includegraphics[height=0.175\textwidth]{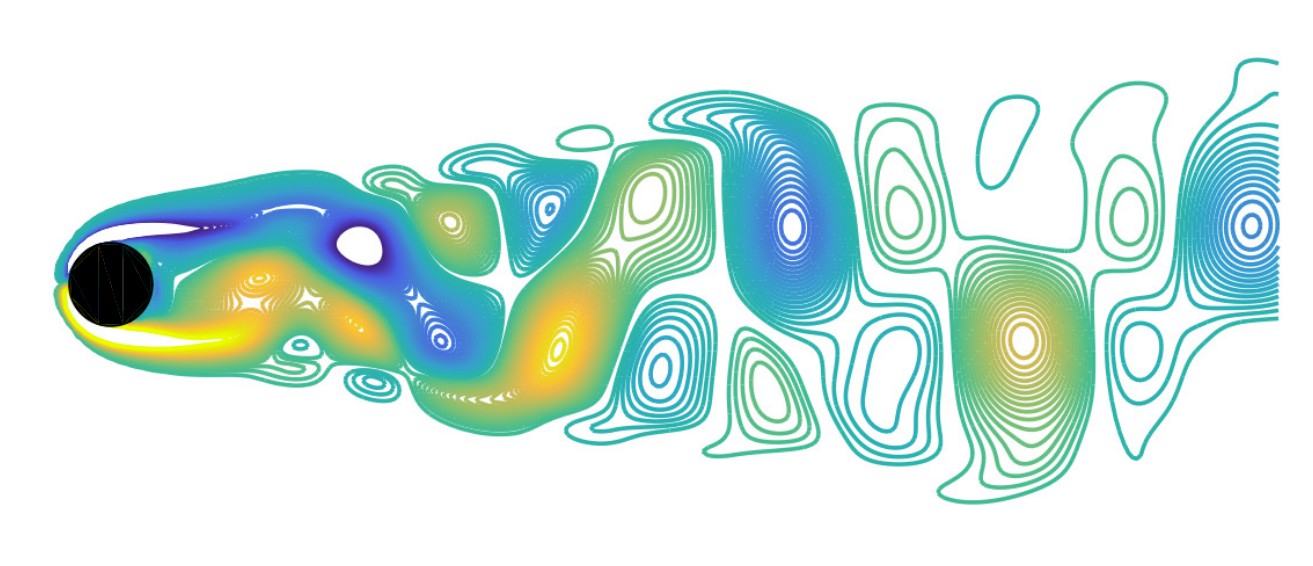}  & ~~~
	\begin{overpic}[height=0.175\textwidth]{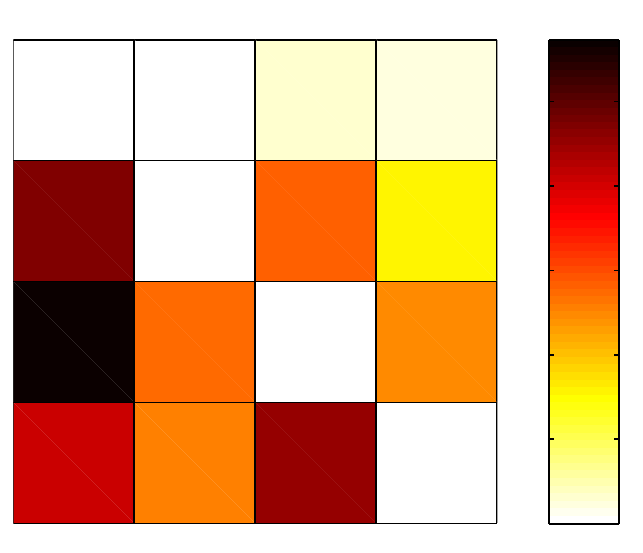} 
	  \put(20,85){$|[\boldsymbol{A}_{\mathcal{G}}]_{mn}|$}	  
	   \put(9,-7){\small I}   \put(25,-7){\small II}   \put(42,-7){\small III}   \put(62,-7){\small IV}
	   \put(-9,65){\small I} \put(-11,48){\small II} \put(-13,30){\small III} \put(-13,10){\small IV}
	   \put(100,2.5){\scriptsize $0$}
	   \put(100,15.5){\scriptsize $0.1$}
	   \put(100,28.5){\scriptsize $0.2$}
	   \put(100,41.5){\scriptsize $0.3$}
	   \put(100,54.5){\scriptsize $0.4$}
	   \put(100,67.5){\scriptsize $0.5$}	
	   \put(40,-20){\scriptsize $n$}
	    \put(-20,40){\scriptsize \rotatebox{90}{$m$}}   
	\end{overpic} 
	\qquad~~
	\begin{overpic}[height=0.175\textwidth]{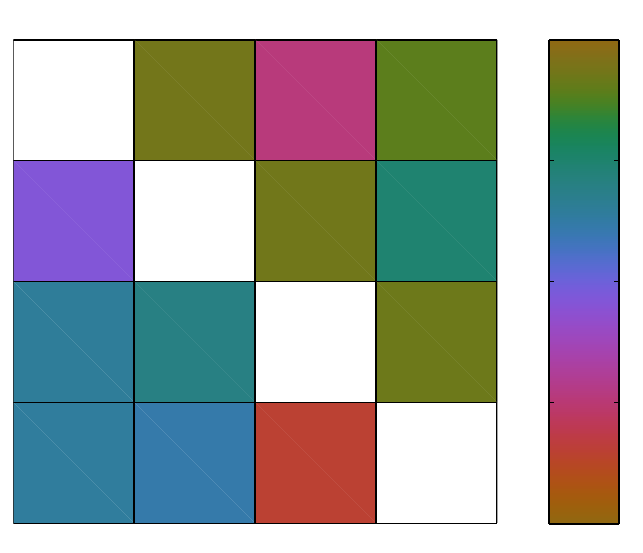} 
	 \put(20,85){$\angle[\boldsymbol{A}_{\mathcal{G}}]_{mn}$}
	   \put(9,-7){\small I}   \put(25,-7){\small II}   \put(42,-7){\small III}   \put(62,-7){\small IV}
	   \put(-9,65){\small I} \put(-11,48){\small II} \put(-13,30){\small III} \put(-13,10){\small IV}
	    \put(100,2.5){\scriptsize $-\pi$}
	   \put(100,21.5){\scriptsize $-\pi/2$}
	   \put(100,39.5){\scriptsize $0$}
	   \put(100,58.5){\scriptsize $\pi/2$}
	   \put(100,77.5){\scriptsize $\pi$}	   
	    \put(40,-20){\scriptsize $n$}
	    \put(-20,40){\scriptsize \rotatebox{90}{$m$}}
	\end{overpic} 	
	\vspace{0.3in}
	\end{tabular}
	\break
	(c) Oscillator dynamics \includegraphics[width=0.2\textwidth]{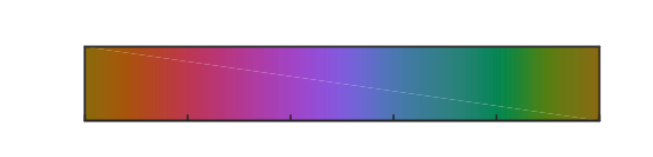}
	\put(-70,-3){\scriptsize $-\pi$}
	   \put(-38,-3){\scriptsize $0$}
	   \put(-10,-3){\scriptsize $\pi$}	   
	\break \break
	\begin{tabular}{c|cccc}	
	\vspace{-0.05in}
	 & ~~oscillator I & oscillator II & oscillator III & oscillator IV \\ \hline
	\rotatebox{90}{\qquad \quad~ DNS} & 
	~~\begin{overpic}[width=0.2\textwidth]{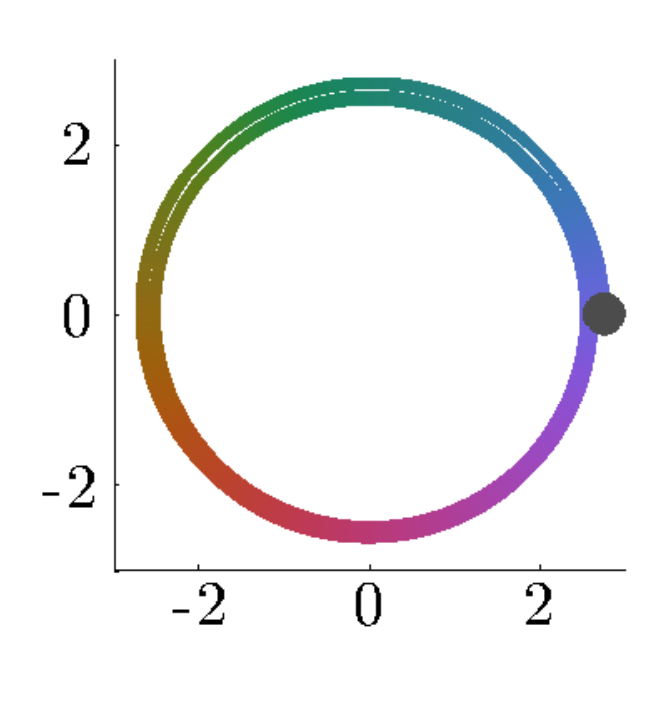}
	   \put(48,-1){\small $a_1$}
	   \put(-5,53){\small $a_2$}
	\end{overpic} &
	\begin{overpic}[width=0.2\textwidth]{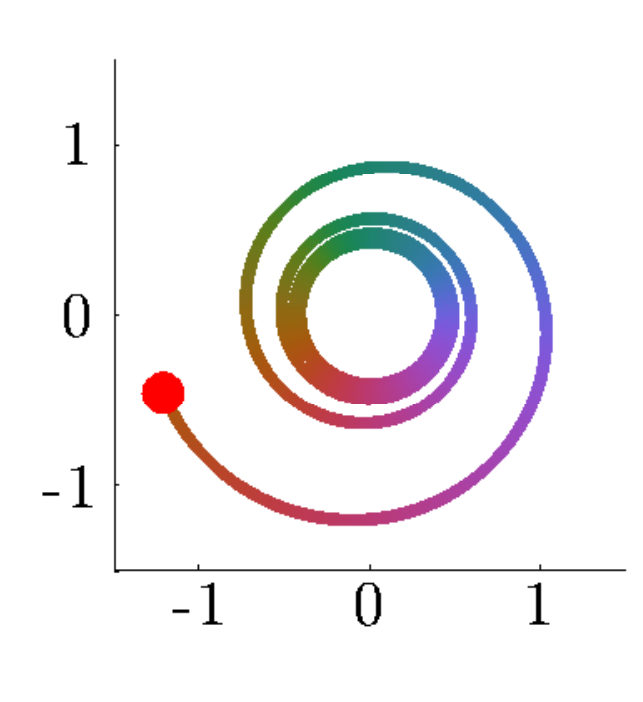}
	\put(48,-1){\small $a_3$}
	\put(-5,53){\small $a_4$}
	\linethickness{3pt}
\put(54,56){\vector(-3,-1){28}}
\put(20,53){\tiny{ $r_\text{II}$}}
	\end{overpic} &	
	\begin{overpic}[width=0.2\textwidth]{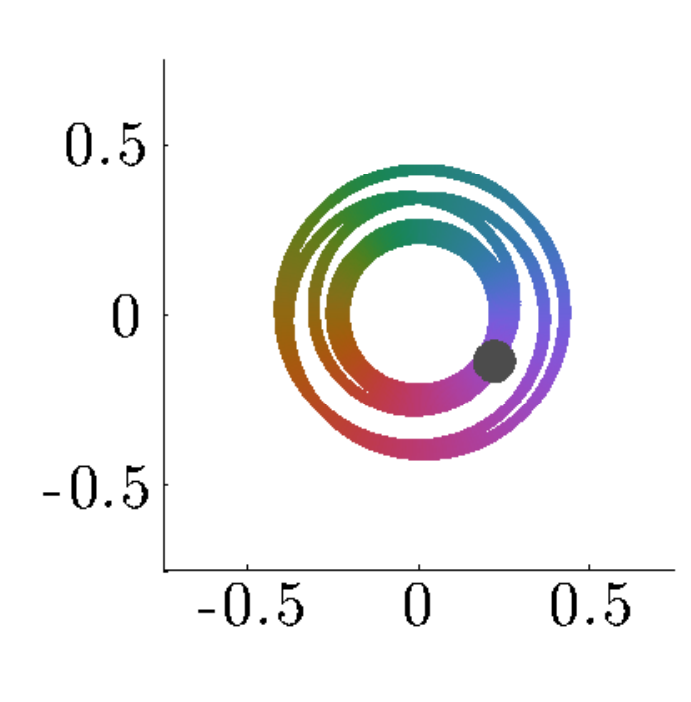}
	\put(55,-1){\small $a_5$}
	\put(-5,53){\small $a_6$}
	\end{overpic} &	
	\begin{overpic}[width=0.2\textwidth]{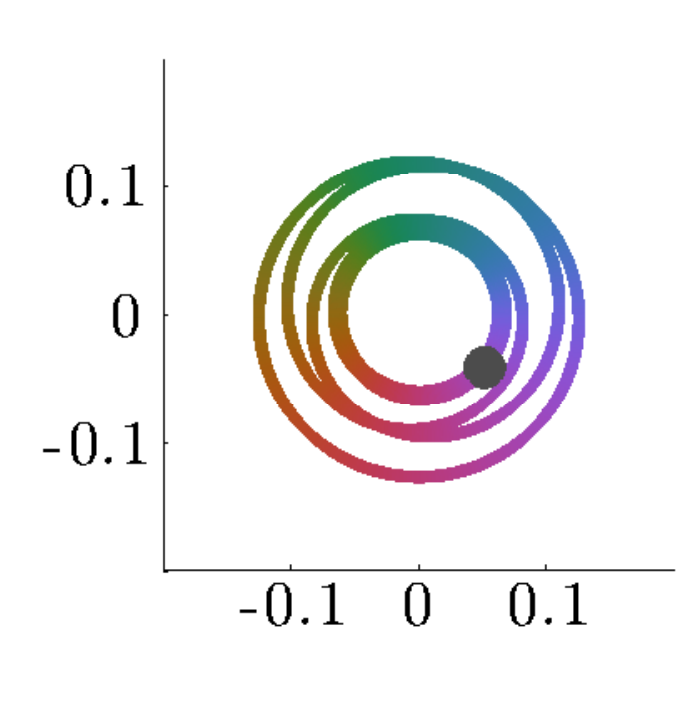}
	\put(55,-1){\small $a_7$}
	\put(-5,53){\small $a_8$}
	\end{overpic} 
	\\
	\rotatebox{90}{\quad ~~Network model} & 
	~~\begin{overpic}[width=0.2\textwidth]{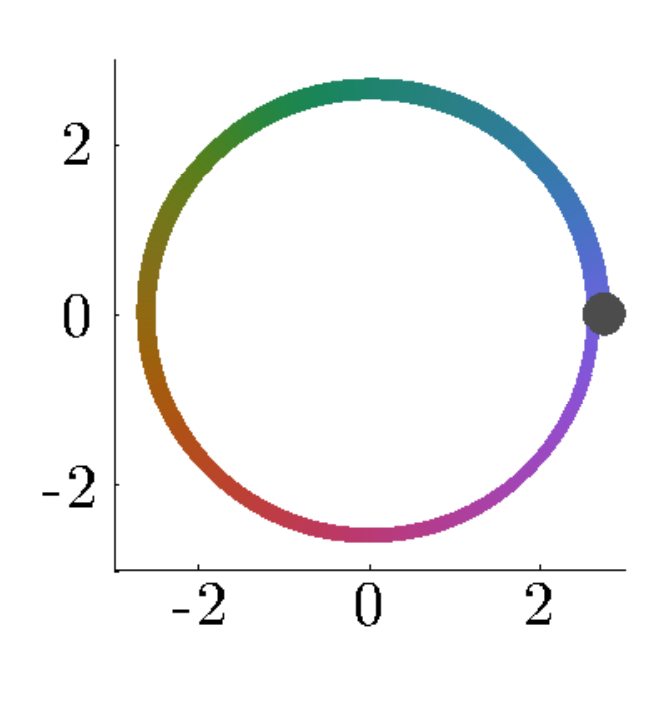}
	   \put(48,-1){\small $a_1$}
	   \put(-5,53){\small $a_2$}
	\end{overpic} &
	\begin{overpic}[width=0.2\textwidth]{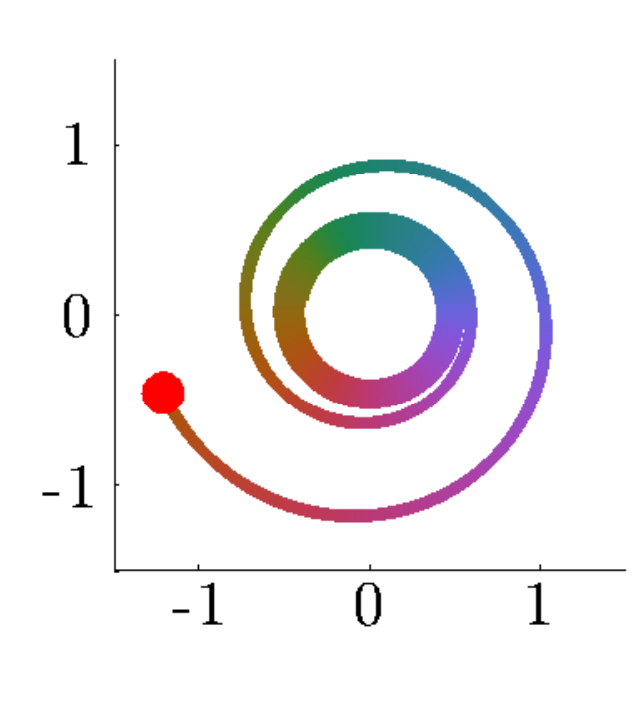}
	\put(48,-1){\small $a_3$}
	   \put(-5,53){\small $a_4$}
	\end{overpic} &	
	\begin{overpic}[width=0.2\textwidth]{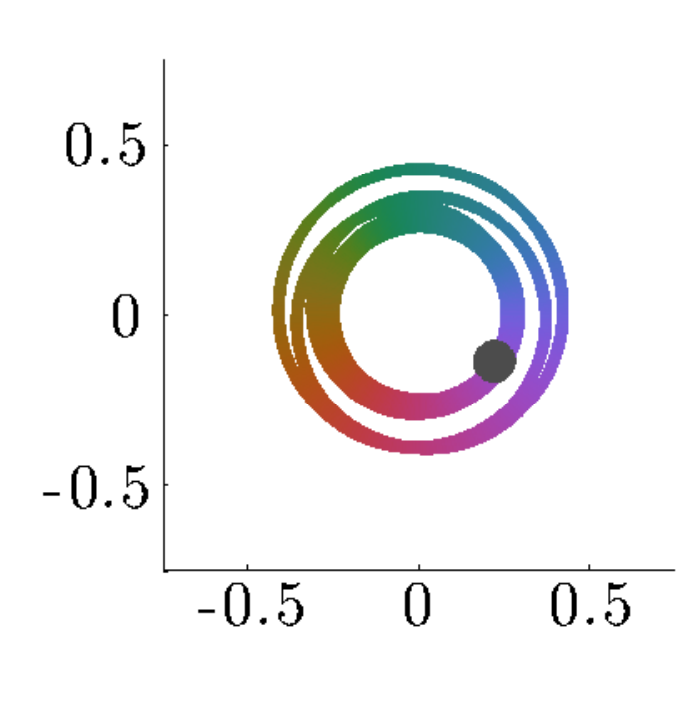}
	\put(55,-1){\small $a_5$}
	   \put(-5,53){\small $a_6$}
	\end{overpic} &	
	\begin{overpic}[width=0.2\textwidth]{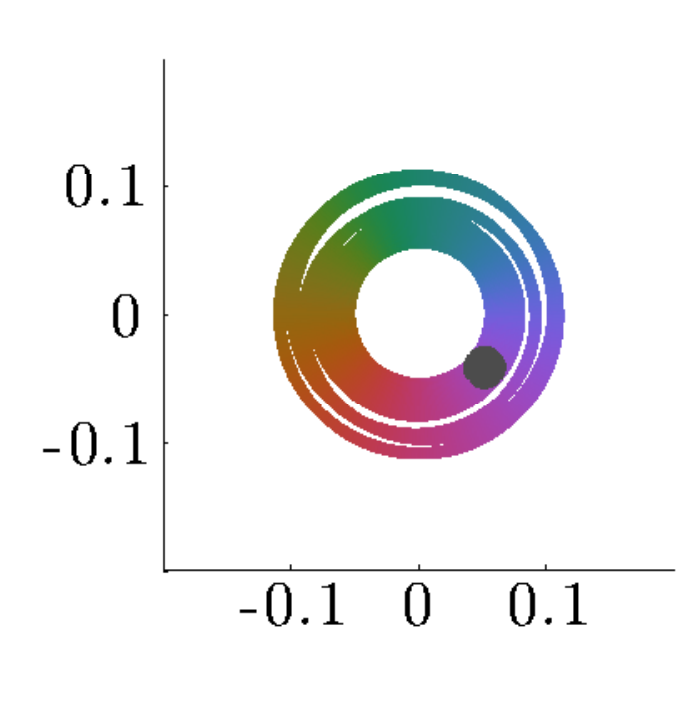}
	\put(55,-1){\small $a_7$}
	   \put(-5,53){\small $a_8$}
	\end{overpic}  \\
	\rotatebox{90}{\quad ~~Galerkin model} & 
	~~\begin{overpic}[width=0.2\textwidth]{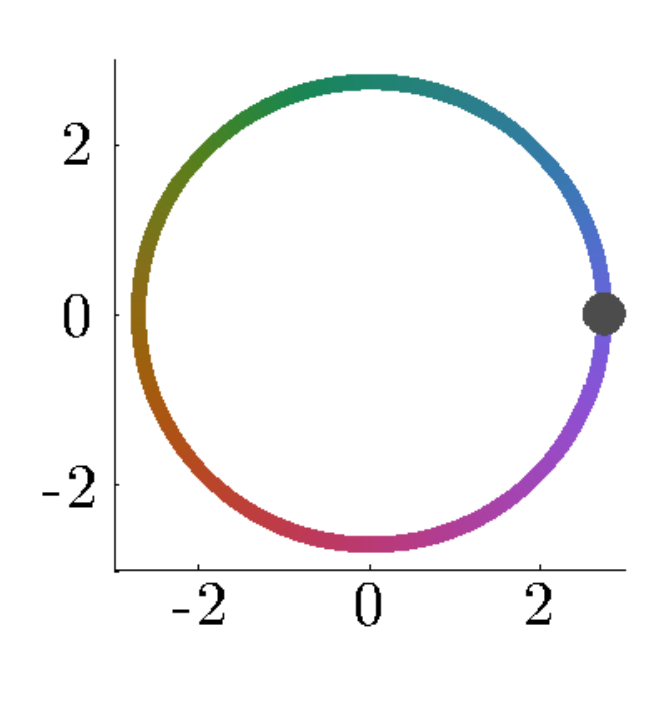}
	   \put(48,-1){\small $a_1$}
	   \put(-5,53){\small $a_2$}
	\end{overpic} &
	\begin{overpic}[width=0.2\textwidth]{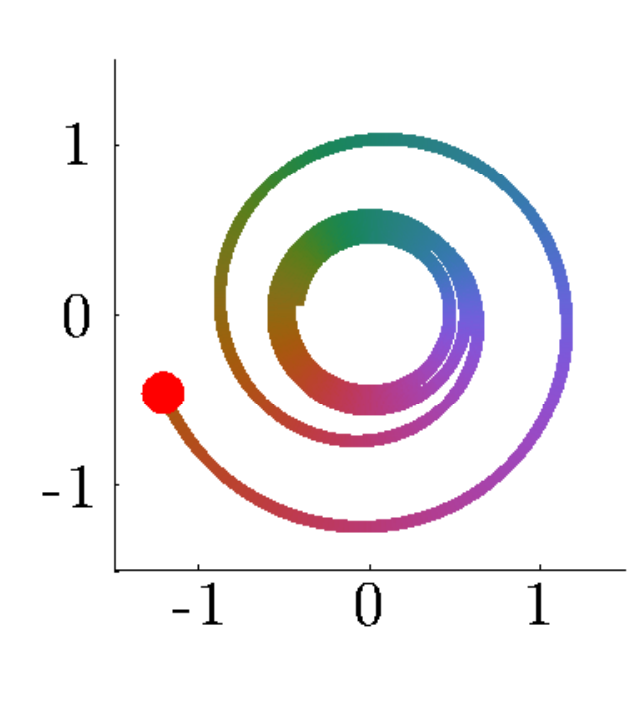}
	\put(48,-1){\small $a_3$}
	\put(-5,53){\small $a_4$}
	\end{overpic} &	
	\begin{overpic}[width=0.2\textwidth]{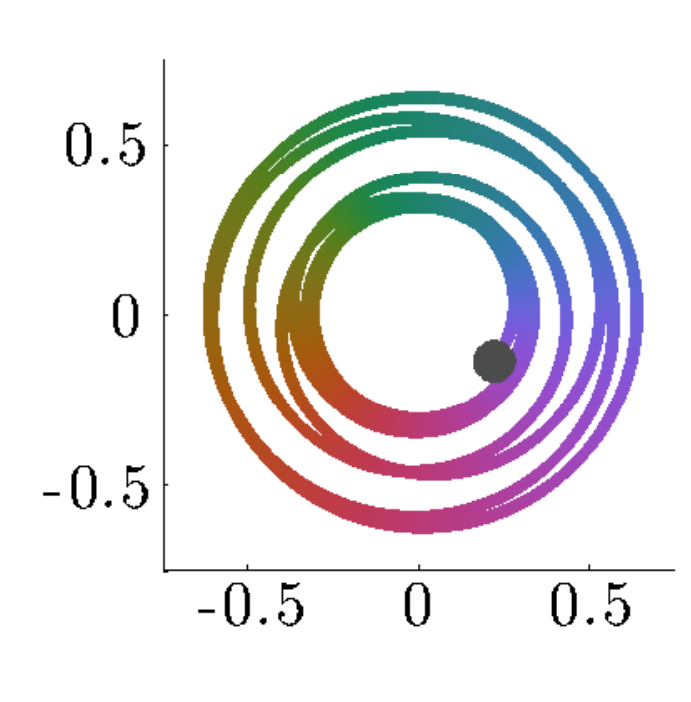}
	\put(55,-1){\small $a_5$}
	\put(-5,53){\small $a_6$}
	\end{overpic} &	
	\begin{overpic}[width=0.2\textwidth]{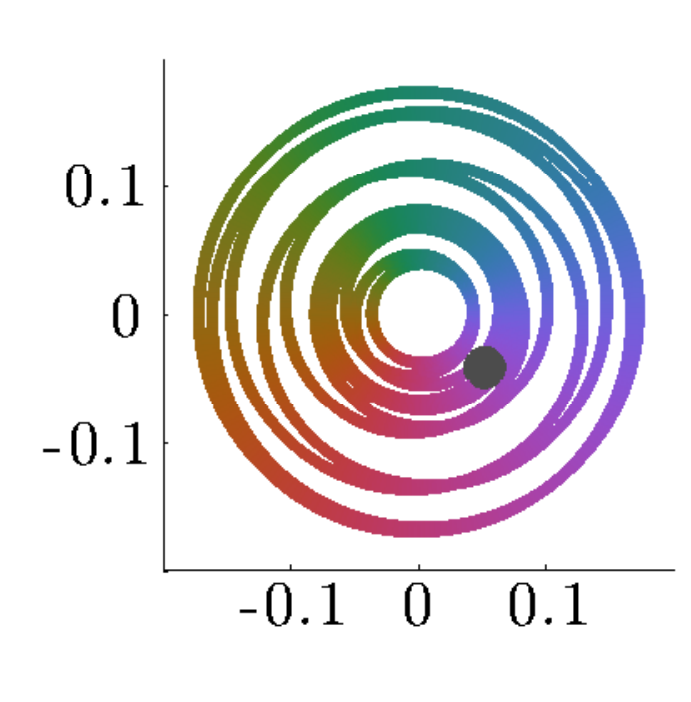}
	\put(55,-1){\small $a_7$}
	\put(-5,53){\small $a_8$}
	\end{overpic}  \\ \hline
	\end{tabular}
  \vspace{2mm}
\caption{Amplitude perturbation introduced at oscillator II. (a) Initial vorticity field corresponding to amplitude perturbation, (b) Adjacency matrix of networked oscillator model.  (c) Oscillator dynamics from DNS, networked oscillator model and POD-Galerkin model (color of temporal dynamics of oscillators indicates phase). Red dot indicates the initial perturbation.}  
\label{fig3}
\end{center}  
\end{figure}

We first introduce an amplitude perturbation to oscillator II, which is associated with the first harmonic of the natural shedding frequency of the flow. Introducing perturbations at this frequency perturbs not only the natural shedding frequency but also the oscillators associated with higher harmonics of the flow due to interactions present in the flow field. Let us consider the addition of $20 \%$ of baseline modal energy to oscillator II ($\beta_\text{II} = 0.2$). No perturbation in phase is added, i.e., $\theta_{\text{II}}^\prime(t_0) = 0$. With these initial perturbations determined, we can construct the initial flow field required for DNS of the perturbed case. The initial vorticity field, as shown in figure \ref{fig3} (a) for the perturbed case, is different from figure \ref{fig2} (a) for the baseline. Here, as oscillator II is energized, we notice the prevalence of spatial vortical structures associated with mode pair ($3,4$). The vorticity magnitude associated with these structures increases due to the added energy and correspondingly appears in the initial vorticity field for the perturbed case. As the introduced perturbation in oscillator II convects downstream in the numerical simulation, it interacts with other oscillators. Using the procedure highlighted above, we construct the normalized fluctuations $\boldsymbol{\zeta}$ and extract the adjacency matrix $\boldsymbol{A}_{\mathcal{G}}$ that highlights network interactions.  

The magnitude and phase of the adjacency matrix are shown in figure \ref{fig3} (b). In this example, the dynamics of oscillator I (leading oscillator) is not affected noticeably by the other oscillators. The dynamics of oscillators II and III show high levels dependence on oscillator I. This is consistent with our expectation as most of the energy in the flow is held in oscillator I and passes down to higher-order oscillators. Also, if we evaluate the strength of the incoming edge weights of each oscillator ($\sum_n w_{mn}$) for this perturbed case, oscillator III shows maximum influence from the other oscillators. The complex matrix structure of the extracted network summarizes the interactions in the flow upon introduction of impulse perturbations. To reveal oscillator phase interactions, we evaluate the phase of the edge weights in figure \ref{fig3} (b). For the dynamics of oscillator II, i.e., mode pair $(3,4)$, the phase of the edge weights indicates that interactions between modes $1 \rightarrow 3$, $2 \rightarrow 4$, $8 \rightarrow 3$ and $7 \rightarrow 4$ are larger than other modal interactions. For this perturbed case, in general, phase advancing effects are predominant for the oscillators affecting each other. 

\begin{figure}
   \begin{center}
    \begin{tabular}{cc} 
    (a) Modal amplitude & (b) Energy\\
     \hspace{-0.175in}
          \begin{overpic}[width=2.75in]{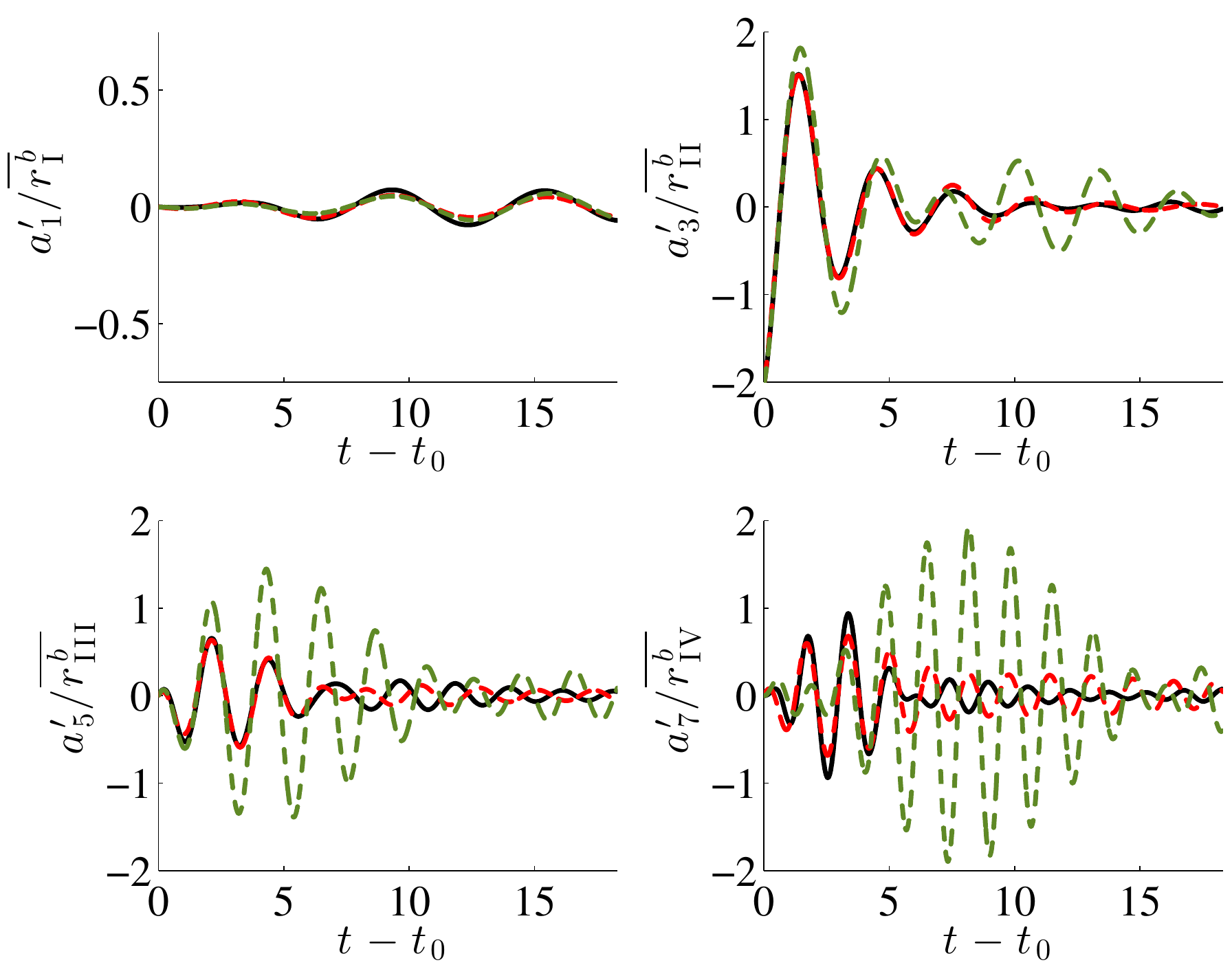}
          \end{overpic} &
          \hspace{-0.1in}
          \begin{overpic}[width=2.75in]{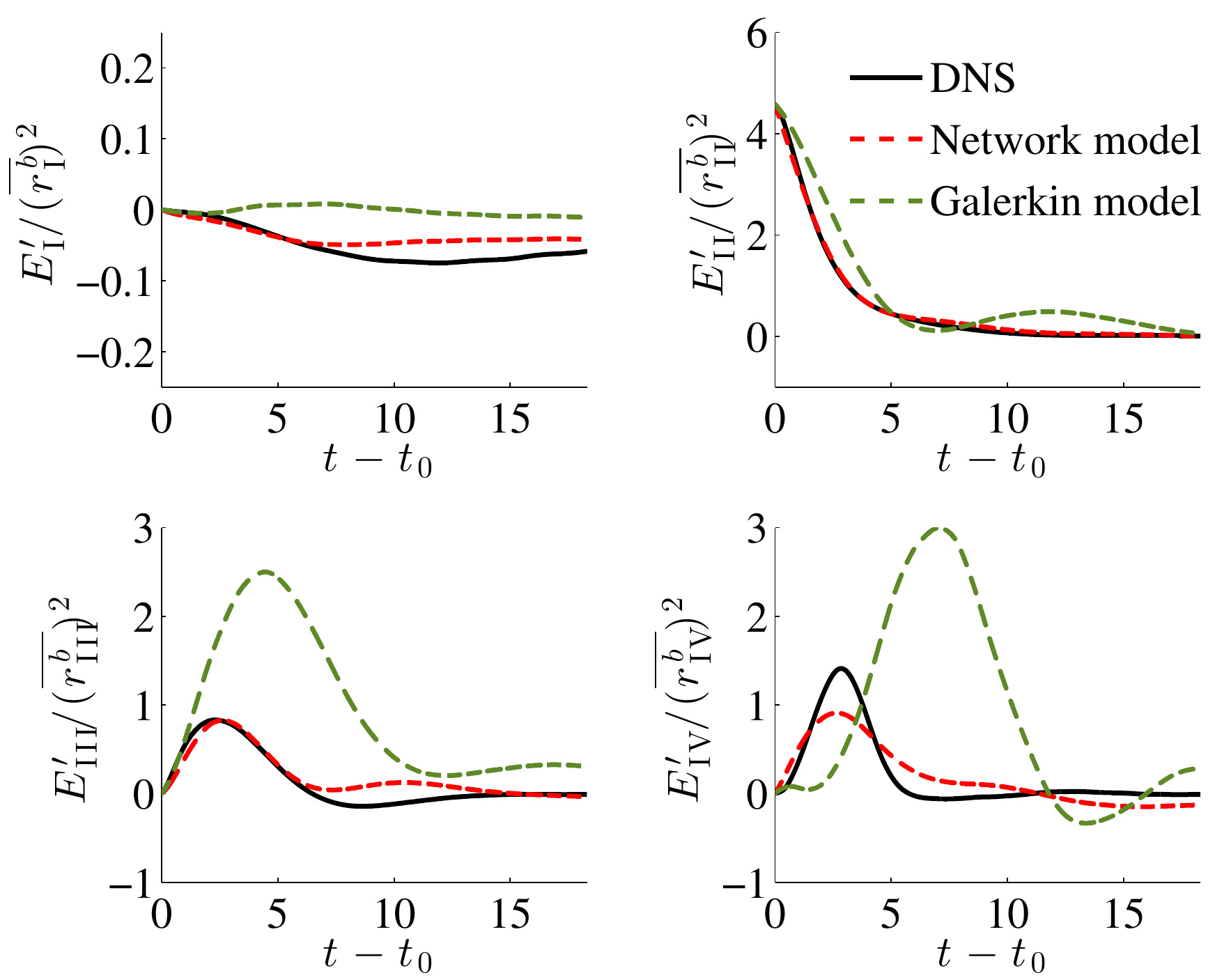}
          \end{overpic}
    \end{tabular}
   \end{center}
   \caption{(a) Modal amplitude and (b) energy tracking for amplitude perturbation introduced to oscillator II.}
   \label{fig4}
\end{figure}

Once the network structure is obtained, we can solve the linear networked oscillator model with the prescribed initial condition corresponding to the perturbation. For comparison, we also construct the empirical Galerkin reduced-order model. The oscillator dynamics from DNS (reference) and those predicted from the networked oscillator and POD-Galerkin models are shown in figure \ref{fig3} (c). The initial amplitude of the perturbed temporal coefficients for oscillator II corresponds to the red dot, $r_\text{II} (t_0)$. However, no phase perturbation is introduced in this example. The initial amplitude and phase of the other oscillators are unchanged, as no perturbations are introduced in these oscillators. It can be seen in figure \ref{fig3} (c) that the oscillator dynamics in the networked oscillator model agree well with DNS trajectories, particularly the dynamics of oscillators I, II and III. The networked oscillator model and DNS show negligible differences in tracking the trajectory of oscillator II from the perturbed initial condition to its final limit cycle. The model trajectory of oscillator IV shows some similarities with DNS. Due to its low energy content, any small deviations in the dynamics of lower-order oscillators causes comparable changes in the trajectory of oscillator IV. On the other hand, the POD-Galerkin model overpredicts the fluctuations in oscillators II, III and IV. It can be seen that the networked oscillator model tracks the fluctuations better than the POD-Galerkin model. 

To further compare the details of the predicted trajectories, we track the fluctuations in modal amplitude and oscillator energy in figure \ref{fig4} (a) and (b), respectively. The networked oscillator model shows excellent agreement with DNS, tracking the amplitude perturbation $(a_{2n-1}^\prime)$ and energy transfers $(E_m^\prime)$. Moreover, we see agreement in the long-time behavior of the fluctuations as the flow returns to the baseline state for the networked oscillator model. In contrast, the POD-Galerkin model is not well-designed for modeling the long-time behavior of modal fluctuations and hence is not expected to work well as time progresses. This is especially true as POD modes are used for the empirical Galerkin formulation. As indicated by the green dashed line, the POD-Galerkin model overpredicts these fluctuations and their associated time scales.


\subsubsection{Phase perturbations}

\begin{figure}  
\begin{center}  
	\begin{tabular}{cc}
	\vspace{0.1in}
	(a) Initial vorticity field & (b) Adjacency matrix \\	
	\includegraphics[height=0.175\textwidth]{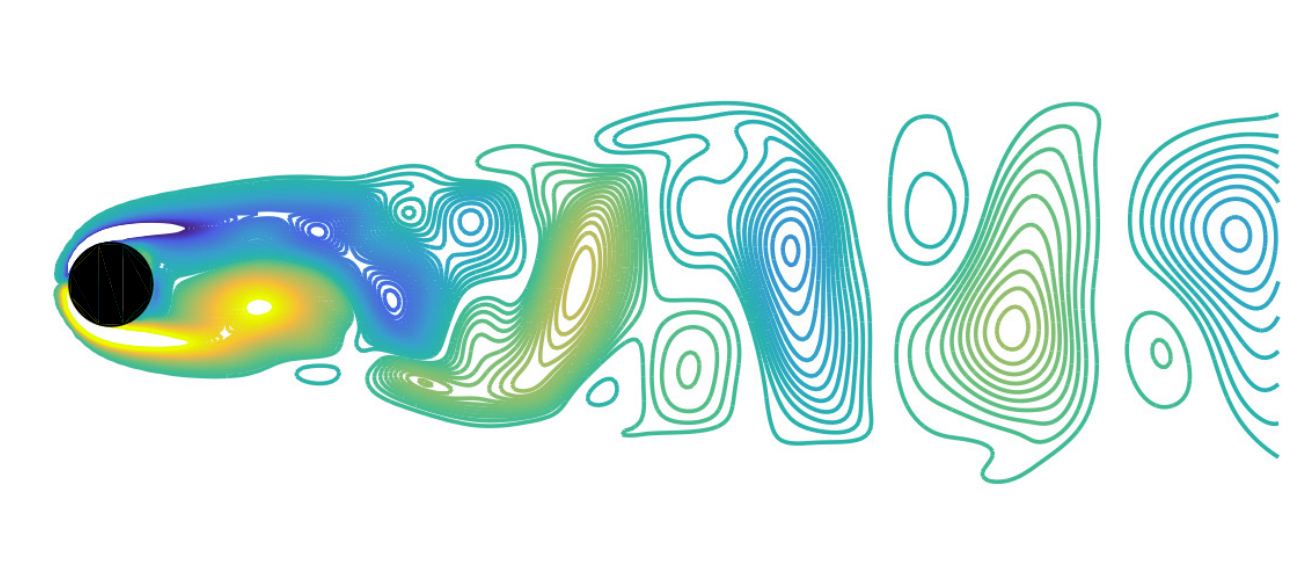}  & ~~~
	\begin{overpic}[height=0.175\textwidth]{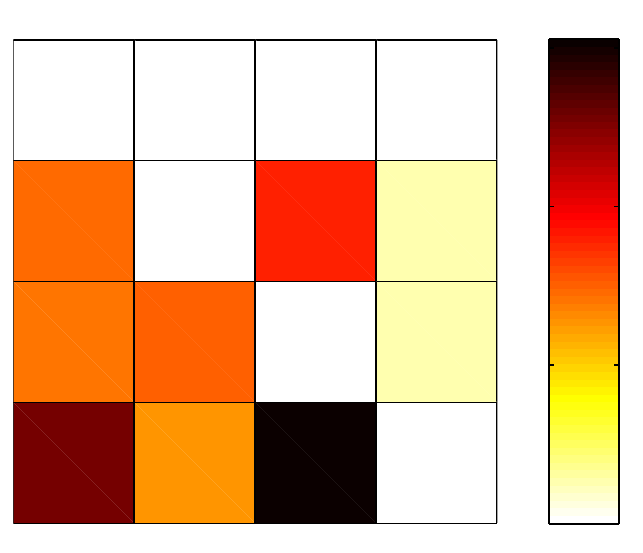} 
	  \put(20,85){$|[\boldsymbol{A}_{\mathcal{G}}]_{mn}|$}	  
	   \put(9,-7){\small I}   \put(25,-7){\small II}   \put(42,-7){\small III}   \put(62,-7){\small IV}
	   \put(-9,65){\small I} \put(-11,48){\small II} \put(-13,30){\small III} \put(-13,10){\small IV}
	   \put(100,2.5){\scriptsize $0$}
	   \put(100,15){\scriptsize $0.1$}
	   \put(100,27.5){\scriptsize $0.2$}
	   \put(100,40){\scriptsize $0.3$}
	   \put(100,52.5){\scriptsize $0.4$}
	   \put(100,65){\scriptsize $0.5$}
	   \put(100,77.5){\scriptsize $0.6$}	
	   \put(40,-20){\scriptsize $n$}
	    \put(-20,40){\scriptsize \rotatebox{90}{$m$}}   
	\end{overpic} 
	\qquad~~
	\begin{overpic}[height=0.175\textwidth]{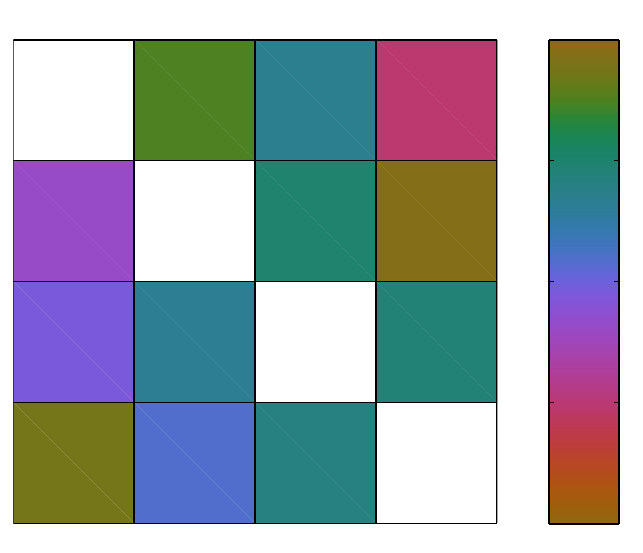} 
	 \put(20,85){$\angle[\boldsymbol{A}_{\mathcal{G}}]_{mn}$}
	   \put(9,-7){\small I}   \put(25,-7){\small II}   \put(42,-7){\small III}   \put(62,-7){\small IV}
	   \put(-9,65){\small I} \put(-11,48){\small II} \put(-13,30){\small III} \put(-13,10){\small IV}
	    \put(100,2.5){\scriptsize $-\pi$}
	   \put(100,21.5){\scriptsize $-\pi/2$}
	   \put(100,39.5){\scriptsize $0$}
	   \put(100,58.5){\scriptsize $\pi/2$}
	   \put(100,77.5){\scriptsize $\pi$}	   
	    \put(40,-20){\scriptsize $n$}
	    \put(-20,40){\scriptsize \rotatebox{90}{$m$}}
	\end{overpic} 	
	\vspace{0.3in}
	\end{tabular}
	\break
	(c) Oscillator dynamics \includegraphics[width=0.2\textwidth]{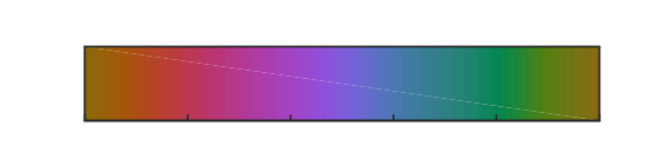}
	\put(-70,-3){\scriptsize $-\pi$}
	   \put(-38,-3){\scriptsize $0$}
	   \put(-10,-3){\scriptsize $\pi$}	   
	\break \break
	\begin{tabular}{c|cccc}	
	\vspace{-0.05in}
	 & ~~oscillator I & oscillator II & oscillator III & oscillator IV \\ \hline
	\rotatebox{90}{\qquad \quad~ DNS} & 
	~~\begin{overpic}[width=0.2\textwidth]{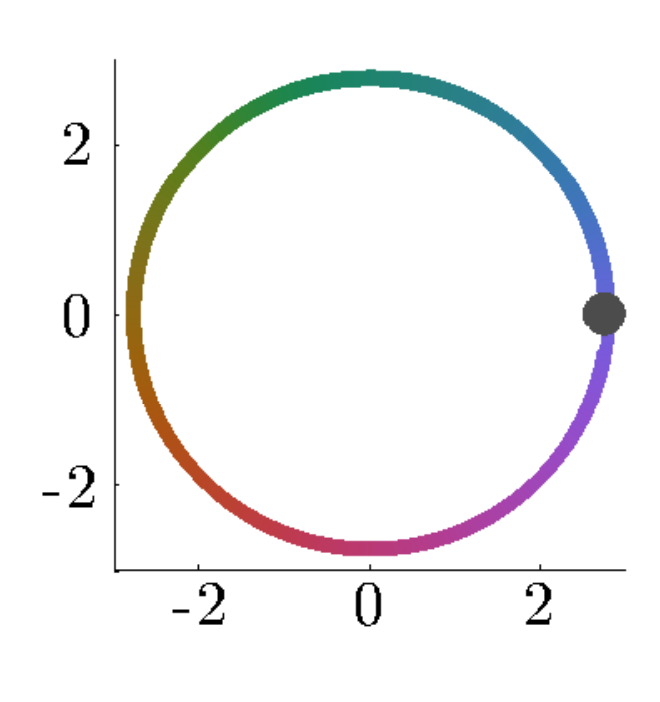}
	   \put(48,-1){\small $a_1$}
	   \put(-5,53){\small $a_2$}
	\end{overpic} &
	\begin{overpic}[width=0.2\textwidth]{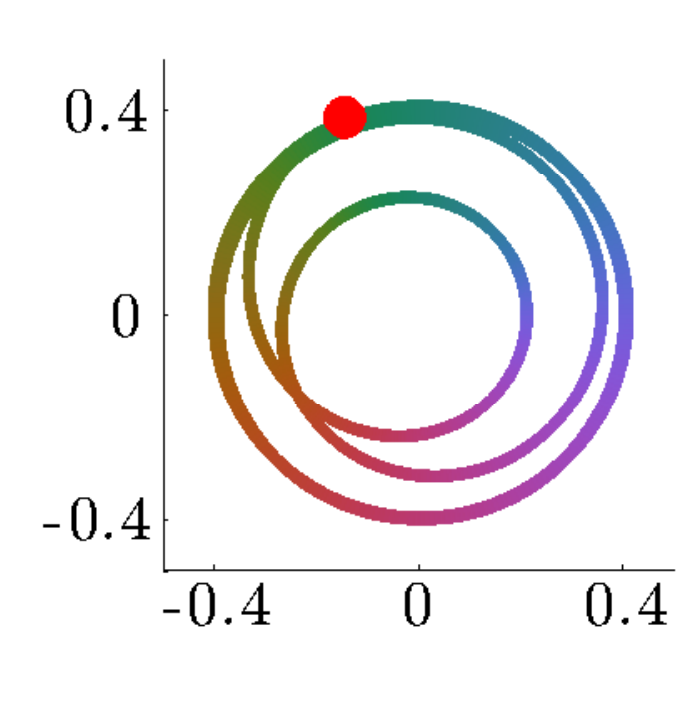}
	\put(55,-1){\small $a_3$}
	\put(-5,53){\small $a_4$}
\put(59,56){\line(-3,-1){26}}
\put(59,56){\line(-1,3){8}}
\put(41,60){\tiny{ $\theta^\prime_\text{II}$}}
	\end{overpic} &	
	\begin{overpic}[width=0.2\textwidth]{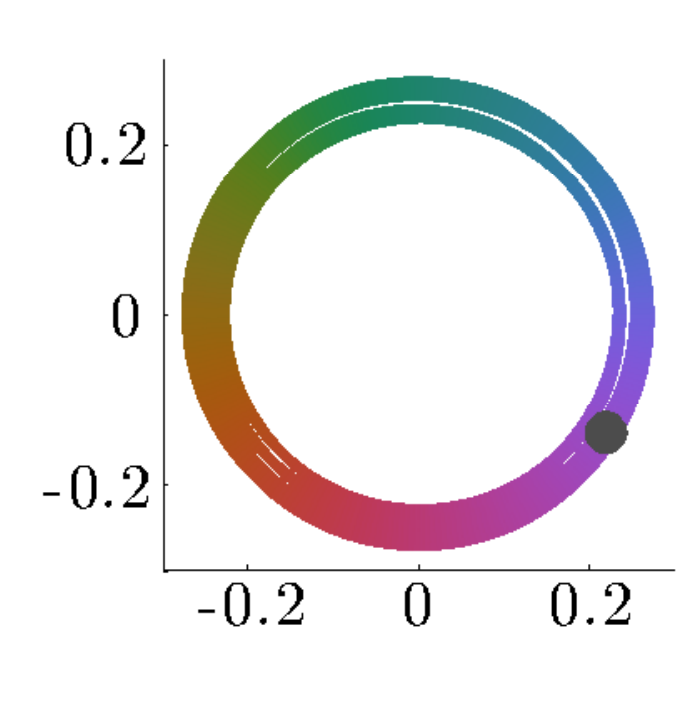}
	\put(55,-1){\small $a_5$}
	\put(-5,53){\small $a_6$}
	\end{overpic} &	
	\begin{overpic}[width=0.2\textwidth]{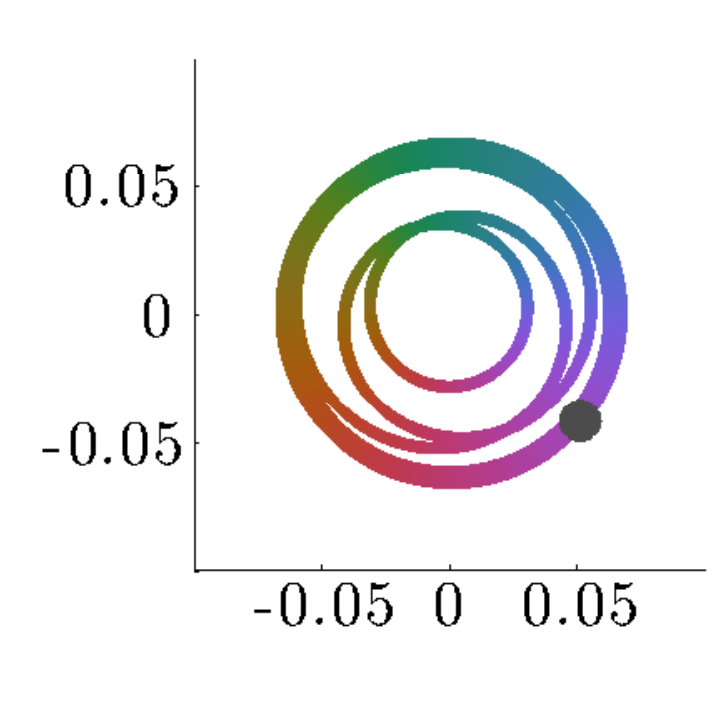}
	\put(57,-1){\small $a_7$}
	\put(-5,53){\small $a_8$}
	\end{overpic} 
	\\
	\rotatebox{90}{\quad ~~Network model} & 
	~~\begin{overpic}[width=0.2\textwidth]{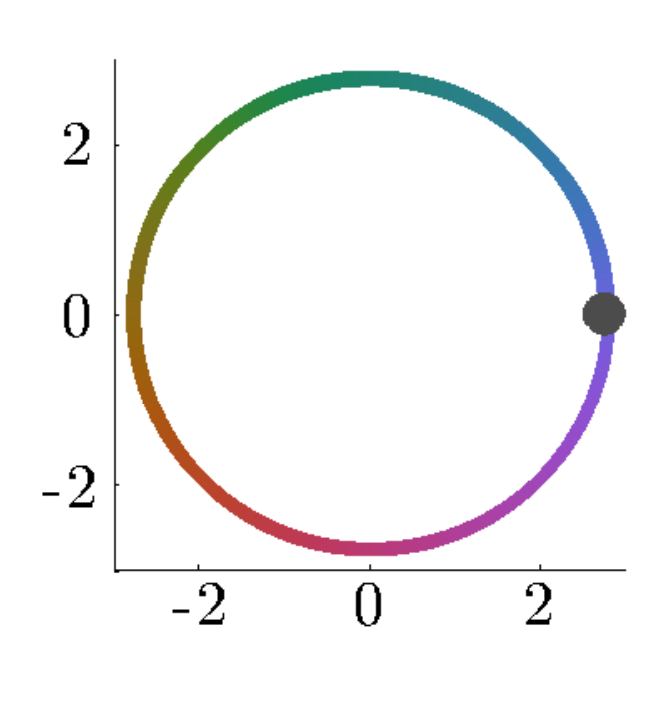}
	   \put(48,-1){\small $a_1$}
	   \put(-5,53){\small $a_2$}
	\end{overpic} &
	\begin{overpic}[width=0.2\textwidth]{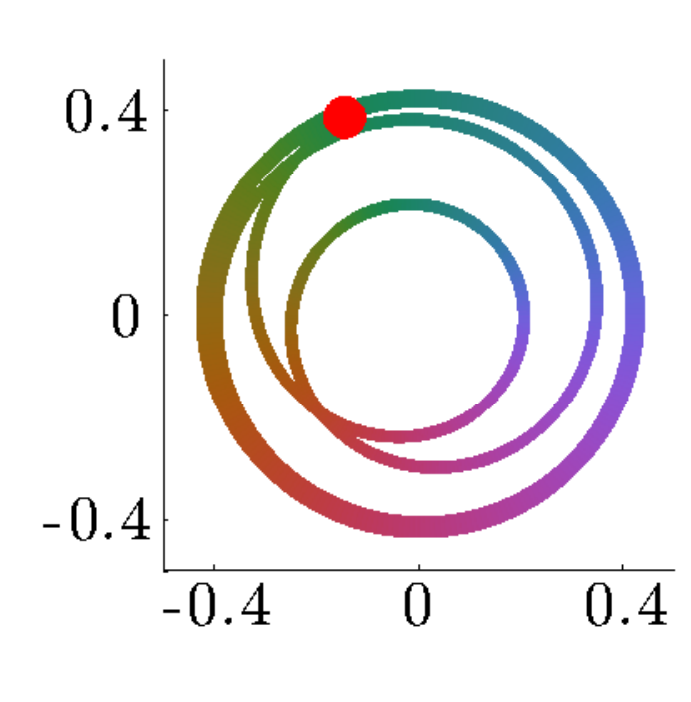}
	\put(55,-1){\small $a_3$}
	   \put(-5,53){\small $a_4$}
	\end{overpic} &	
	\begin{overpic}[width=0.2\textwidth]{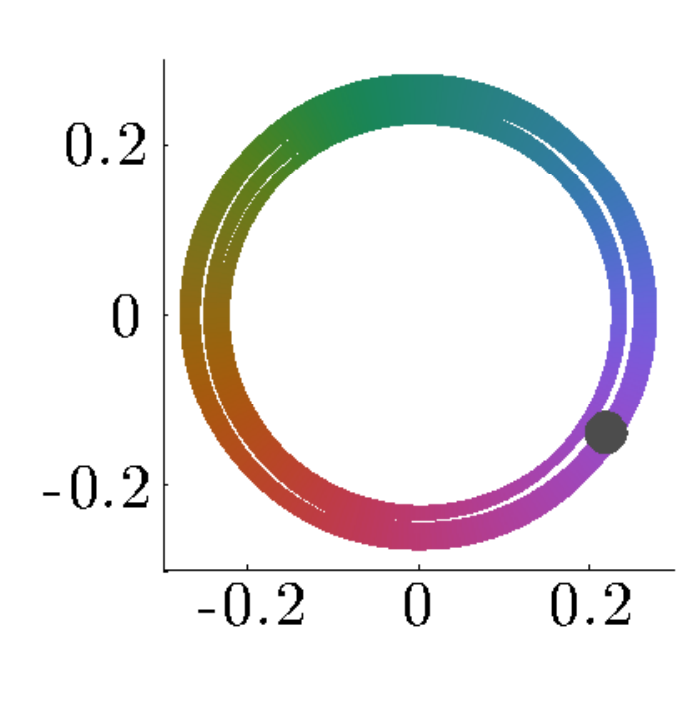}
	\put(55,-1){\small $a_5$}
	   \put(-5,53){\small $a_6$}
	\end{overpic} &	
	\begin{overpic}[width=0.2\textwidth]{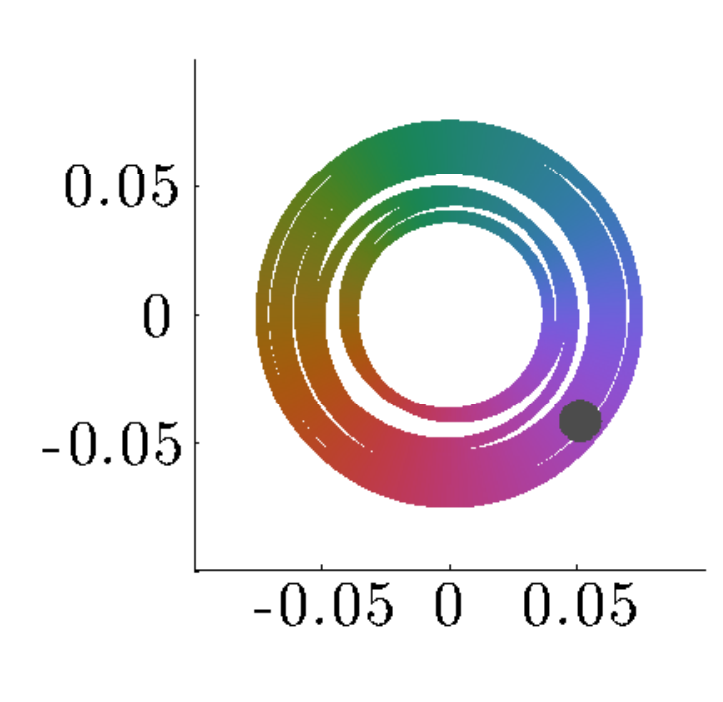}
	\put(57,-1){\small $a_7$}
	   \put(-5,53){\small $a_8$}
	\end{overpic}  \\
	\rotatebox{90}{\quad ~~Galerkin model} & 
	~~\begin{overpic}[width=0.2\textwidth]{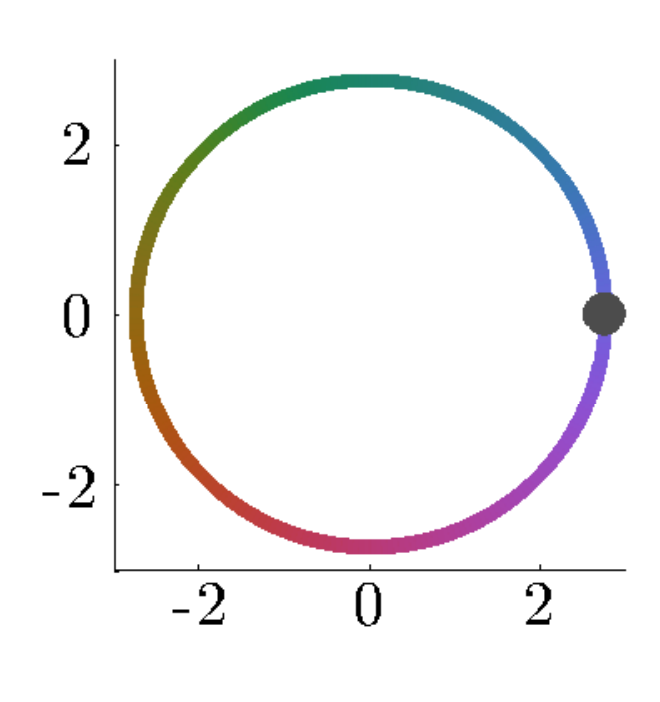}
	   \put(48,-1){\small $a_1$}
	   \put(-5,53){\small $a_2$}
	\end{overpic} &
	\begin{overpic}[width=0.2\textwidth]{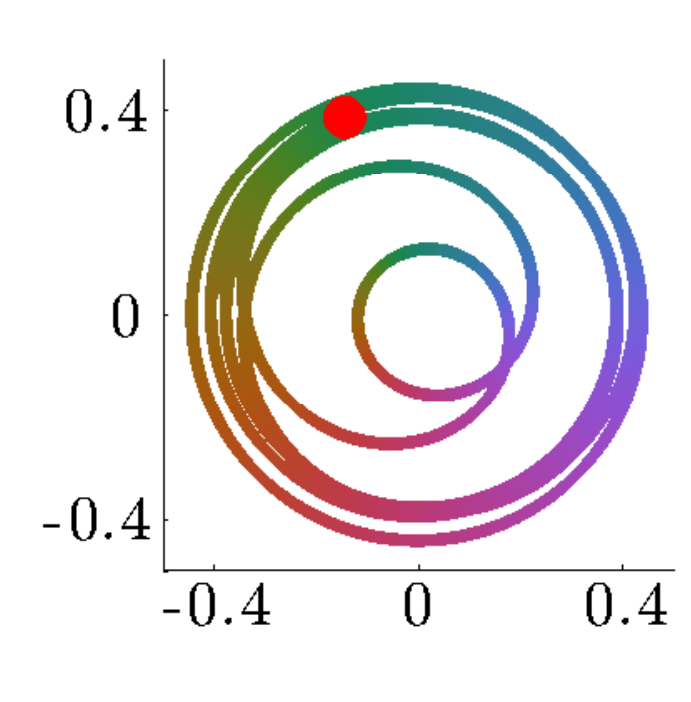}
	\put(55,-1){\small $a_3$}
	\put(-5,53){\small $a_4$}
	\end{overpic} &	
	\begin{overpic}[width=0.2\textwidth]{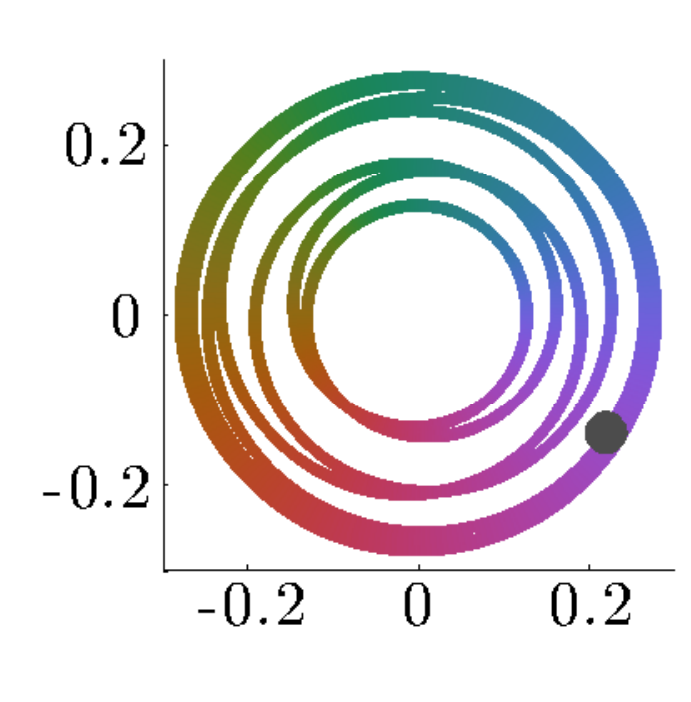}
	\put(55,-1){\small $a_5$}
	\put(-5,53){\small $a_6$}
	\end{overpic} &	
	\begin{overpic}[width=0.2\textwidth]{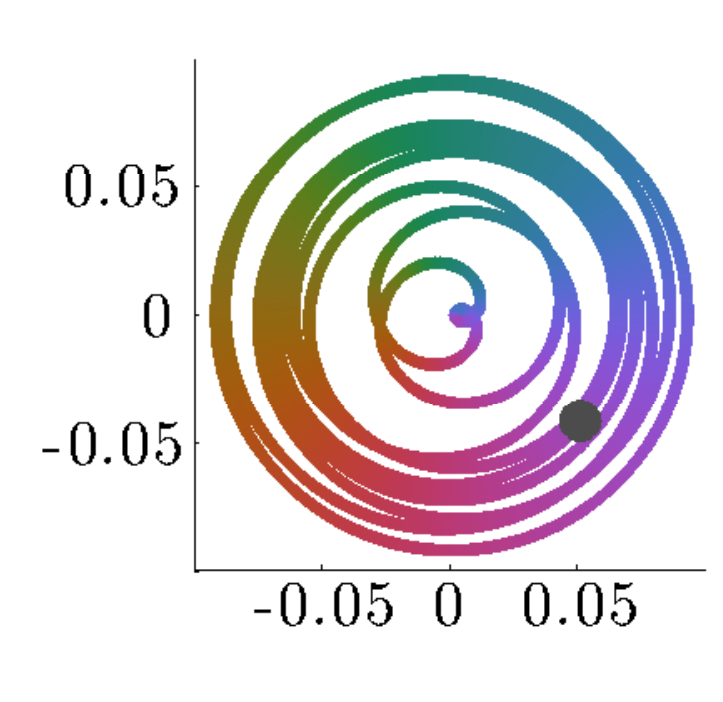}
	\put(57,-1){\small $a_7$}
	\put(-5,53){\small $a_8$}
	\end{overpic}  \\ \hline
	\end{tabular}
  \vspace{2mm}
 \caption{Phase perturbation introduced at oscillator II. (a) Initial vorticity field corresponding to phase perturbation, (b) Adjacency matrix of networked oscillator model.  (c) Dynamics of DNS, networked oscillator model and POD-Galerkin model (color of oscillators indicates phase). Red dot indicates perturbed state.}
   \label{fig5}
\end{center}  
\end{figure}
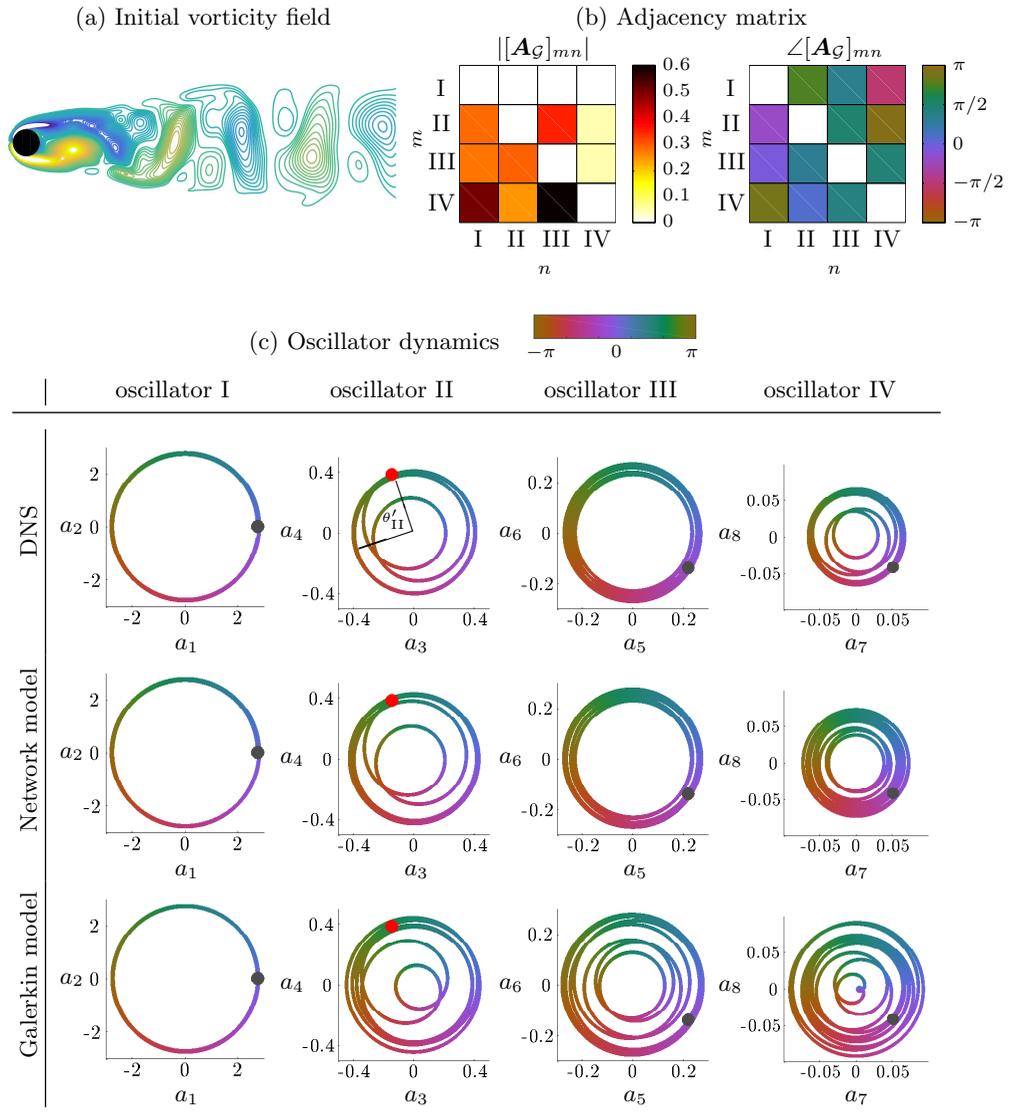

Similar to tracking amplitude perturbations, we can also analyze the flow response to phase perturbations. These phase perturbations are introduced by shifting the initial phase of the oscillators at $t_0$. As energy input is related to the amplitude of the perturbations, phase perturbations do not introduce any additional energy to the flow. Here, let us introduce a phase perturbation of size $\theta_\text{II}^\prime(t_0) = -\pi/2$ by shifting the phase associated with oscillator II at $t_0$ in the baseline case in the clockwise direction by $\pi/2$. The initial vorticity field resulting from the phase perturbation is shown in figure \ref{fig5} (a). Over time, both the amplitude and phase of the oscillators are affected with the introduction of phase perturbations in DNS. These fluctuations are tracked and the network structure is extracted as shown in figure \ref{fig5} (b). As seen from the magnitude of the edge weights, similar to the amplitude perturbation case, oscillator I dynamics are negligibly affected by the other oscillators. Furthermore, oscillator IV does not contribute much to the dynamics of oscillators II and III. Deductions of modal interactions can be made from the phase corresponding to the edge weights.

\begin{figure}
   \begin{center}
    \begin{tabular}{c} 
    \hspace{-0.175in}
          \begin{overpic}[width=5.5in]{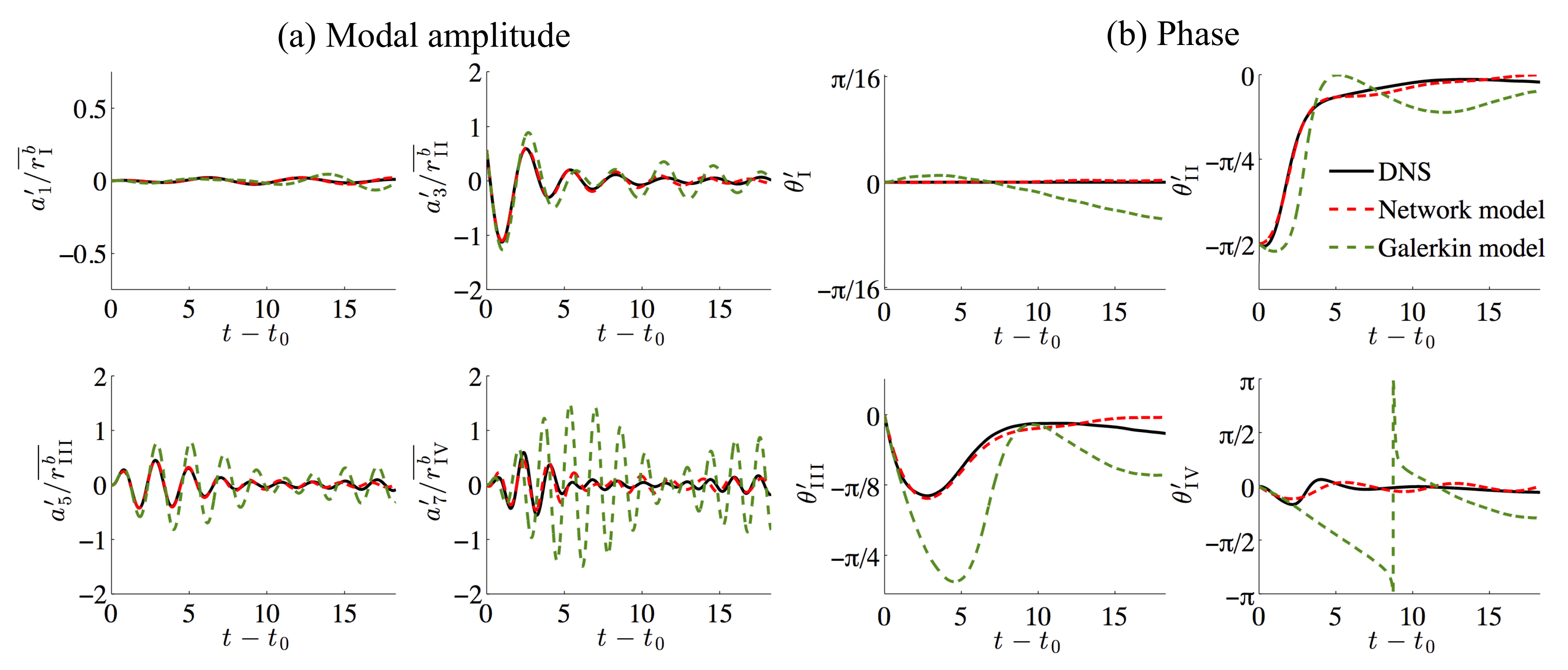}
          \end{overpic} 
    \end{tabular}
   \end{center}
   \caption{(a) Modal amplitude and (b) phase tracking for amplitude perturbation introduced at oscillator II.}
   \label{fig6}
\end{figure}

The oscillator dynamics for the phase perturbation case are shown in figure \ref{fig5} (c). We notice that the phase perturbation introduced in oscillator II at $t_0$ results in a phase shift to an initial condition indicated by the red dot. Similar to the amplitude perturbation case, the dynamics in the networked oscillator model show better agreement with the DNS than the POD-Galerkin model. This is confirmed on tracking the amplitude fluctuations $(a_{2n-1}^\prime)$ of the individual modes and corresponding phase dynamics $(\theta_m^\prime)$ as shown in figure \ref{fig6} (a) and (b), respectively. The POD-Galerkin model is unable to correctly track the phase of the oscillators and hence overpredicts the amplitude of the fluctuations. 

\subsubsection{Aggregate network model}

The perturbed flow cases examined above are limited to amplitude and phase perturbations introduced to oscillator II only. Similar analyses can be performed for oscillators I, III and IV with various combinations of amplitude and phase perturbations introduced to each oscillator. For each case, the network structure of interactions can be extracted individually by tracking the trajectory of the oscillator fluctuations from DNS as demonstrated in the approach above. While the corresponding linear networked oscillator models can be built for each specific case, the resulting network structure obtained depends on the initial perturbations introduced. Thus, one may argue that such individually tuned model does not necessarily capture the overall interactions in a generally effective manner for any perturbation. Such a general network interaction model is desired not only to capture interactions for any combination of perturbations but also to design effective strategies for flow control. In this section, we describe our approach to build an aggregate network model that captures the fluid flow response to a variety of perturbation inputs.     

We consider a range of perturbed flow cases by varying the amplitude and phase perturbation size and to which combination of oscillators to perturb. For a particular oscillator perturbed, we vary the energy input ($\beta_m$) from $0.05$ to $1$ ($5\%$ to $100\%$ additional baseline modal energy in the flow). We also vary the phase perturbation size ($\theta_m^\prime$) from $-2\pi/3$ to $2\pi/3$ to cover a broad coverage of plausible initial perturbations introduced. The range of amplitude and phase perturbations considered here are quite large. These large perturbations lead to the emergence of strong nonlinear interactions between the oscillators. However, we are interested in extracting a linear networked oscillator model to approximate these nonlinear interactions for the purpose of control. For the multitude of perturbation cases, we collect the oscillator fluctuation data from DNS. For the aggregate network model, we concatenate the trajectories of all the perturbed cases obtained from DNS. We then compute the normalized fluctuation $\zeta_m$ and its time derivative $\dot{\zeta}_m$, instead of individually tracking perturbations and monitoring the oscillator fluctuations in each perturbed case. 

\begin{figure}
   \begin{center}
    \begin{tabular}{c} 
          \begin{overpic}[width=5.2in]{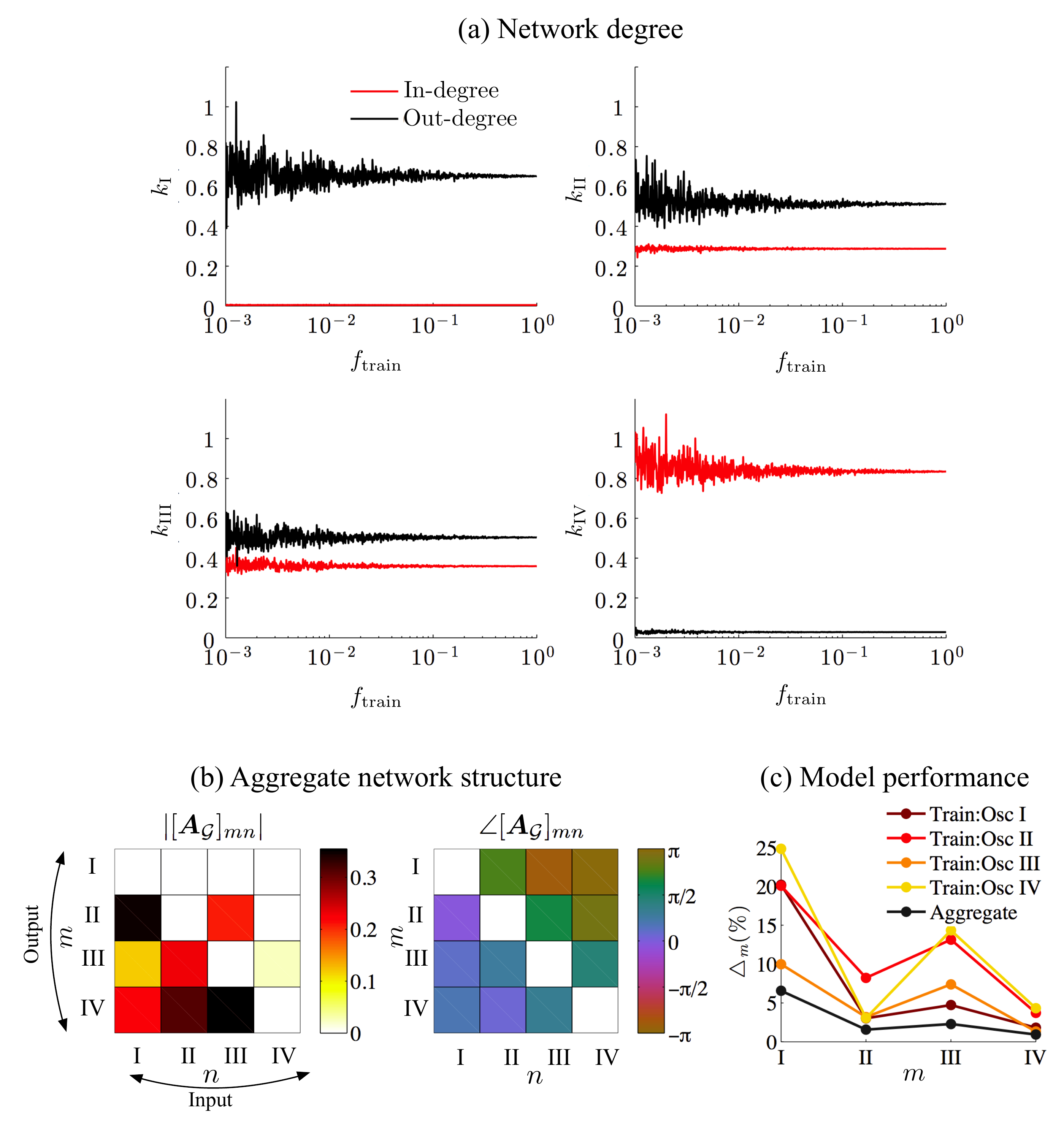}
          \end{overpic} 
    \end{tabular}
   \end{center}
   \caption{(a) Network degree for varying fractions of training data. (b) Adjacency matrix of aggregate network model extracted and (c) associated model performance.}
   \label{fig7}
\end{figure}

We then segregate the combined data into training and test sets for performing cross-validation and evaluating the predictive capabilities of the network model. Varying fractions of the combined input-output data are randomly chosen as the training sets. For each training set, regression analysis is performed to extract a corresponding network model. We then examine the in-degree ($k_m = \sum_{n=\text{I}}^M [{\boldsymbol{A}}_\mathcal{G}]_{mn}$) and out-degree ($k_n = \sum_{m=\text{I}}^M [{\boldsymbol{A}}_\mathcal{G}]_{mn}$) of the network nodes for each model extrated. The variation of the network degrees with respect to the fraction of the chosen training set is shown in figure \ref{fig7} (a). We observe that as the fraction of the training data used increases, convergence of the network degree is obtained. We also learn that the average in-degree increases from low-order oscillators to higher-order oscillators and the average out-degree decreases. Oscillators I and IV have maximum out-degree and in-degree respectively so that oscillator I influences the other oscillators most while oscillator IV is the most influenced. Oscillators II and III have a more balanced in and out-degree indicating more balanced energy transfer for each mode.

Based on the network degree convergence, a fraction of training data $f_\text{train} = 0.8$ corresponding to $80 \%$ of the combined input-output data is randomly chosen as the training set and the remaining $20\%$ of the data is used to assess the performance of the network model extracted. Using the training set, the aggregate network model extracted is shown in figure \ref{fig7} (b). The magnitude of the network reveals that the lower-order oscillators have more influence on the dynamics of the higher-order oscillators. This follows from our earlier discussion regarding the network node degrees and also agrees with our intuition that energy is passed from lower-order oscillators to higher-order oscillators. Alternatively, we can construct the aggregate network model by considering complete trajectories of $80 \%$ of randomly chosen perturbed cases yielding a very similar aggregate network model.

We also compare the aggregate network model with network models built with non-randomly chosen training sets. In particular, we consider trajectories of perturbed cases of individual oscillators. Combining the perturbed trajectories for various  amplitude and phase perturbation sizes for each oscillator, we build network models corresponding to oscillators I, II, III and IV. We use the same test data set as before to assess the performance of each of these models along with the aggregate model. We predict the time derivative of the normalized fluctuation $\dot{\zeta}_m$ with each of the models on the test data set and compare with DNS reference values. In figure \ref{fig7} (c), we assess the prediction error through a normalized root mean square deviation ($\triangle_m$), given by 
 \begin{equation}
\begin{split}
\triangle_m = \frac{1}{\text{max}(|\dot{\zeta}_m|)}\sqrt{\frac{\sum_{k=1}^{n_t}  |\hat{\dot{\zeta}}_m (k) - \dot{\zeta}_m (k)|^2}{n_t}},
\end{split}
 \end{equation}
where $n_t$ is the length of test data set. Here, $\dot{\zeta}_m(k)$ is the time derivative of oscillator $m$ obtained from DNS and $\hat{\dot{\zeta}}_m(k)$ is the predicted time derivative value based on oscillator network model for the $k$th test data set. The root mean square deviation is normalized by the range of the measured data. We can observe from figure \ref{fig7} (c) that the aggregate model achieves the least error in predicting the normalized fluctuation time derivative. As the aggregate model contains more information of the general interactions based on various oscillator perturbations, it yields enhanced predictive capabilities for energy transfer amongst all oscillators.  

Using the aggregate network structure shown in figure \ref{fig7} (b), we evaluate the performance of the networked oscillator model in tracking the fluctuations of the perturbation cases examined in this work. We compare the DNS trajectories of the oscillator dynamics with the aggregate network oscillator model with introduction of amplitude perturbations corresponding to $\beta_m = 0.2$ in figure \ref{fig8} in the Appendix. The aggregate networked oscillator model is able to track the amplitude and phase dynamics for each case with reasonable accuracy. We can see that the introduction of an amplitude perturbation in oscillator I does not cause large deviations in the higher-order oscillators. Moreover, the prescribed energy input requires only slight deviation from the limit cycle orbit for perturbations introduced on oscillator I as it possesses high level of energy to begin with. A perturbation introduced to oscillator III results large fluctuations in oscillators II and IV, while a perturbation introduced to oscillator IV results in comparable fluctuations in oscillator III but only small deviations in oscillator I and II dynamics. These features are well captured by the aggregate networked oscillator model. 


We also show the comparison of predicted trajectories of the aggregate networked oscillator model for phase perturbation cases shown in figure \ref{fig9} of the Appendix. As oscillator I contains large amounts of energy, a phase perturbation of $\theta_\text{I}^\prime$ causes an effective phase perturbation of $m\theta_n^\prime$ for oscillators II, III and IV. Thus, the phase perturbation on oscillator I induces a combined phase perturbation on the higher-order oscillators as they synchronize their phases with respect to oscillator I. Phase perturbations introduced in oscillator III cause fluctuations in oscillators II and IV, while phase perturbations introduced in oscillator IV do not produce any fluctuations in lower-order oscillators. As with the amplitude perturbation cases, the aggregate network model also captures the features associated with the phase perturbation cases. Thus, one aggregate network structure is able to capture a multitude of perturbation cases, yielding a global oscillator interaction model.  

The aggregate network model is not only useful to gain insights into energy transfer, but is also used to design feedback control laws to manipulate the flow unsteadiness. In this work, we only characterize the interactions between the baseline POD modes in the flow induced by the perturbations introduced. However, these perturbations result in the emergence of residual flow structures not captured by the baseline POD modes. Appendix B discusses how to capture the emerging modes. Based on the modal interactions characterized, we design flow control strategies to suppress modal fluctuations. In the next section, we discuss such flow control design strategies using the networked oscillator model in conjunction with a linear quadratic regulator (LQR), and discuss how the residual behavior not captured by the modal dynamics influence the controlled flow.     


\subsection{Feedback control}
\label{subsec:ctrlflow}

With the modal network interactions characterized and the aggregate network model developed in the previous section, we now design a feedback control strategy to suppress modal oscillations. As mentioned previously, modal oscillations in the flow are reflected in the temporal coefficients associated with the modes. Suppression of these modal oscillations is critical in reducing the wake unsteadiness. \cite{roshkodrag} and \cite{mao2015nonlinear} reported the strong relationship between unsteadiness in the wake and the drag force acting on the bluff body. If the application of control forces the modal temporal coefficients to zero, the flow will approach the mean flow. However, the mean flow is not a steady solution to the Navier--Stokes equation in general, and hence the flow diverges from the mean towards the unstable steady state in this cylinder flow case.  

\begin{figure}
   \begin{center}
    \begin{tabular}{c} 
          \begin{overpic}[width=5in]{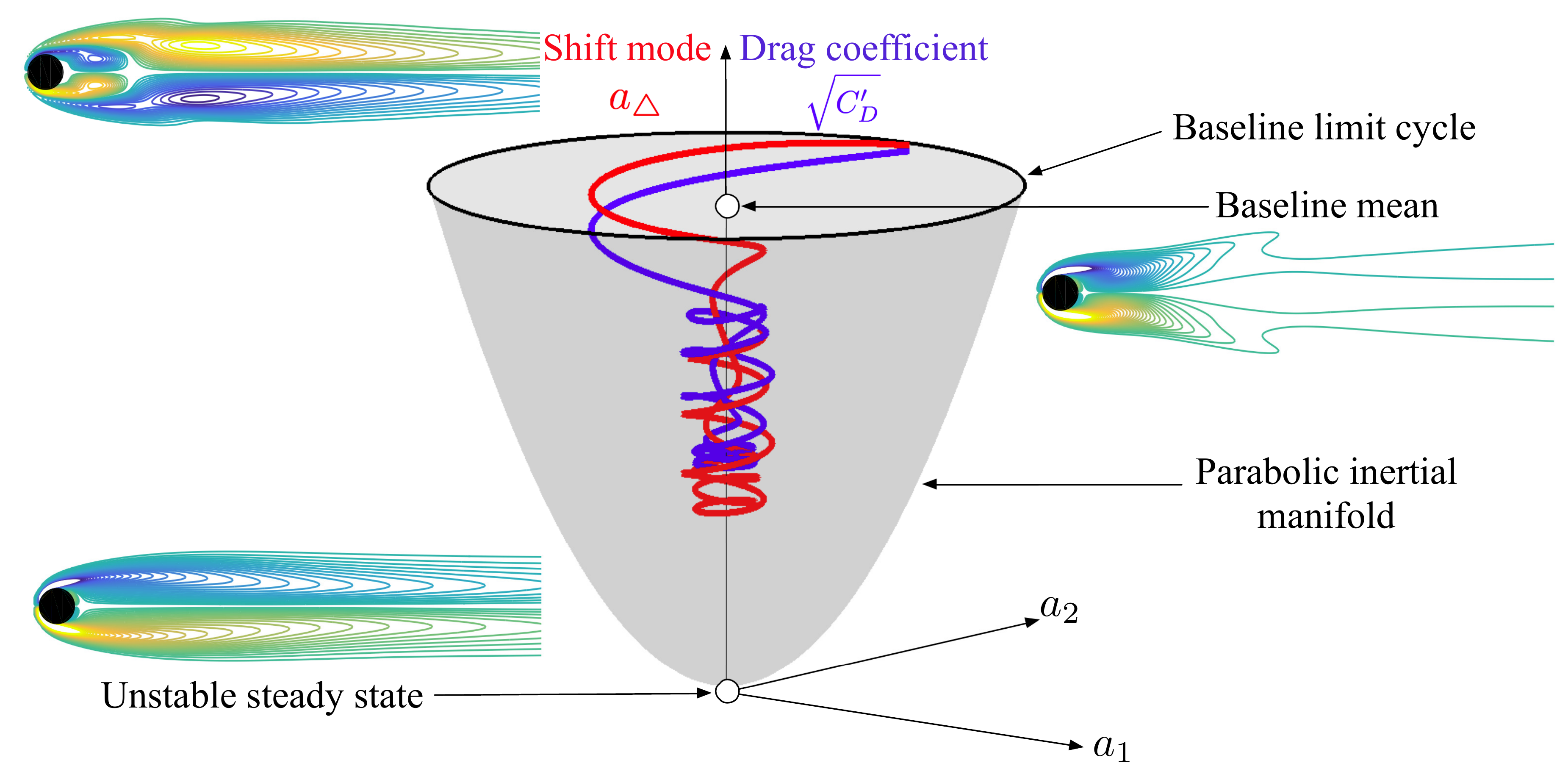}
          \end{overpic} 
    \end{tabular}
   \end{center}
   \caption{Shift mode temporal coefficient (red) and drag coefficient (blue) variation with application of control. Parabolic inertial manifold shown in grey.}
   \label{fig15}
\end{figure}

In the seminal work of \cite{Noack:JFM03}, it was shown that the mean flow and the unstable steady state are connected by a shift mode. The shift mode for the cylinder flow problem is shown in figure \ref{fig15} (top left). This shift mode captures the transient dynamics between the onset of vortex shedding near the unstable steady state shown in figure \ref{fig15} (bottom left) and the baseline mean flow of the periodic von K\'{a}rm\'{a}n vortex street in the globally stable limit cycle. This evolution takes place along the parabolic inertial manifold shown in grey in figure \ref{fig15}. Here, the limit cycle of oscillator I (modes $1$ and $2$) is illustrated in the $a_1$-$a_2$ plane. The vertical axis represents the change in the temporal dynamics associated with the shift mode (as shown in red) with application of control. The drag coefficient also varies between the mean flow and unstable fixed point. A minimum drag coefficient is attained at the unstable steady state $C_D^*$, which also gives zero lift force on the cylinder. The kinetic energy associated with the shift mode varies as $a_\triangle^2$ and the drag force on the cylinder in the mean shift regime scales as $\sqrt{C_D^\prime} \propto a_\triangle$, as shown in figure \ref{fig15} (blue) where, $C_D^\prime = C_D - C_D^*$. Thus, a realization of mean shift towards the unstable steady state yields a reduction in drag force.

We first develop a low-dimensional control framework based on the networked oscillator model to attenuate perturbations in the flow. We then extend the formulation to suppress the overall flow unsteadiness. In the present work, the aggregate networked oscillator model is a linear ordinary differential equation given by Eq. (\ref{eq:zeta}) governing the modal perturbation dynamics in the flow. As the networked oscillator model is linear, we can exploit the use of the LQR to control the modal fluctuations. Adding a forcing input to the networked oscillator model (in vector form), we arrive at
\begin{equation}
\dot{\boldsymbol{\zeta}} = -\boldsymbol{L}_{\mathcal{G}} \boldsymbol{\zeta} + \boldsymbol{B}\boldsymbol{v},
\label{eq:zetaco1}
\end{equation}
where $\boldsymbol{\zeta} = [\zeta_\text{I}, \zeta_\text{II} \dots \zeta_M]^T$, $\boldsymbol{v} \in \mathbb{C}^{M \times 1}$ is the forcing input and $\boldsymbol{B} \in \mathbb{R}^{M \times 1}$ is the actuation input matrix. The $m^{th}$ entry of $\boldsymbol{B}$ corresponds to forcing added to oscillator $m$. 

We implement optimal full-state control with $\boldsymbol{v} = -\boldsymbol{K}\boldsymbol{\zeta}$ such that 
\begin{equation}
\dot{\boldsymbol{\zeta}} = (-\boldsymbol{L}_{\mathcal{G}}   - \boldsymbol{B}\boldsymbol{K}) \boldsymbol{\zeta},
\label{eq:zetaco}
\end{equation}
where the gain matrix $\boldsymbol{K}$ is determined from the Riccati equation for LQR. An optimal control strategy using LQR minimizes the quadratic cost function of the form, 
\begin{equation}
J = \int_0^\infty \left[\boldsymbol{\zeta}(t)^T \boldsymbol{Q} \boldsymbol{\zeta}(t) + \boldsymbol{v}(t)^T \boldsymbol{S} \boldsymbol{v}(t)\right]\rm{d}{\it t},
\label{eq:costco}
\end{equation}
where $\boldsymbol{Q}$ and $\boldsymbol{S}$ are the state deviation and input usage weights respectively. In the control studies considered henceforth, we set $\boldsymbol{Q} = \boldsymbol{I}$ and $\boldsymbol{S} = \sigma \boldsymbol{I}$ and consider a range of values for $\sigma$. The case corresponding to $\sigma=1$ weighs the state deviation cost and input usage cost equally while $\sigma < 1$ weighs the state deviation cost more. 

\begin{figure}
   \begin{center}
    \begin{tabular}{c} 
    \hspace{-0.2in}
          \begin{overpic}[width=5.5in]{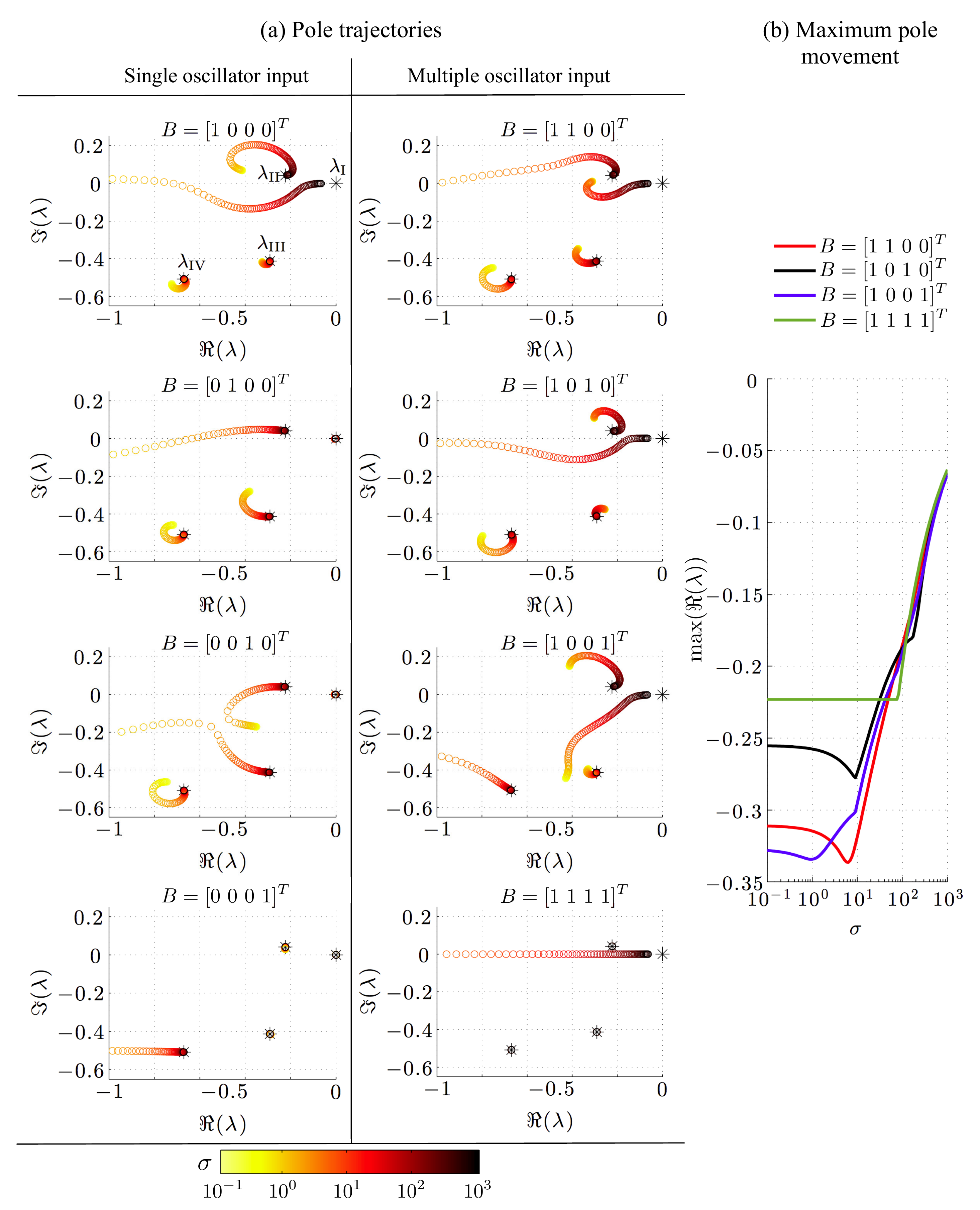}
          \end{overpic} 
    \end{tabular}
   \end{center}
   \caption{(a) Pole trajectories with application of different control inputs to the aggregate network model for a range of $\sigma$. Stars indicate the eigenvalues of the Laplacian matrix. (b) Maximum pole movement for the multiple oscillator input cases. }
   \label{fig16}
\end{figure}

We can force the individual oscillators or a combination of oscillators (reflected in $\boldsymbol{B}$). To aid the selection of which oscillators to force, we examine the movement of the eigenvalues of the Laplacian matrix with and without control for different combinations of input matrix $\boldsymbol{B}$ as shown in figure \ref{fig16} (a). Eigenvalues of Laplacian matrix ($-\boldsymbol{L}_{\mathcal{G}}$) reveal the dynamical characteristics of the system. As four oscillators are considered in this work, we obtain four eigenvalues ($\lambda_\text{I} \dots \lambda_\text{IV}$) denoted by the use of star symbols in figure \ref{fig16} (a) for varying $\sigma$. Note that $\lambda_m \le 0$ for all eigenvalues which is a characteristic feature of Laplacian-based systems. It should be noted that the presence of $\lambda_m = 0$ poses difficulties in computing the empirical Grammians to quantify the degrees of controllability. This issue requires careful assessment for network-based control design and will be investigated in upcoming studies. 

We compare the response for input matrix $\boldsymbol{B}$ for both single oscillator input and some multiple oscillator input cases. For each input case, we determine the LQR gain matrix $\boldsymbol{K}$ for values of $\sigma$ ranging from $0.1$ to $1000$ and examine the movement of eigenvalues of  ($-\boldsymbol{L}_{\mathcal{G}}   - \boldsymbol{B}\boldsymbol{K}$) which govern the behavior of the controlled system. For the single-oscillator input cases, we observe large movement in the eigenvalue corresponding to the forced oscillator as $\sigma$ is decreased, i.e., the real part of $\lambda_m$ decreases when more forcing input is provided to oscillator $m$. We also notice that an input in oscillator I affects all eigenvalues while inputs to the higher-order oscillators do not move the $\lambda_\text{I}$ eigenvalue. This is expected as oscillator I has maximum out-degree and correspondingly has the highest influence in the network. 

As input on oscillator I is required to affect the $\lambda_\text{I}$ eigenvalue, we consider multiple oscillator input cases including oscillator I. A noticeable movement in the eigenvalues is observed with inputs on oscillators I and IV. For forcing inputs added to all oscillators, only $\lambda_\text{I}$ eigenvalue is affected with no influence on the other system eigenvalues. To summarize the effectiveness of forcing input on controlling the system behavior, we track the $\text{max}(\Re(\lambda))$ for multiple oscillator input cases in figure \ref{fig16} (b). We observe that for small $\sigma$, oscillator input combinations of I and IV outperform the other input combinations while, for mid-range values of $\sigma$, oscillator combinations of I and III exhibit better control performance. 

We design the LQR controller such that we move system eigenvalues towards the left side of the complex plane. Thus, we select the input matrix $\boldsymbol{B} = [1~0~0~1]^T$ which adds forcing input to oscillators I and IV. Considering full-state feedback control with this choice of $\boldsymbol{B}$, we compute the control gain matrix $\boldsymbol{K}$ using LQR. In the following, we examine two control scenarios. First, we attenuate modal perturbations introduced in the flow. Then, we apply the  LQR-based feedback control to suppress all modal amplitudes altogether. This entails reducing the unsteadiness in the flow due to modal oscillations and achieve drag reduction.


\subsubsection{Disturbance rejection}

\begin{figure}  
\begin{center}  
	\break
	Oscillator dynamics \includegraphics[width=0.2\textwidth]{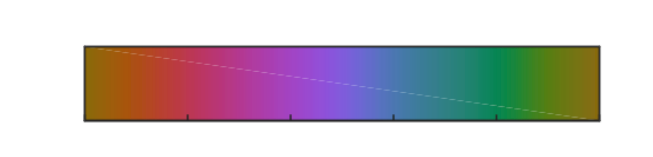}
	\put(-70,-3){\scriptsize $-\pi$}
	   \put(-38,-3){\scriptsize $0$}
	   \put(-10,-3){\scriptsize $\pi$}	   
	\break \break
	\begin{tabular}{c|cccc}	
	\vspace{-0.05in}
	 & ~~~~oscillator I & ~~~~oscillator II & ~~~~oscillator III & ~~~~oscillator IV \\ \hline
	\rotatebox{90}{\qquad \quad~ DNS} & 
	~~\begin{overpic}[width=0.2\textwidth]{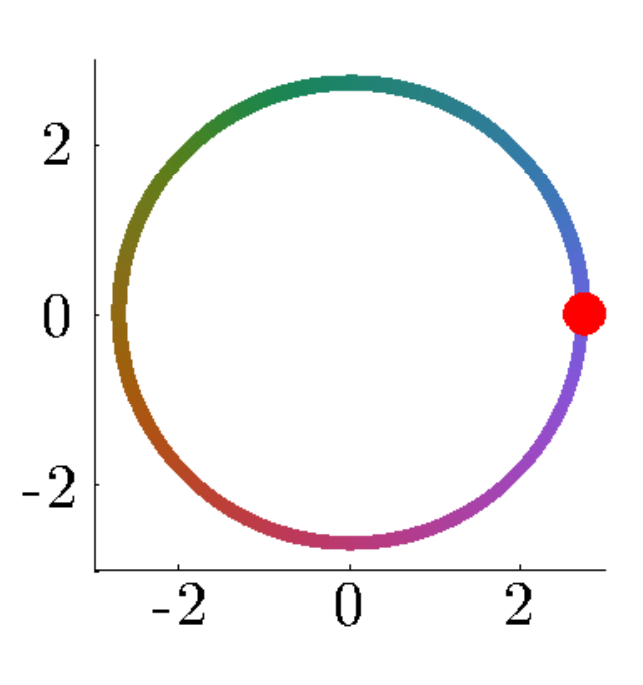}
	   \put(48,-1){\small $a_1$}
	   \put(-7,53){\small $a_2$}
	\end{overpic} &
	\begin{overpic}[width=0.2\textwidth]{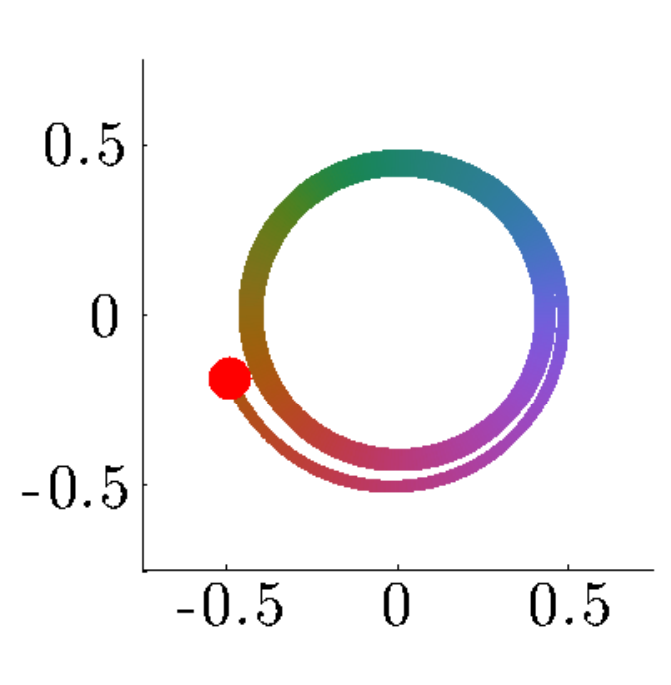}
	\put(54,-1){\small $a_3$}
	\put(-5,53){\small $a_4$}
	\linethickness{3pt}
	\end{overpic} &	
	\begin{overpic}[width=0.2\textwidth]{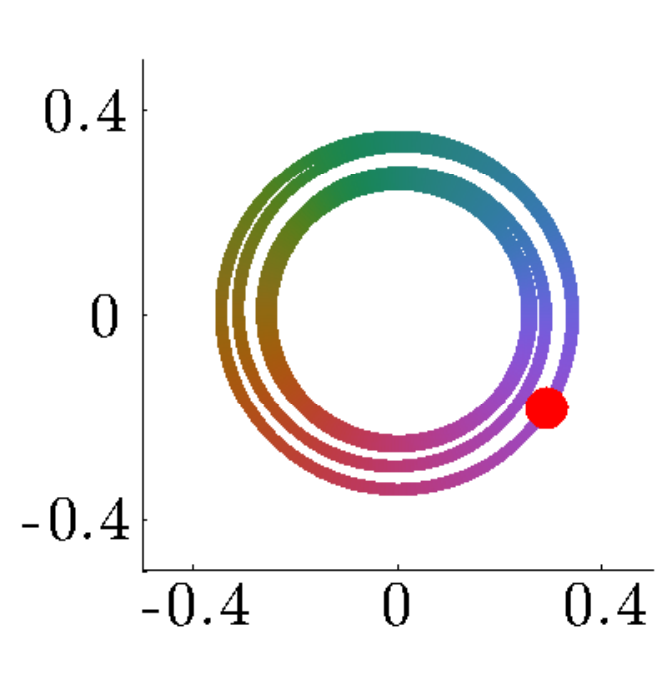}
	\put(54,-1){\small $a_5$}
	\put(-5,53){\small $a_6$}
	\end{overpic} &	
	\begin{overpic}[width=0.2\textwidth]{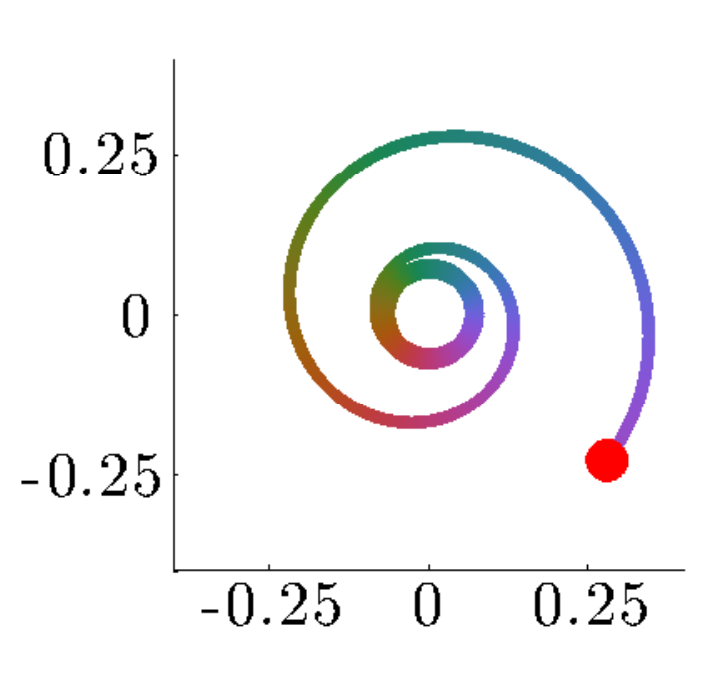}
	\put(56,-1){\small $a_7$}
	\put(-5,53){\small $a_8$}
	\end{overpic} 
	\\
        \rotatebox{90}{\quad ~~DNS with control }
	\rotatebox{90}{\quad ~~~~~~~~~($\sigma = 0.1$)}& 
	~~\begin{overpic}[width=0.2\textwidth]{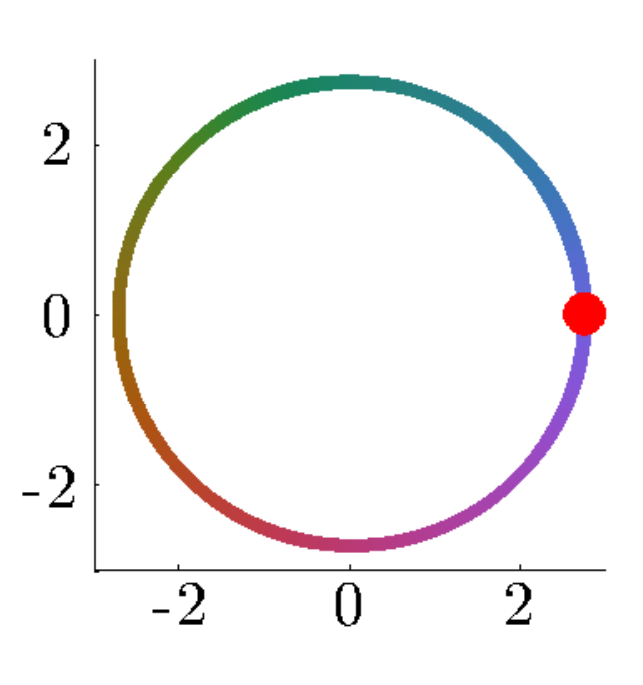}
	   \put(48,-1){\small $a_1$}
	   \put(-7,53){\small $a_2$}
	\end{overpic} &
	\begin{overpic}[width=0.2\textwidth]{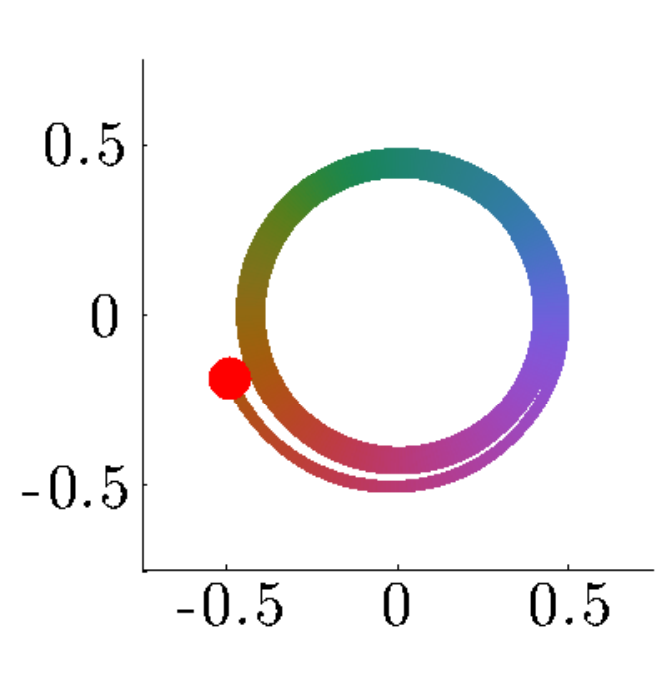}
	\put(54,-1){\small $a_3$}
	\put(-5,53){\small $a_4$}
	\end{overpic} &	
	\begin{overpic}[width=0.2\textwidth]{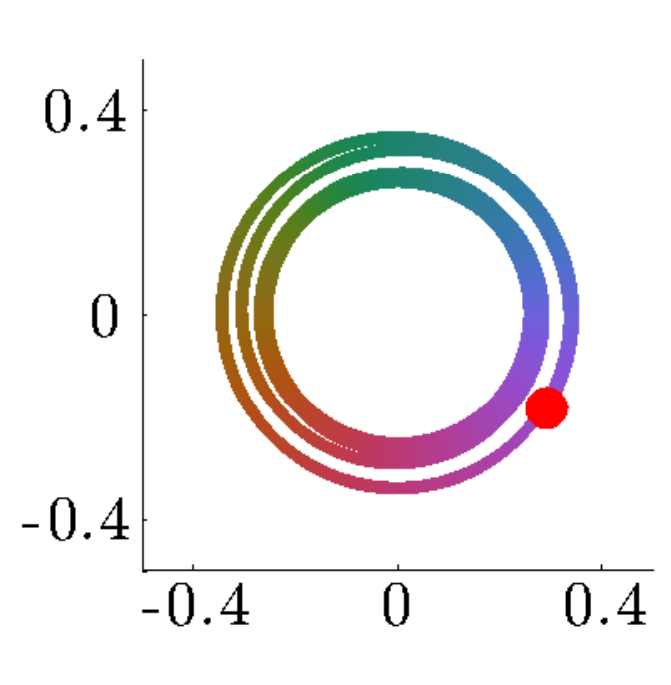}
	\put(54,-1){\small $a_5$}
	\put(-5,53){\small $a_6$}
	\end{overpic} &	
	\begin{overpic}[width=0.2\textwidth]{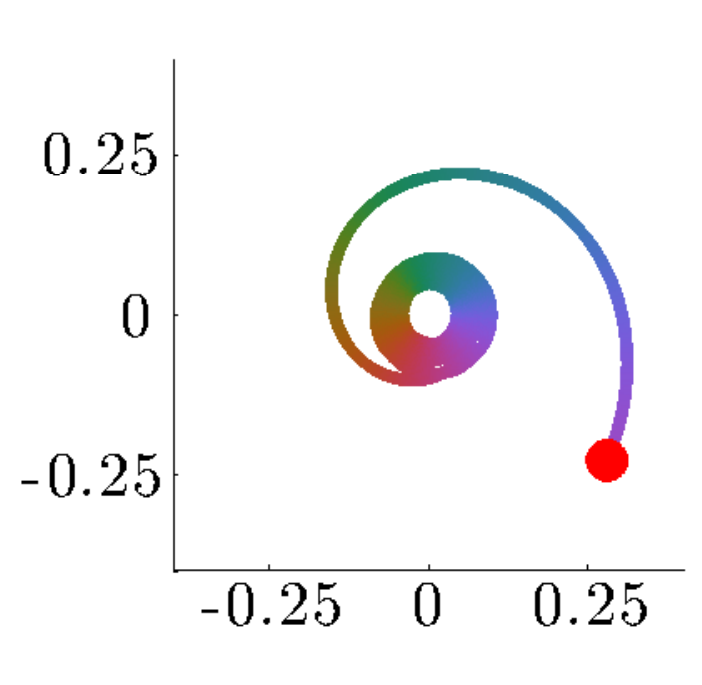}
	\put(56,-1){\small $a_7$}
	\put(-5,53){\small $a_8$}
	\end{overpic}  \\ \hline
	\end{tabular}
 \caption{Oscillator dynamics without control and with control implemented in DNS for suppressing perturbations to limit cycle.}
   \label{fig11}
\end{center}  
\end{figure}
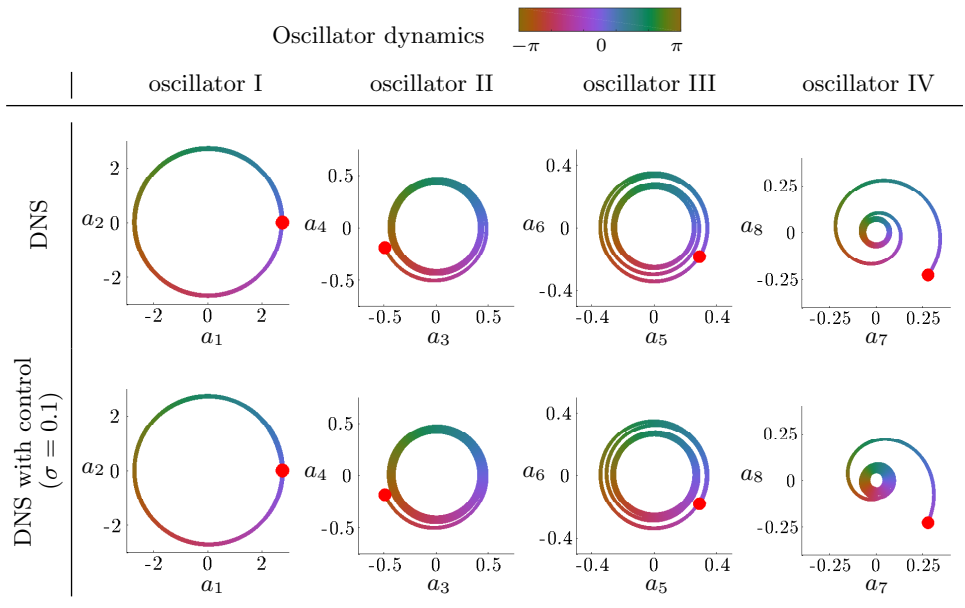

We first illustrate the control of modal disturbances introduced to the cylinder flow. This entails the control of fluctuations in the modal temporal coefficients ($z_m^\prime$). To demonstrate the control, we consider random amplitude perturbations added to all oscillators with $\text{max}(\beta_m) = 0.02$. The temporal coefficients of the oscillators for this perturbed case obtained from DNS are shown in figure \ref{fig11} (top). The networked oscillator control system given by Eq. (\ref{eq:zetaco}) describes control of temporal coefficients associated with the modes. To implement control design in DNS, a body force  corresponding to $- \boldsymbol{B}\boldsymbol{K}\boldsymbol{\zeta}$ is added to the momentum  equation with the spatial modal information is incorporated along with the temporal coefficients. The addition of body force is The results with the application of control in DNS for $\sigma = 0.1$ are shown in figure \ref{fig11} (bottom). We can drive the oscillators to the natural limit cycle much faster with control. We notice that as control is introduced in oscillators I and IV, the effectiveness of control is more pronounced with these oscillators. This also follows from figure \ref{fig16} (a) where the correspondence between the Laplacian eigenvalues and oscillator inputs was discussed. Similar analysis can be performed for controlling a multitude of modal perturbations introduced. 

\subsubsection{Total oscillation control}

In this section, we consider the control of total modal oscillations in the flow associated with wake unsteadiness. This requires the control of the temporal coefficients associated with the oscillators ($z_m$) instead of just the modal fluctuations ($z_m^\prime$). Though the oscillator interactions characterized in this work are based on fluctuations with respect to the natural limit cycle (baseline) state, we consider these interactions to be characteristic of the modal oscillations in the flow. In fact, the energy transfer mechanism should be similar for $z_m^\prime$ and $z_m^\prime + z_m^b$. We realize that by suppressing $z_m$, the flow is attracted towards the unstable steady state. Since we can assume that the above networked oscillator based LQR control is applicable near the baseline limit cycle, we expect that there is some region of validity of control to achieve drag reduction. Inhibiting nonlinear energy transfer should remove the energy input to sustain wake oscillations. Hence, we expect to use the same aggregate model extracted to suppress the limit cycle along with fluctuations of the modes to some degree. 

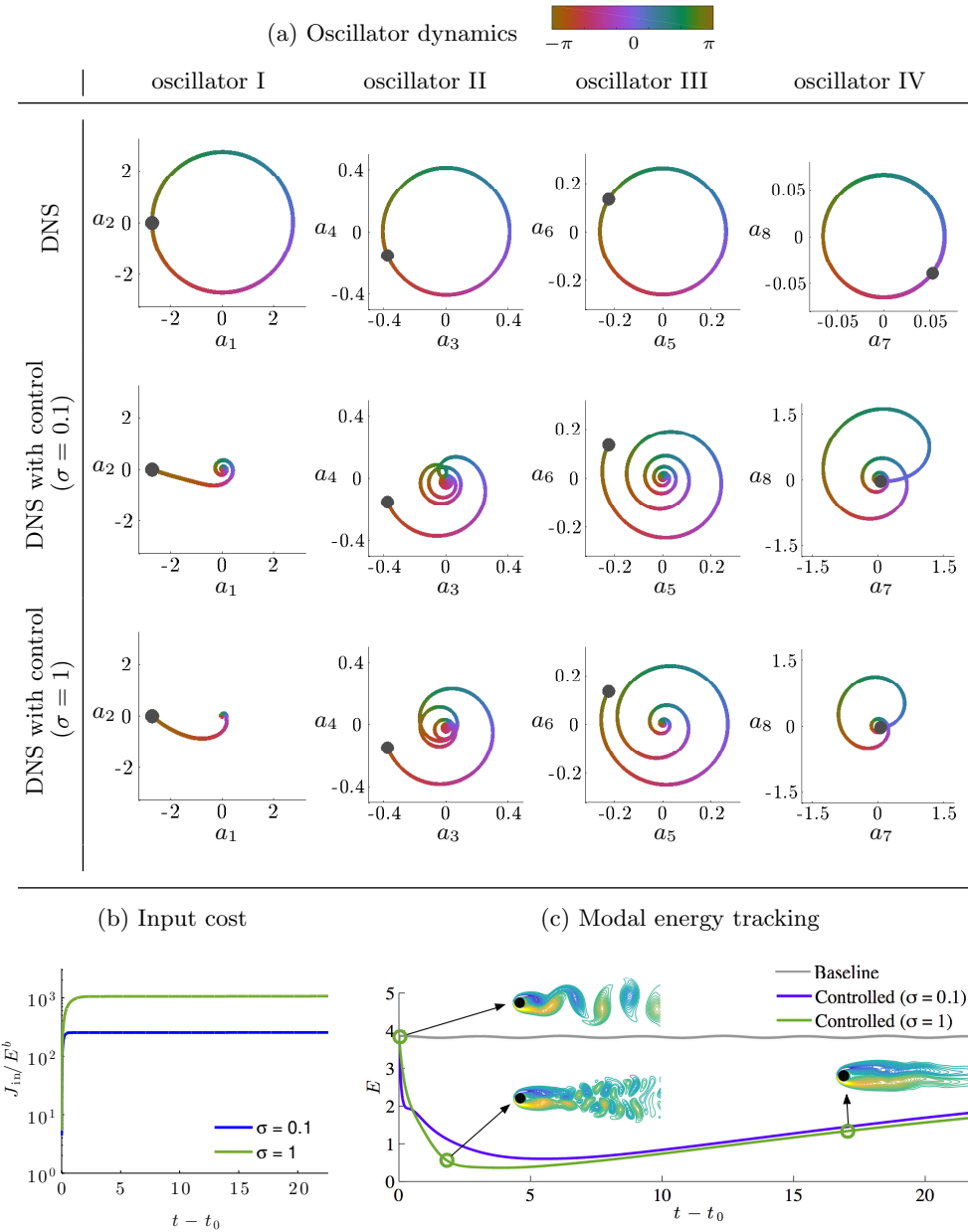
\begin{figure}  
\begin{center}  

	\break
	(a) Oscillator dynamics \includegraphics[width=0.2\textwidth]{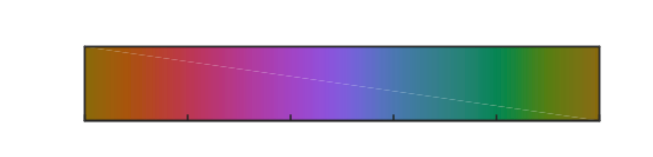}
	\put(-70,-3){\scriptsize $-\pi$}
	   \put(-38,-3){\scriptsize $0$}
	   \put(-10,-3){\scriptsize $\pi$}	   
	\break \break
	\begin{tabular}{c|cccc}	
	\vspace{-0.05in}
	 & ~~oscillator I & oscillator II & oscillator III & oscillator IV \\ \hline
	\rotatebox{90}{\qquad \quad~ DNS} & 
	~~\begin{overpic}[width=0.2\textwidth]{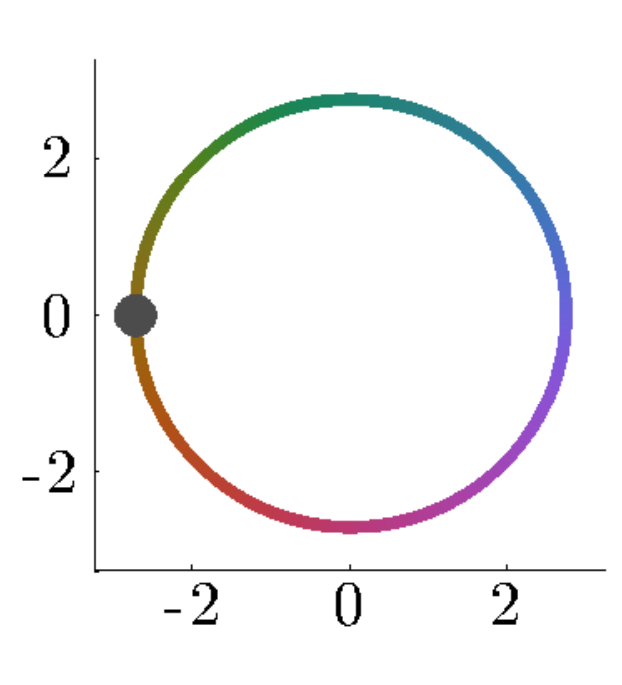}
	   \put(48,-1){\small $a_1$}
	   \put(-7,53){\small $a_2$}
	\end{overpic} &
	\begin{overpic}[width=0.2\textwidth]{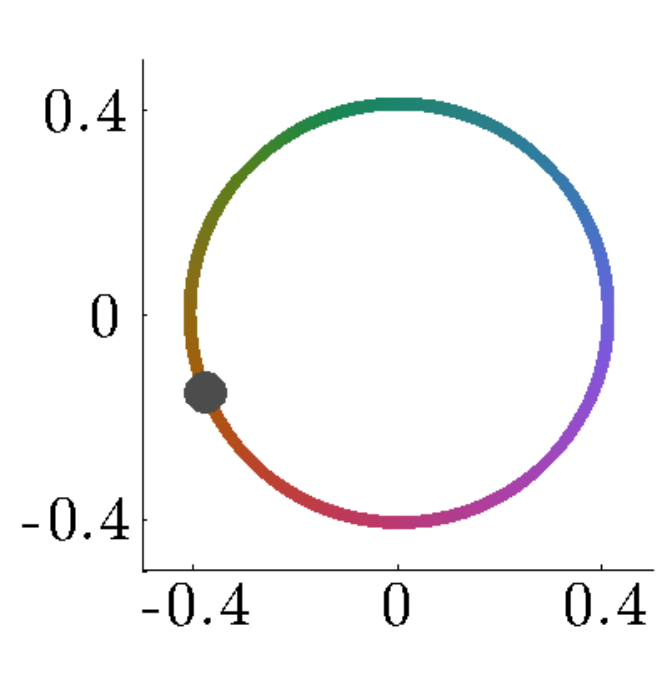}
	\put(54,-1){\small $a_3$}
	\put(-5,53){\small $a_4$}
	\linethickness{3pt}
	\end{overpic} &	
	\begin{overpic}[width=0.2\textwidth]{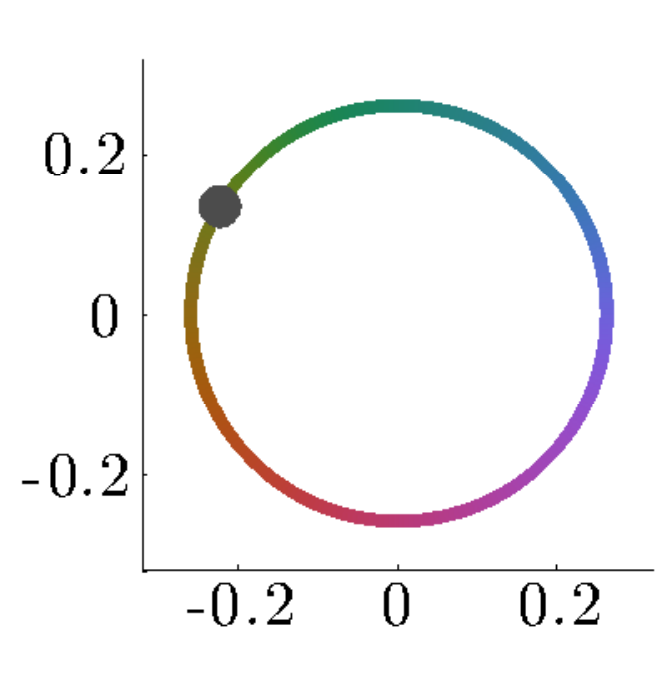}
	\put(54,-1){\small $a_5$}
	\put(-5,53){\small $a_6$}
	\end{overpic} &	
	\begin{overpic}[width=0.2\textwidth]{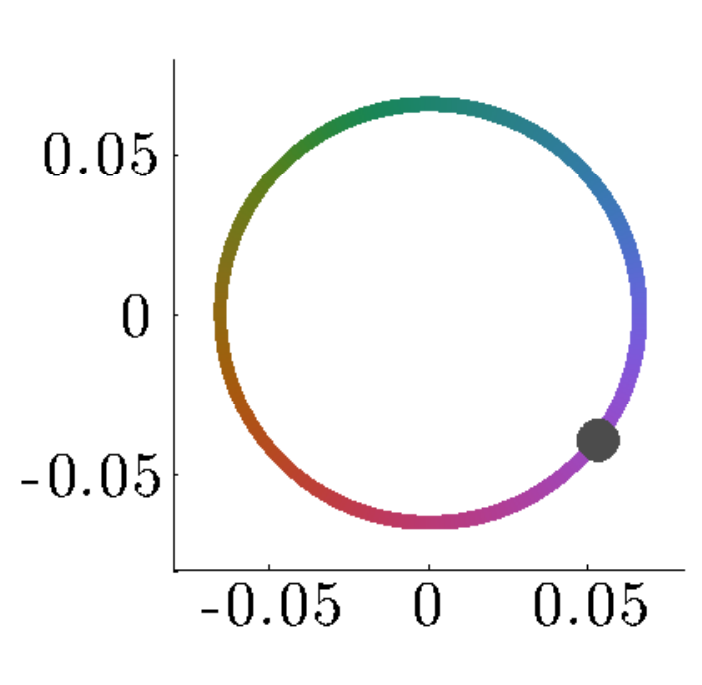}
	\put(54,-1){\small $a_7$}
	\put(-5,53){\small $a_8$}
	\end{overpic} 
	\\
	\rotatebox{90}{\quad ~~DNS with control }
	\rotatebox{90}{\quad ~~~~~~~~~($\sigma= 0.1$)}& 
	~~\begin{overpic}[width=0.2\textwidth]{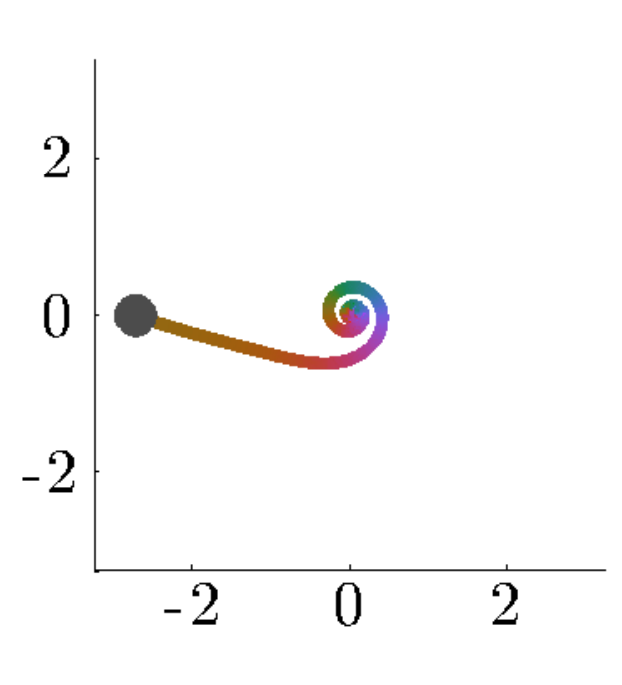}
	   \put(48,-1){\small $a_1$}
	   \put(-7,53){\small $a_2$}
	\end{overpic} &
	\begin{overpic}[width=0.2\textwidth]{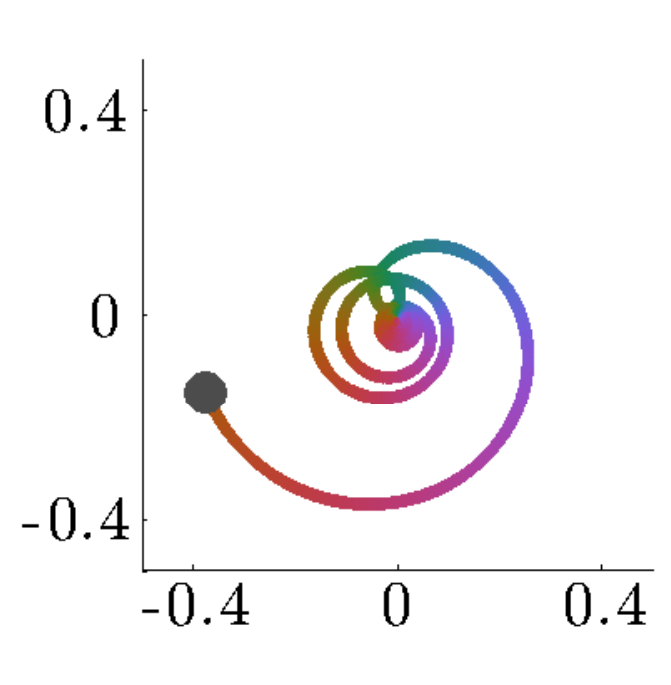}
	\put(54,-1){\small $a_3$}
	\put(-5,53){\small $a_4$}
	\end{overpic} &	
	\begin{overpic}[width=0.2\textwidth]{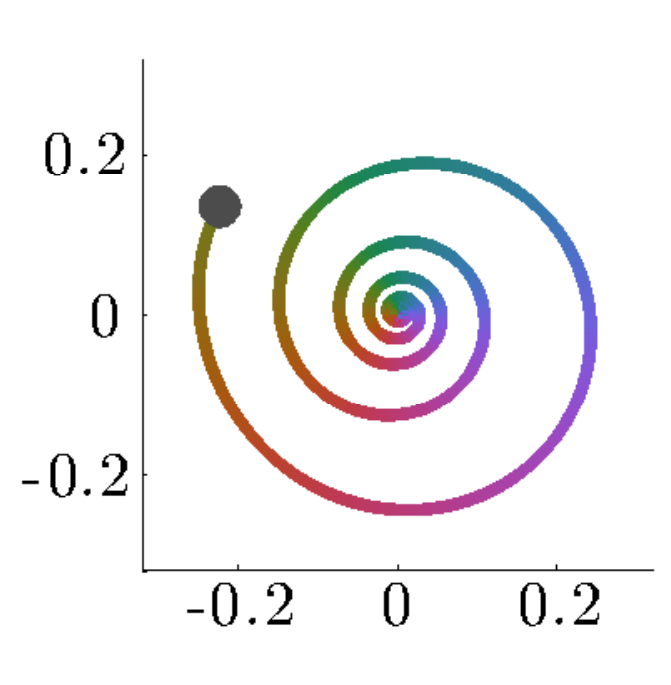}
	\put(54,-1){\small $a_5$}
	\put(-5,53){\small $a_6$}
	\end{overpic} &	
	\begin{overpic}[width=0.2\textwidth]{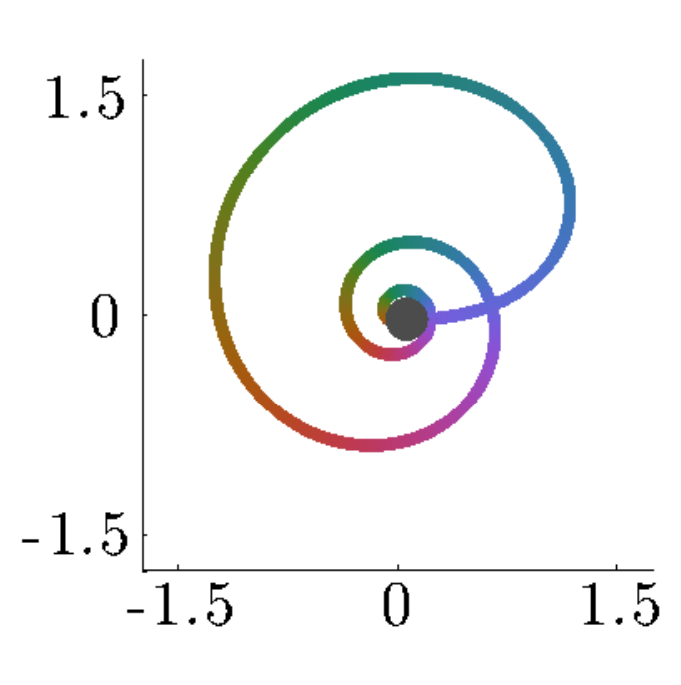}
	\put(54,-1){\small $a_7$}
	\put(-5,53){\small $a_8$}
	\end{overpic}  \\
	\rotatebox{90}{\quad ~~DNS with control }
	\rotatebox{90}{\quad ~~~~~~~~~($\sigma = 1$)}& 
	~~\begin{overpic}[width=0.2\textwidth]{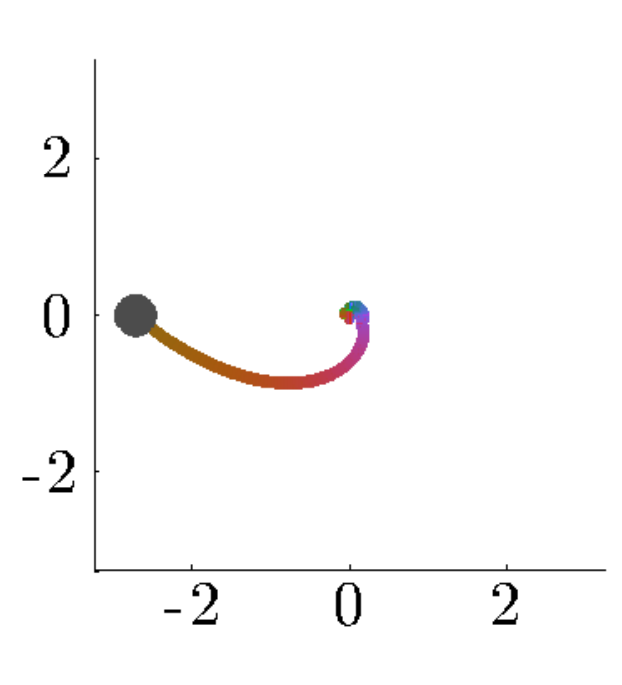}
	   \put(48,-1){\small $a_1$}
	   \put(-7,53){\small $a_2$}
	\end{overpic} &
	\begin{overpic}[width=0.2\textwidth]{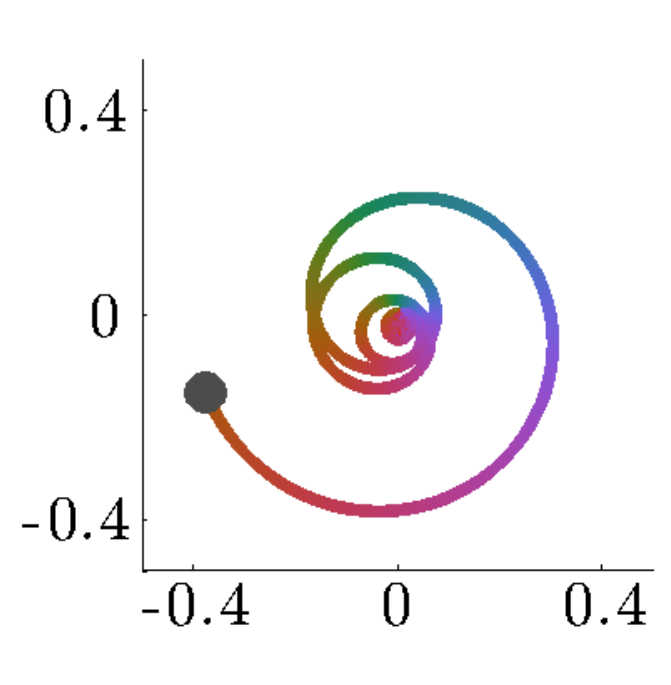}
	\put(54,-1){\small $a_3$}
	   \put(-5,53){\small $a_4$}
	\end{overpic} &	
	\begin{overpic}[width=0.2\textwidth]{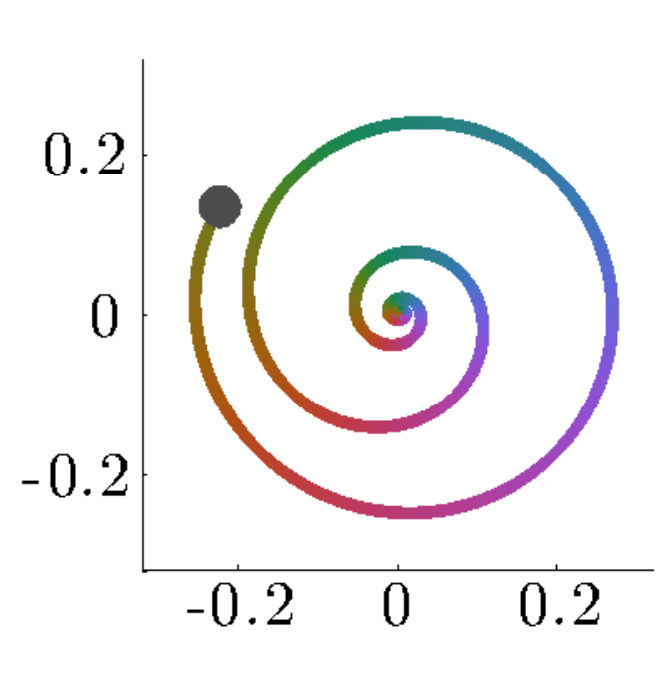}
	\put(54,-1){\small $a_5$}
	   \put(-5,53){\small $a_6$}
	\end{overpic} &	
	\begin{overpic}[width=0.2\textwidth]{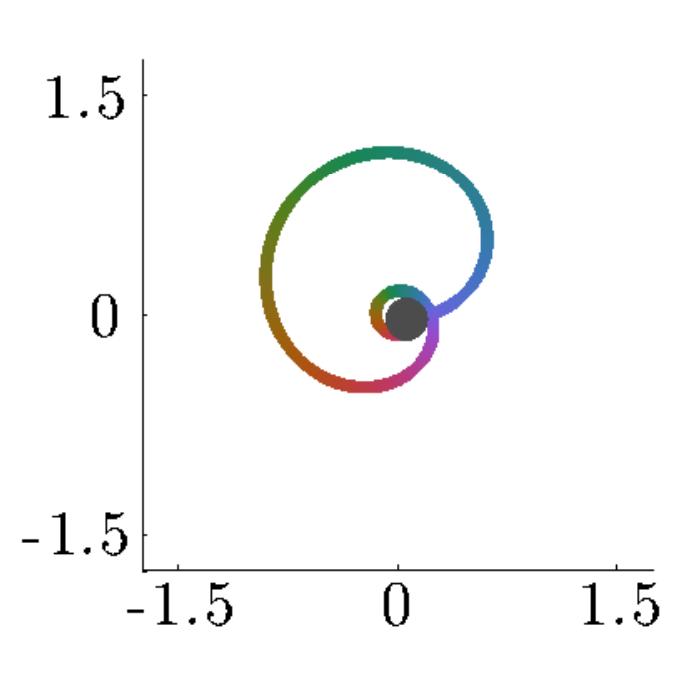}
	\put(54,-1){\small $a_7$}
	   \put(-5,53){\small $a_8$}
	\end{overpic}  \\

	\\ \hline
	\end{tabular}
	\break
	\begin{tabular}{cc}
	\vspace{0.1in}
	(b) Input cost & (c) Modal energy tracking \\	
	\includegraphics[height=0.275\textwidth]{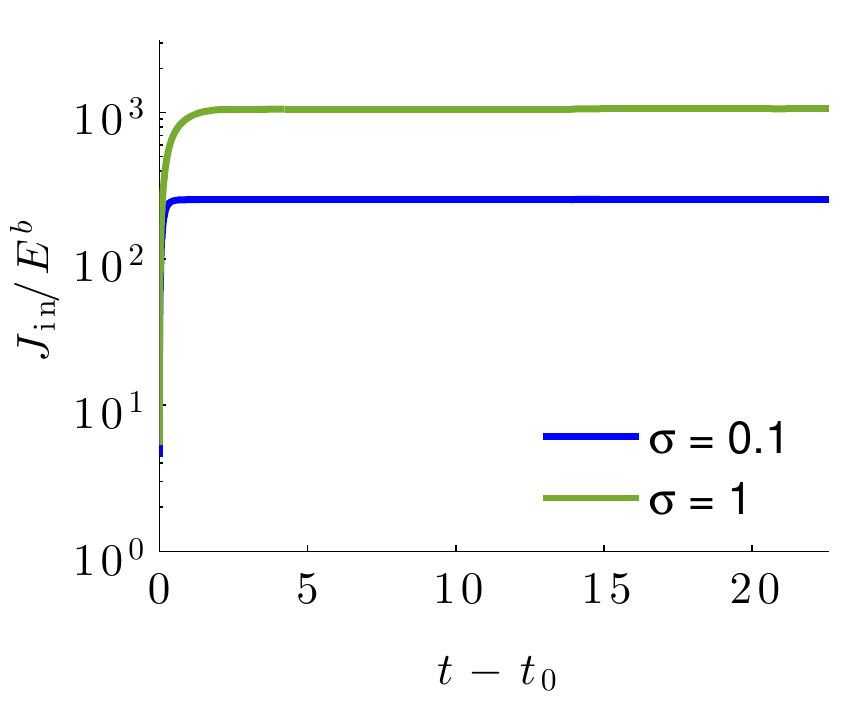}  & 
	\includegraphics[height=0.275\textwidth]{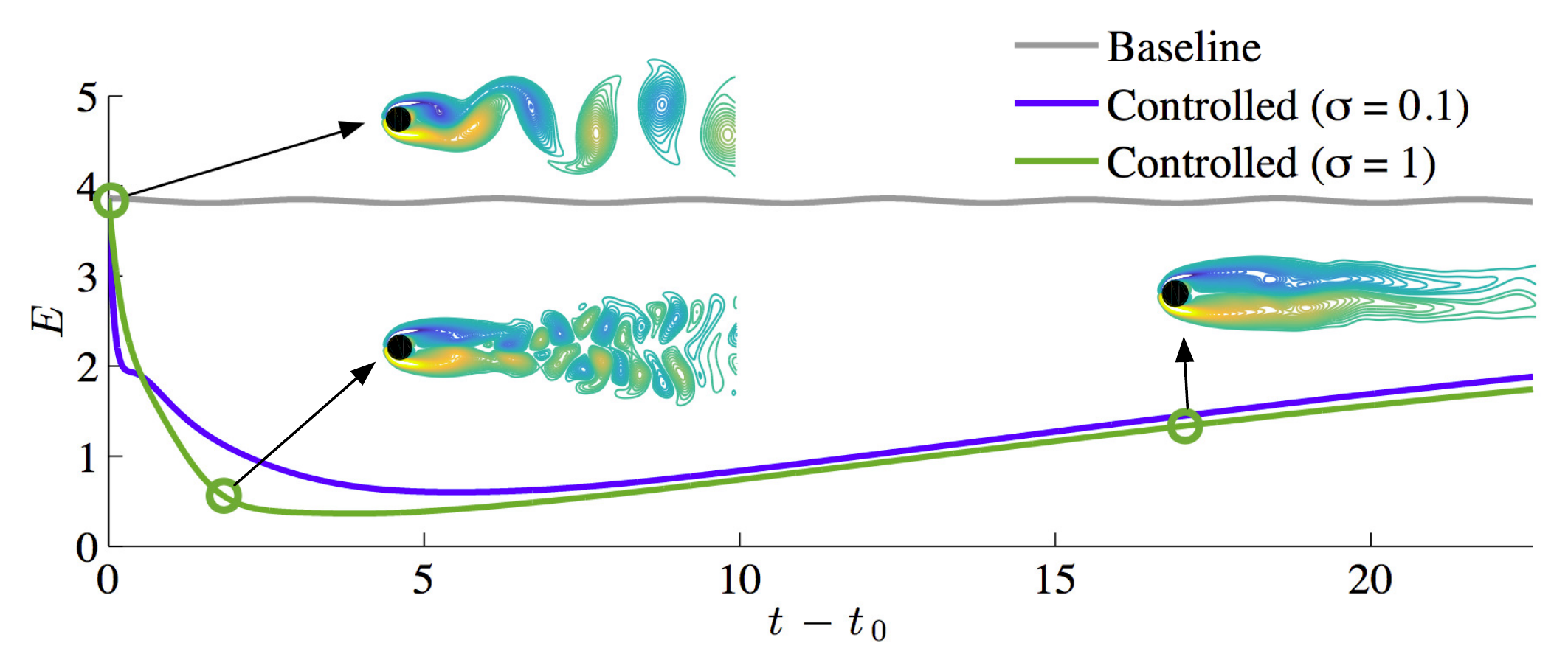}	
	\end{tabular}
 \caption{(a) Oscillator dynamics without control and with control implemented in DNS for suppressing perturbations to limit cycle and (b) associated input control cost and (c) total modal energy tracking for the controlled cases in comparison to baseline energy.}
   \label{fig12}
\end{center}  
\end{figure}

Our objective is to control the temporal coefficients associated with the modes, yielding a control system of the form,
\begin{equation}
\dot{z}_m =  -\sum_{n=\text{I}}^{M} [\boldsymbol{L}_{\mathcal{G}}  - \boldsymbol{B}\boldsymbol{K}]_{mn} z_n.
\label{eq:zetaco1}
\end{equation}
In DNS, this amounts to adding a body force corresponding to $- \boldsymbol{B}\boldsymbol{K}\boldsymbol{z}$. Note that this control is similar to perturbation control, except here the target state for the oscillator temporal dynamics is zero as opposed to the baseline limit cycle. Hence, this control input steers the modal amplitudes corresponding to the baseline limit cycle to zero as shown in figure \ref{fig12} (a). Here, we show the control performance for $\sigma = 0.1$ and $1$. With control, the temporal coefficient associated with oscillator I first decays to zero followed by the higher-order oscillators. The energy lost by oscillator I is compensated with an initial increase in energy associated with oscillator IV. This is expected as oscillator IV is the most influenced node in the network. This highlights the energy transfers from lower-order oscillators to higher-order oscillators. As control input is added to oscillator IV, we suppress the corresponding modal oscillations. We show the input cost of control in figure \ref{fig12} (b). We see that the input cost for control, $J_\text{in} = \int_0^\infty\boldsymbol{v}(t)^T \boldsymbol{S} \boldsymbol{v}(t)\rm{d}{\it t}$,  is lower for the $\sigma = 0.1$ case compared to the $\sigma = 1$ case.

We show the suppression of total modal energy with control in figure \ref{fig12} (c). Initially, we see that modal energy decreases faster with $\sigma =  0.1$. However, higher level of modal energy is suppressed with $\sigma = 1$ using higher input cost. This is also supported by our observation of the maximum pole movement in figure \ref{fig16} (b). Once the modal amplitudes are forced to zero, the total modal energy reduces to zero. That is, the flow returns to the mean flow along with remnants of small structures corresponding to oscillator IV associated with increase in energy as shown in figure \ref{fig12} (c). This is expected as the unsteady flow field is decomposed into mean flow and modal components, so that forcing the modal components to zero reduces the flow to the mean. However, the mean flow is not an equilibrium and a shift in the mean flow is observed as time progresses. This mean flow deformation is attributed to the Reynolds stress generated by the modal fluctuations which modifies the base flow  \citep{Brunton2015amr}. The change in the base flow leads to a corresponding decrease in modal energy until equilibrium is achieved. In the case of the cylinder flow problem, the flow tends towards the unstable steady state.

\begin{figure}
   \begin{center}
    \begin{tabular}{c} 
          \begin{overpic}[width=4.5in]{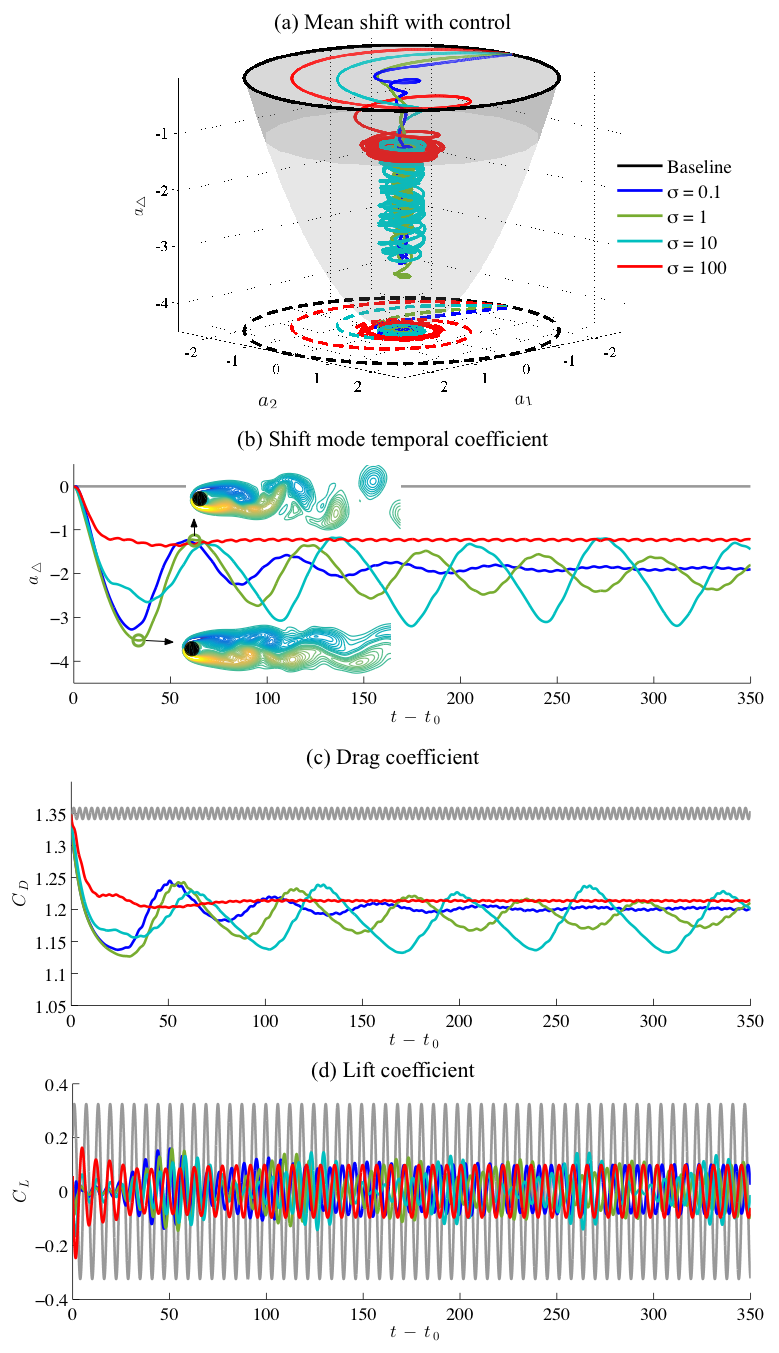}
          \end{overpic} 
    \end{tabular}
   \end{center}
   \caption{(a) Shift mode temporal dynamics on parabolic inertial manifold for a range of $\sigma$ and (b) shift mode temporal dynamics in the time domain. (c) Drag and (d) lift forces compared to the baseline.}
   \label{fig13}
\end{figure}

As discussed earlier, the mean shift between the mean flow and the unstable steady state can be described by the shift mode. The temporal coefficient ($a_\triangle$) corresponding to the shift mode ($\boldsymbol{u}_\triangle$) can be obtained by projection as $a_\triangle = \left<\boldsymbol{u}-\bar{\boldsymbol{u}},\boldsymbol{u}_\triangle\right>$. We also perform control considering $\sigma= 10$ and $100$. The variation of shift mode temporal coefficients with respect to temporal coefficients of the dominant POD modes for the range of $\sigma$ considered is shown in figure \ref{fig13} (a). For reference, the evolution of the flow from unstable steady state to the mean flow that follows a parabolic inertial manifold is shown in grey \citep{Noack:JFM03}. The darker grey region of the manifold indicates the region of effective control using the networked oscillator model. In this region, the nonlinear energy transfers are inhibited successfully to reduce wake oscillations. Thus, in suppressing the modal oscillations, we achieve a mean shift in the flow exhibited by the shift mode temporal coefficient.

\begin{figure}
   \begin{center}
    \begin{tabular}{cc} 
    (a) Drag power & (b) Actuation power \\
     \hspace{-0.175in}
          \begin{overpic}[width=2.7in]{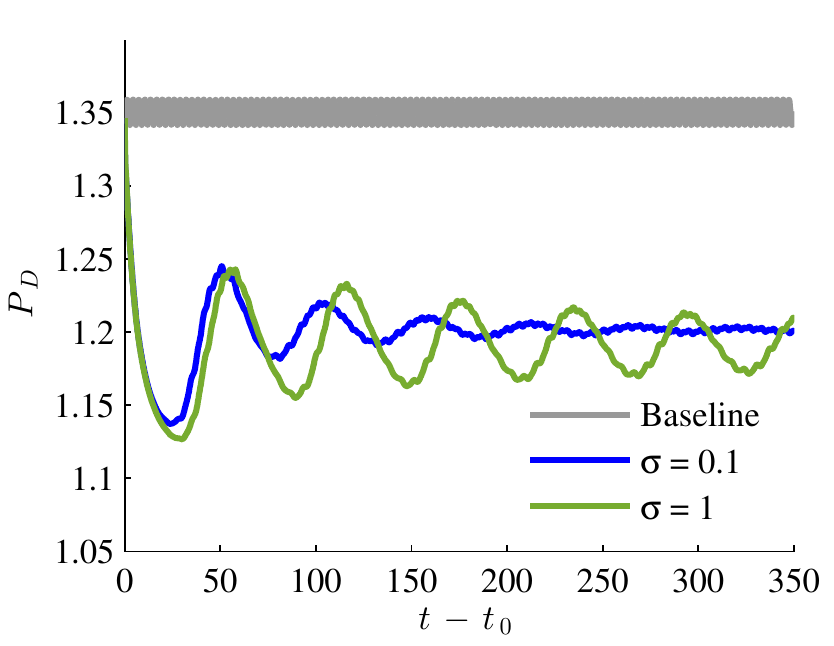}
          \end{overpic} &
          \hspace{-0.1in}
          \begin{overpic}[width=2.7in]{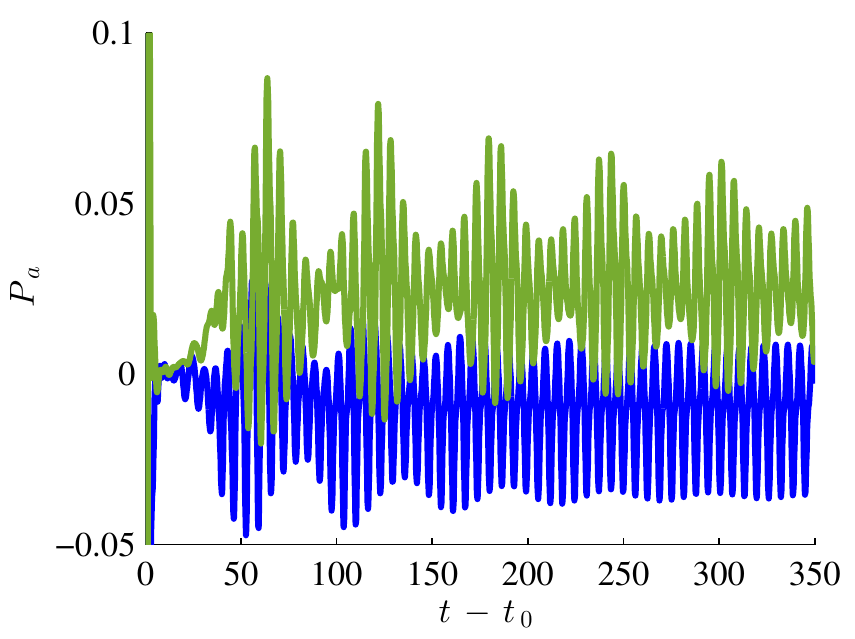}
          \end{overpic}
    \end{tabular}
   \end{center}
   \caption{(a) Drag power and (b) actuation power required for total oscillation control.}
   \label{fig18}
\end{figure}

We also present the temporal dynamics associated with the shift mode in figure \ref{fig13} (b). It can be observed that initially $a_\triangle = 0$ in the limit cycle and as control is applied, the shift mode temporal coefficient decreases and approaches the unstable steady state. As $\sigma$ decreases, we achieve a steeper decrease in $a_\triangle$.  Examining the long time history of the temporal variation of the shift mode, the choice of lower $\sigma$ results in low frequency oscillations in $a_\triangle$. These low frequency oscillations are attributed to the exchange of energy between the low and moderate drag states \citep{Munday:PF13}. For control effort corresponding to $\sigma = 100$, a steady state with application of control is achieved more rapidly. 

We also evaluate the drag power $P_D$ and the invested actuation power for control $P_a$ given by
\begin{equation}
P_D = \frac{F_D U}{\frac{1}{2}\rho U^3 d} \quad \text{and} \quad P_a = \frac{\left<\boldsymbol{f} \cdot \boldsymbol{u}\right>}{\frac{1}{2}\rho U^3 d},
\label{powercoeff}
\end{equation}
where $\boldsymbol{f}$ is the actuation force added in DNS. The drag and actuation power required for $\sigma = 0.1$ and $1$ are shown in figure \ref{fig18} (a) and (b), respectively. As expected, the drag power for the controlled cases are lower than the baseline. We observe that the actuation power needed at steady state with control is significantly lower that the drag power. This results from suppression of the modal oscillations associated with the flow. 

The networked oscillator control framework considered here is based on interactive dynamics of the baseline POD modes and does not include any other modes. We note that the shift mode itself can be incorporated into the controller \citep{noack2011reduced,luchtenburg2009generalized}. The shift mode is not added to the current formulation but such extension can be realized. To force the flow completely to the unstable steady state, additional feedback controller about the unstable steady state can be introduced. Details on such a patch controller design is given in appendix B.

 We evaluate the unsteady forces on the cylinder with application of control compared to the baseline drag and lift coefficient variation in figures \ref{fig13} (c) and (d), respectively. The variation in the drag coefficient is similar to the shift mode variation as discussed previously. Almost a $12\%$ reduction in drag and $70\%$ reduction in lift fluctuation are achieved with the application of control. We can thus reduce unsteady forces on the cylinder by suppressing modal oscillations in the flow and inhibiting energy transfers therein. Using a networked oscillator model in conjunction with optimal control, we can control energy transfer dynamics effectively for unsteady fluid flows. 


\section{Concluding Remarks}
 We construct a networked oscillator model for describing modal interactions in unsteady fluid flows. The modal oscillators comprised of POD conjugate mode pairs constitute the nodes on the modal network. The interactions between the oscillators form edges of the network, which are characterized by analyzing impulse-response to the fluid flow. Perturbations are introduced in the modal oscillators and their transfer of perturbation energy over the network is studied. The amplitude and phase perturbations introduced in the Navier--Stokes equations are tracked using regression techniques to develop the networked oscillator framework. The networked oscillator model is a linear approximation to the nonlinear modal interactions in unsteady fluid flows. 
 
Using a canonical example of unsteady flow over a cylinder, the energy transfer dynamics are highlighted. Agreement of the model with DNS is observed for both amplitude and phase perturbation cases. The networked oscillator model captures the  modal perturbation dynamics better compared to the empirical Galerkin formulation.  An aggregate network model capturing general oscillator interactions in the flow is obtained using combined regression techniques over a collection of perturbed cases. The aggregate network model is attributed with the least error in terms of prediction of the temporal dynamics of the modal oscillators. The degree of the nodes corresponding to the aggregate network structure provides insights on the importance of individual oscillators to energy transfers and the overall system dynamics.  
 
With the knowledge of network interactions between oscillators, an optimal feedback control strategy is designed to suppress oscillator fluctuations with respect to the natural limit cycle faster. A judicious choice of input oscillators forced is made by examining the movement of the graph Laplacian poles with LQR design. On control of modal perturbations in the flow, faster suppression of the oscillator temporal dynamics compared to the baseline limit cycle is observed. Controlling the overall fluctuations of the oscillators resulted not only in the suppression of the modal oscillations but also in a mean shift towards the unstable steady state exhibited by the shift mode. As energy transfers that sustain wake oscillations were inhibited, a mean shift of the flow to a lower energy state inside the parabolic inertial manifold is observed. The mean shift correspondingly led to a reduction in the unsteady forces on the cylinder. A $12\%$ reduction in drag force and a $70\%$ reduction in lift force are achieved with feedback flow control. In the long-time behavior of the controlled flow, a competition between low and moderate drag states are observed. 

The networked oscillator modeling and control approach shown here has leveraged the knowledge of modal interactions providing insights beyond traditional approaches. Controlling the modal interactions at a fundamental level motivates analogous studies using localized actuators and limited sensors for modeling and controlling  unsteady fluid flows.   
\label{sec:cr}

%
\section*{Acknowledgements}

We are grateful for the support by the U.S. Air Force Office of Scientific Research (Award number: FA9550-16-1-0650; Program Manager: Dr. Douglas R.~Smith).

\renewcommand{\thefigure}{A\arabic{figure}}
\setcounter{figure}{0}
\section*{Appendix A}

\begin{figure}
   \begin{center}
    \begin{tabular}{c} 
          \begin{overpic}[width=5in]{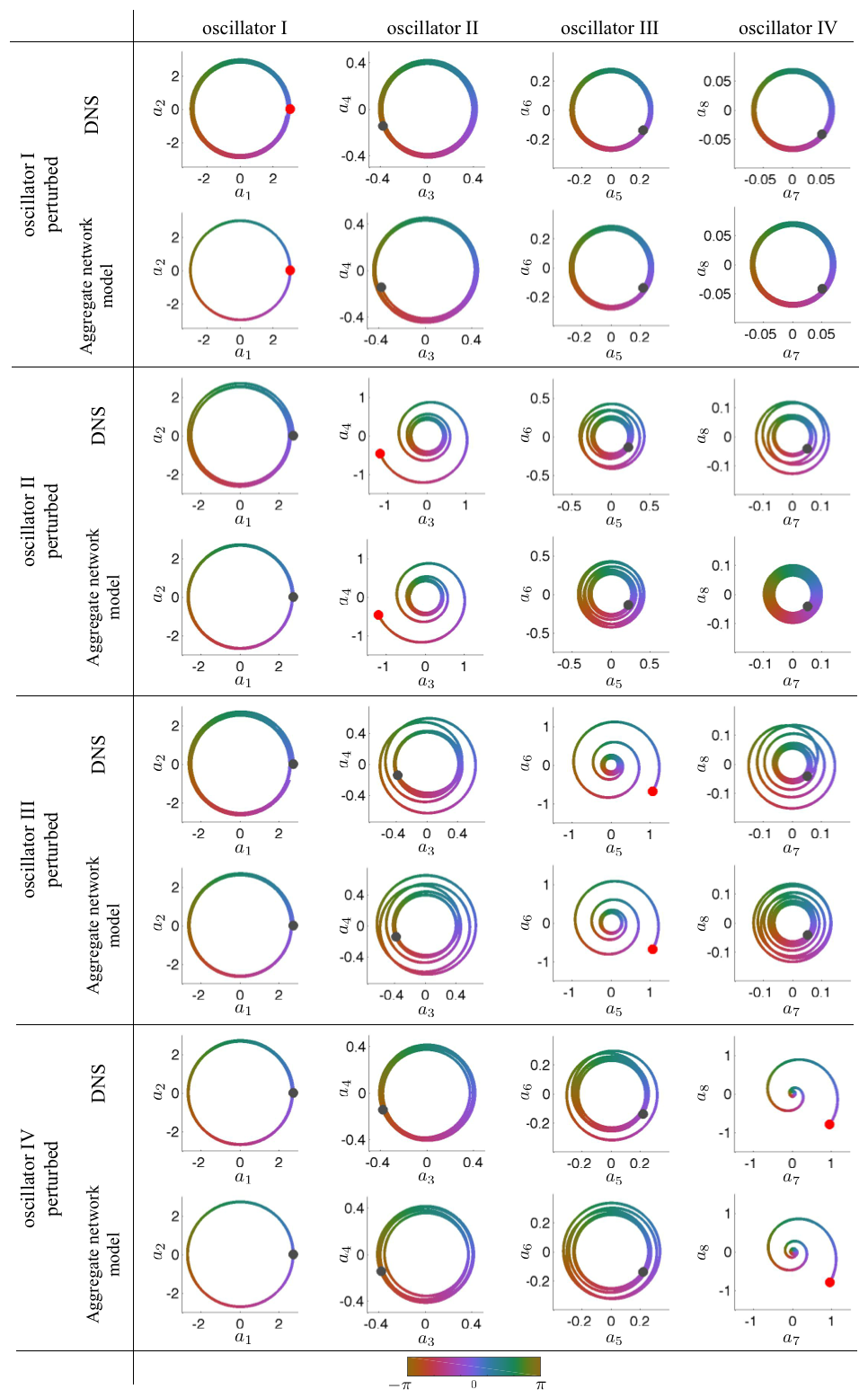}
          \end{overpic} 
    \end{tabular}
   \end{center}
   \caption{Oscillator dynamics of DNS and aggregate networked oscillator model for amplitude perturbations to oscillators corresponding to energy input of $\beta_m = 0.2$ with $\theta_m^\prime(t_0) = 0$.}
   \label{fig8}
\end{figure}

\begin{figure}
   \begin{center}
    \begin{tabular}{c} 
          \begin{overpic}[width=5in]{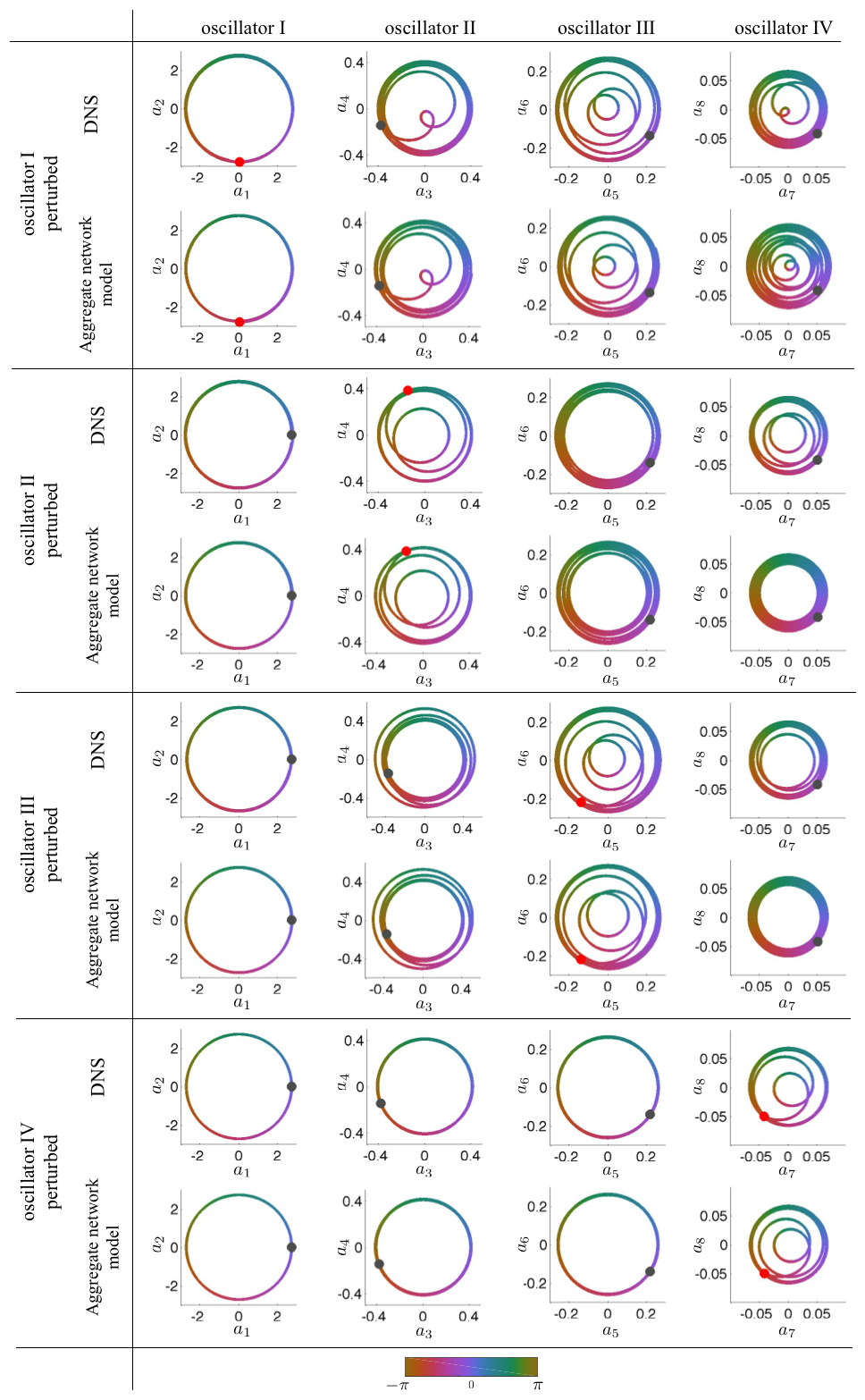}
          \end{overpic} 
    \end{tabular}
   \end{center}
   \caption{Oscillator dynamics of DNS and aggregate networked oscillator model for phase perturbations to oscillators corresponding to $\theta_m^\prime(t_0) = -\pi/2$ with $\beta_m = 0$.}
   \label{fig9}
\end{figure}

The performance of the aggregate model is shown in figures \ref{fig8} and \ref{fig9}. DNS comparison of the oscillator dynamics using the aggregate network oscillator model with introduction of amplitude perturbations corresponding to $\beta_m = 0.2$ on each oscillation is shown in figure \ref{fig8}. The comparison of predicted trajectories for phase perturbation of $\theta_m^\prime(t_0) = -\pi/2$ on each oscillator is shown in figure \ref{fig9}.

\renewcommand{\thefigure}{B\arabic{figure}}
\setcounter{figure}{0}

\begin{figure}
   \begin{center}
    \begin{tabular}{c} 
          \begin{overpic}[width=4.5in]{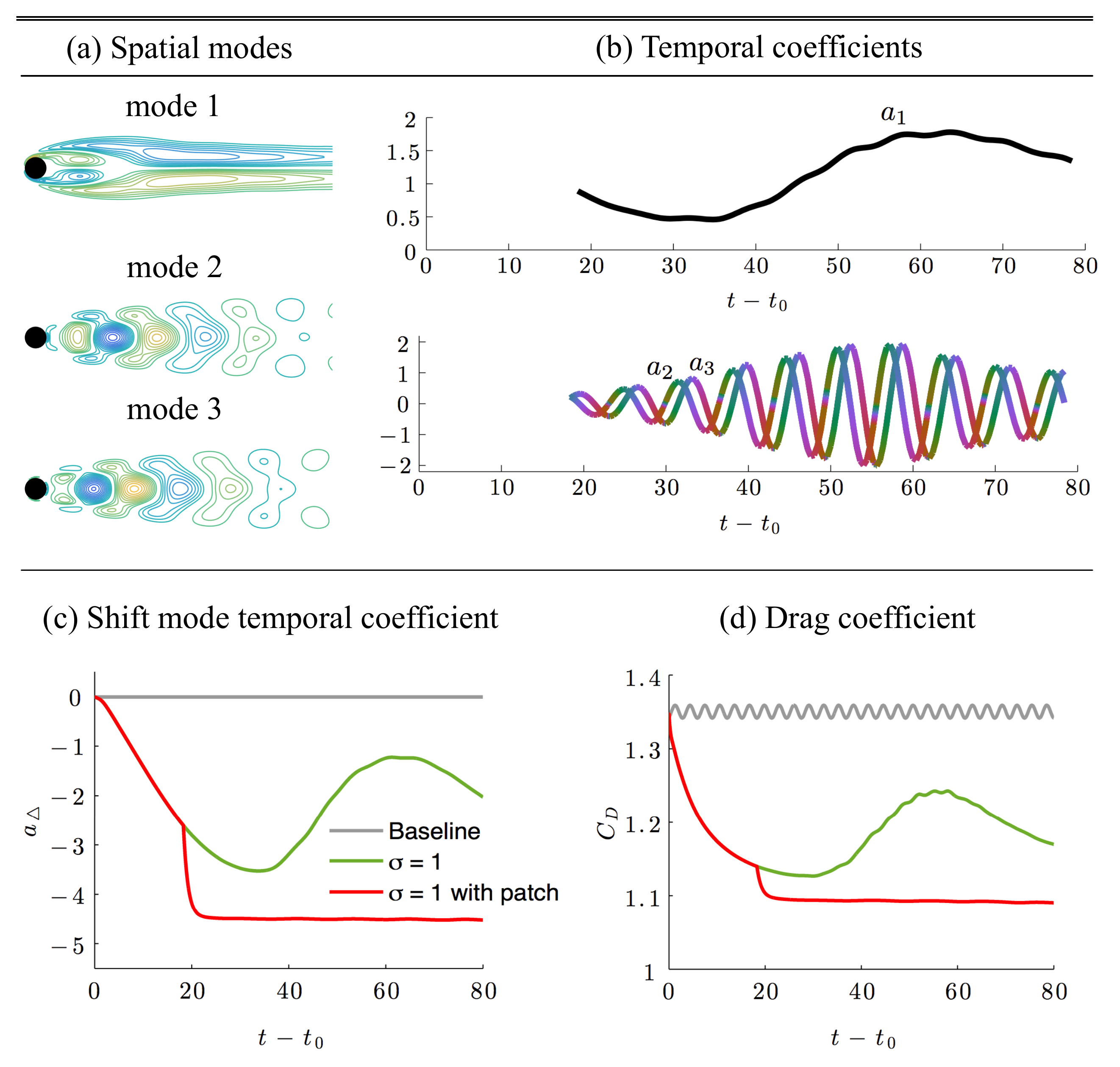}
          \end{overpic} 
    \end{tabular}
   \end{center}
   \caption{(a) Spatial modes and (b) corresponding temporal coefficient from POD decomposition near unstable steady state  for $\sigma = 1$ controlled case. (c) Shift mode temporal coefficient and (c) drag coefficient with application of second patch controller.}
   \label{fig17}
\end{figure}

\section*{Appendix B}

To force the flow to the unstable steady state, we first perform a POD decomposition of the collected velocity snapshots $\boldsymbol{u}$ after the application of control near the unstable steady state $\boldsymbol{u}^*$ as
 \begin{equation}
   \boldsymbol{u}(\boldsymbol{x},t)  = \boldsymbol{u}^*(\boldsymbol{x}) + \sum_{j=1}^N a_j(t) \boldsymbol{\phi}_j^{\boldsymbol{u}}(\boldsymbol{x}).
   \end{equation}

In particular, here we consider the $\sigma= 1$ case and collect velocity snapshots in the time interval between $t = 19.3$ and $80$. During this time interval, we start to observe oscillations in the shift mode temporal coefficient. The extracted spatial modes and temporal coefficients are shown in figure \ref{fig17} (a) and (b) respectively. Here, we show the dominant POD modes. We can see that mode $1$ shares similarities with the shift mode. Although modes $2$ and $3$  correspond to a frequency similar to oscillator I, the spatial modal structure is different. Here, we consider the first $10$ modes for control. We can construct a linearized model by curve-fitting the temporal coefficients. Designing an additional controller (patch controller) based on these coefficients, we can force the flow to the unstable steady state and reach minimum drag performance as shown in figure \ref{fig17} (c) and (d).

\bibliographystyle{jfm}
\bibliography{oscillators_ref}

\end{document}